\documentclass[preprint,12pt,authoryear]{elsarticle}

\usepackage{amssymb}
\usepackage{amsthm}
\usepackage{wrapfig}
\usepackage{hyperref}
\usepackage{natbib}
\usepackage{amsmath}
\usepackage{gensymb}
\usepackage{tikz}
\usepackage{pgfplots}
\usetikzlibrary{calc}

\usepackage[ruled]{algorithm2e} 

\SetAlFnt{\small}
\SetAlCapFnt{\small}
\SetAlCapNameFnt{\small}
\SetAlCapHSkip{0pt}

\usepackage{array}

\usepackage{dsfont}
\usepackage{amsthm}
\usepackage{overpic}
\usepackage{contour}
\contourlength{1pt}
\usepackage{xspace}
\usepackage{sidecap}
\usepackage{rotating}
\usepackage[normalem]{ulem}
\usepackage{booktabs} 
\usepackage{caption}
\usepackage{subcaption}
\usepackage{microtype}
\usepackage{booktabs}
\usepackage{enumitem}
\usepackage{amsthm}
\usepackage{wrapfig}
\usepackage{multirow}
\usepackage{tabularx}
\usepackage{nicefrac}
\usepackage{algpseudocode}
\usepackage{chngcntr}
\usepackage{apptools}
\usepackage[super]{nth}
\AtAppendix{\counterwithin{lemma}{section}}

\usepackage{ulem}

\newcommand{\RR}{\mathbb{R}}

\newcommand{\QQQ}{\mathcal{Q}}

\newcommand{\ZZ}{\mathbb{Z}}

\newcommand{\MM}{\mathcal{M}}
\newcommand{\VV}{\mathcal{V}}

\newcommand{\EE}{\mathcal{E}}

\newcommand{\FF}{\mathcal{F}}

\newlength{\indentlaenge}
\newlength{\mylength}
\newlength{\mylengthzwei}
\usepackage{enumitem}
\makeatletter
\newcommand{\myenumlabel}[2]{#2\def\@currentlabel{#2}\label{#1}}
\makeatother

\makeatletter
\newcommand{\specialcell}[1]{\ifmeasuring@#1\else\omit$\displaystyle#1$\ignorespaces\fi}
\makeatother
\newcommand{\pushright}[1]{\ifmeasuring@{#1}\else\omit\hfill$\displaystyle{#1}$\fi\ignorespaces}

\renewcommand{\eqref}[1]{Eq.~(\ref{#1})}
\newcommand{\figref}[1]{Fig.~\ref{#1}}
\newcommand{\secref}[1]{Sec.~\ref{#1}}

\def\cput(#1,#2)#3{\put(#1,#2){\hbox to 0pt{\hss{#3}\hss}}}
\def\lput(#1,#2)#3{\put(#1,#2){\hbox to 0pt{\hss{#3}}}}
\def\rput(#1,#2)#3{\put(#1,#2){\hbox to 0pt{{#3}\hss}}}

\journal{}

\begin{document}

\begin{frontmatter}


\title{Fabrication-Aware Strip-Decomposable Quadrilateral Meshes}

\author[1]{Ioanna Mitropoulou}
\author[2]{Amir Vaxman}
\author[3]{Olga Diamanti}
\author[1]{Benjamin Dillenburger}

\affiliation[1]{organization={ETH Zurich},
            country={Switzerland}}

\affiliation[2]{organization={University of Edinburgh},
            country={United Kingdom}}

\affiliation[3]{organization={TU Graz},
            country={Austria}}

\begin{abstract}
Strip-decomposable quadrilateral (SDQ) meshes, i.e., quad meshes that can be decomposed into two transversal strip networks, are vital in numerous fabrication processes; examples include woven structures, surfaces from sheets, custom rebar, or cable-net structures. However, their design is often challenging and includes tedious manual work, and there is a lack of methodologies for editing such meshes while preserving their strip decomposability. We present an interactive methodology to generate and edit SDQ meshes aligned to user-defined directions, while also incorporating desirable properties to the strips for fabrication. 
Our technique is based on the computation of two coupled transversal tangent direction fields, integrated into two overlapping networks of strips on the surface.
As a case study, we consider the fabrication scenario of robotic non-planar 3D printing of freefrom shell surfaces and apply the presented methodology to design and fabricate non-planar print paths.

\end{abstract}

\begin{graphicalabstract}
\begin{figure}[h]
  \centering
  \includegraphics[width=\linewidth]{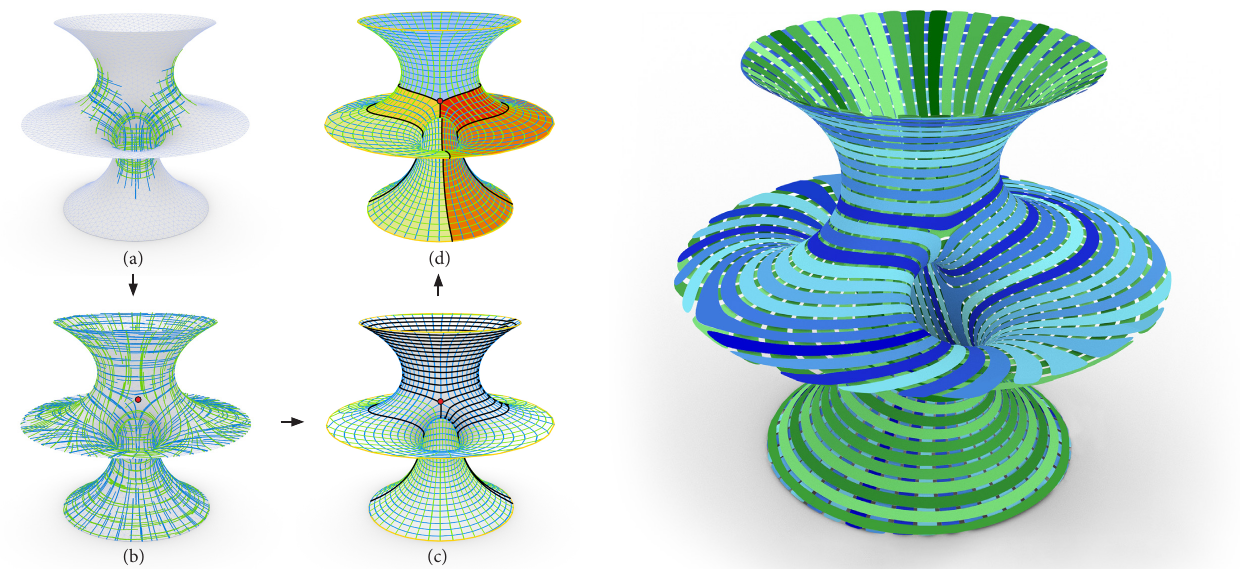}
  \caption{We generate strip-decomposable quad (SDQ) meshes, i.e., meshes that can be decomposed into two coupled transversal strip networks, with embedded fabrication constraints. We exemplify our method on the Costa minimal surface. (a) The input consists of a triangle mesh and a sparse set of tangent directional constraints: 10\% of input faces are constrained to principal curvature directions. (b) We optimize two transversal $2$-vector fields for fabrication constraints and integrability. (c) The fields are integrated into a SDQ mesh. Misaligned singularities cause winding strips (black) that make the fabrication infeasible. (d) We address this with a few topological editing operations that lead to a good patch layout. Right: Visualization of the two final strip networks in blue and green. }
  \label{fig:artefacts}
\end{figure}
\end{graphicalabstract}

\begin{highlights}
\item A methodology for generating strip-decomposable quad (SDQ) meshes by integrating a pair of transversal tangent 2-vector fields.
\item A set of editing operations for improving the strips topology and patch layout of an SDQ mesh. 
\item The application of SDQ meshes for robotic non-planar 3D printing of shells.
\end{highlights}

\begin{keyword}
strip-decomposable quad mesh \sep strip network \sep vector field \sep fabrication-aware \sep print path design \sep robotic 3D printing \sep non-planar 
\end{keyword}

\end{frontmatter}



\section{Introduction}
\label{sec:intro}

The representation of free-form surfaces using \emph{strips} has attracted significant attention for their various applications in architecture and manufacturing, such as the approximation of architectural envelopes using single-curved panels \cite{Pottmann_2008_Freeform_Surfaces_from_Single_Curved_Panels}, the fabrication of surfaces from flat sheets \cite{Takezawa_2016_Fabrication_of_Freeform_Objects_by_Principal_Strips}, the covering of surfaces with textiles \cite{Hilo_2017_prototype_website}, or the design and fabrication of corrugations/pleats \cite{Fornes_2017_Chrystalis_Amphitheater}. A free-form surface can be partitioned into strips using a parametrization along one direction discretized on a regular interval. We denote the collection of strips and their topology as a \emph{strip network}.

\begin{figure}[h]
  \centering
  \includegraphics[width=\linewidth]{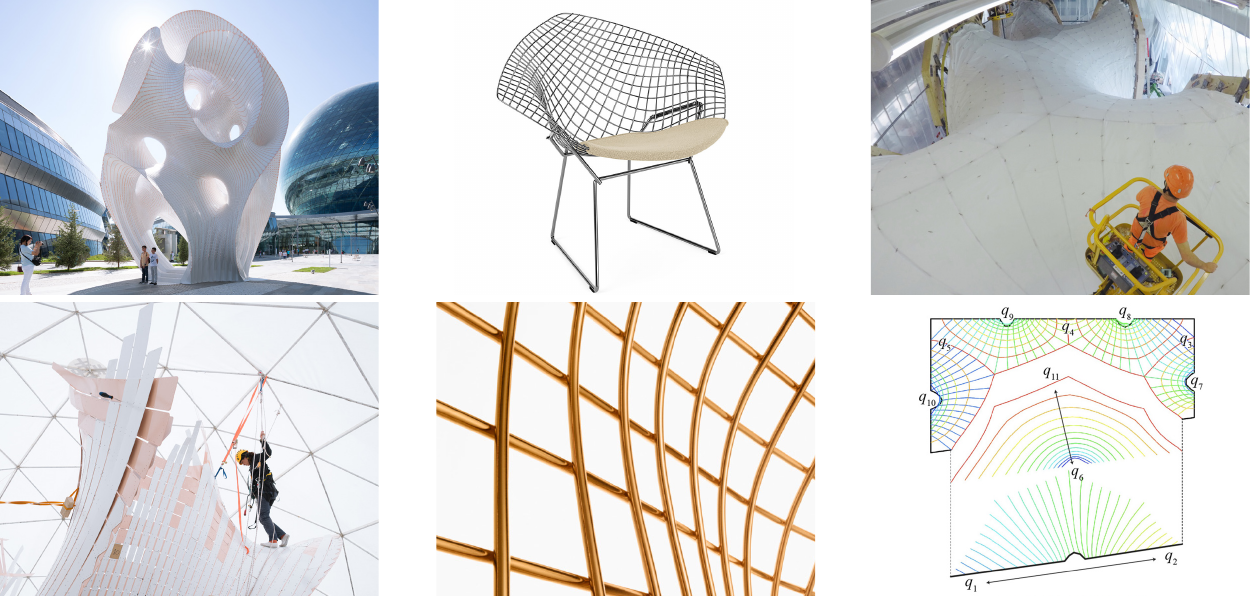}
  \caption{Examples of applications of SDQ meshes. Left: Minima Maxima world expo pavilion, by TheVeryMany, Marc Fornes \cite{Fornes_2017_Minima_MAxima}. Middle: Bertoia diamond chair, by Harry Bertoia \cite{Bertoia_1952_Diamond_chair}. Right: NEST HiLo roof - Full-scale construction prototype, by Block Research Group \cite{Veenendaal2015_Hilo, VanMele2022_Hilo_cable_net, Hilo_2017_prototype_website}.}
  \label{fig:applications_of_SDQ_meshes}
\end{figure}

A one-dimensional parametrization, however, overlooks information pertaining to the orthogonal direction, which can be instrumental for various design and fabrication scenarios. Examples of such scenarios include woven structures, surfaces from sheets in crossing directions, warp-weft fabrication, custom rebar, cable-net structures, and the design of secondary transversal structural elements on structural strip patterns. In \emph{Minima-Maxima} \cite{Fornes_2017_Minima_MAxima} (\figref{fig:applications_of_SDQ_meshes}-left) a freeform surface is assembled from flat metal sheets forming two transversal strip networks that do not intermix. The \emph{Diamond chair} \cite{Bertoia_1952_Diamond_chair} (\figref{fig:applications_of_SDQ_meshes}-middle), which is fabricated using custom rebar, relies on the arrangement of metal rods along a crossing pattern where each transversal direction is realized with continuous rods. Finally, the \emph{NEST HiLo roof} \cite{Veenendaal2015_Hilo, VanMele2022_Hilo_cable_net, Hilo_2017_prototype_website} (\figref{fig:applications_of_SDQ_meshes}-right) is built with a cable-net and fabric formwork system, where two transversal strip networks are used both in the design for the assignment of force densities (bottom) and in the fabrication for the covering of the structure using fabric strips (top). More examples of applications of transversal strip networks are discussed in \cite{Oval_BRG_2021_Two-Colour_Topology_Finding_of_Quad-Mesh_Patterns}.

In this work, we present a methodology for generating and editing transversal strip networks aligned with user-defined directions, while also enabling the incorporation of desirable properties for the fabrication of the strips. The type of geometries we address is relatively smooth geometries with open boundaries and without significant surface detail features.

The control over strip alignment has diverse applications. For instance, aligning strips with stress directions can enhance the mechanical properties of the resulting layout \cite{Schiftner_2010_Statics-Sensitive_Layout_of_Planar_Quadrilateral_Meshes}. Alignment with principal curvatures can yield a natural appearance, accentuate features, and facilitate the creation of nearly planar quads \cite{Liu_2006_Conical_meshes_and_developable_surfaces}. Moreover, alignment with user-defined curves enables a higher degree of design customization. Finally, maintaining alignment with boundaries ensures strips are consistently orthogonal or parallel to the boundary lines.

\begin{figure}[h]
  \centering
  \includegraphics[width=\linewidth]{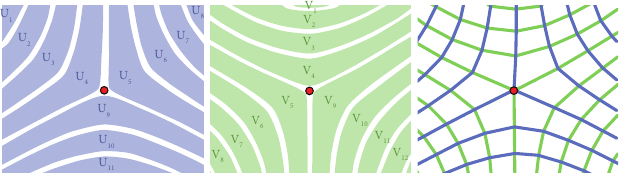}
  \caption{An SDQ mesh (right) is the overlay of two transversal topologically-coupled strip networks.}
  \label{fig:green_blue}
\end{figure}

\cite{Pottmann_2008_Freeform_Surfaces_from_Single_Curved_Panels} studied strips as semi-discrete representations of freeform surfaces, using a two-dimensional surface parametrization with one continuous and one discrete parameter. We differ from this approach conceptually in that we discretize both directions. 
In particular, we consider the overlay of two transversal and topologically-coupled strip networks, such that each network discretizes its counterpart, producing what we call a \emph{strip-decomposable quad} (SDQ) mesh, as shown in \figref{fig:green_blue}. In this configuration, edges can be two-colored, i.e., we assign a color to every edge so that no two neighboring edges sharing a quad have the same color \cite{Oval_BRG_2021_Two-Colour_Topology_Finding_of_Quad-Mesh_Patterns}. Consequently, all vertices have even valences, and all faces are quadrilaterals, except for boundary faces. An SDQ mesh is therefore a quad mesh that can be unambiguously decomposed into two independent discrete strip networks that share the shame topology and discretization. 

This fully discrete representation exhibits various advantages. Firstly, it is a compact representation that can be represented as a conventional quad mesh, a widely recognized standard across various modeling software. Additionally, as we show in this work, coupling two strip networks in this way enables editing and reconfiguring the strips using elementary topological operations, thereby providing greater flexibility and adaptability to the dual strip network.

Although SDQ meshes have numerous applications, there is a lack of techniques for their generation. Conventional quadrangulation methods, such as \cite{bommes2009_mixed-integer-quadrangulation,kalberer2007_quadcover---surface-parameterization-using}, are typically unsuitable, as they produce quad meshes with vertices of odd valence, making it impossible to two-color their edges unambiguously. A common workflow for creating SDQ meshes involves manually constructing low-resolution initial connectivity, either as a polygonal mesh \cite{2022_Tutorial_strip_morphologies_course, 2022_Tutorial_Digital_Fabrication_Workflow}, or as a skeletal graph \cite{Veenendaal2015_Hilo, 2023_Tutorial_with_stripper}, which is then subdivided until the desired resolution is reached. Furthermore, after generating an SDQ mesh, modifying the connectivity of its vertices, especially the combinatorics of the singularities, is challenging as local changes at the level of single quads can invalidate the strip decomposability. 

As a case study, we consider robotic non-planar 3D printing of freeform shell surfaces as a fabrication scenario that benefits from such a dual strip network, and apply our methodology for designing and fabricating non-planar print paths.

\paragraph{Contributions}
We offer a pipeline for designing SDQ meshes while considering fabrication-related properties of strips. In particular, we contribute;
\begin{itemize}
    \item An algorithm to compute a directional field that considers fabrication objectives and integrates into two coupled transversal parameterizations corresponding to two transversal strip networks (\secref{sec:Stripe_parametrizations}).
    \item A set of simple and intuitive connectivity editing operations that enable extensive user control over the topology of the strips (\secref{sec:mesh_editing}).
\end{itemize}
\section{Related Work}
\label{sec:Related Work}

\subsection{Fabrication-aware strips}
Strip meshes have been studied extensively for their various applications in design and fabrication. 
\cite{Pottmann_2008_Freeform_Surfaces_from_Single_Curved_Panels} studied the refinement and convergence properties of strips and presented a method that partitions freeform surfaces into single curved strips. Spiraling, developable strips were used in \emph{Zippables} \cite{schuller2018_shape-representation-by-zippables} to zip a surface together from flat ribbons, while \cite{verhoeven2022_dev2pq:-planar-quadrilateral-strip-remeshing} extracted a quad dominant strip pattern of developable surfaces. \cite{Differentiable_Stripe_Patterns_Maestre_2023} use strips to represent bi-material distributions with high stiffness contrast and compute patterns that lead to an optimal approximation of high-level performance goals. 

The transversal strip network is introduced in \cite{Takezawa_2016_Fabrication_of_Freeform_Objects_by_Principal_Strips} for the fabrication of freeform objects from flat paper pieces using principal strips aligned with the principal curvature directions. 
However, this approach relies on tracing streamlines, which  are sensitive to local perturbations and can thus  only be applied to smooth surfaces with disc or annulus topology. In contrast, we apply our approach to more complex surfaces with a higher genus and more boundaries.

\cite{Oval_BRG_2021_Two-Colour_Topology_Finding_of_Quad-Mesh_Patterns} proposed an alternative way of generating a two-colored quad mesh topology, starting with fixed singularities and then combinatorially connecting them into patches that are subdivided to generate the final mesh.
This approach relies on pre-defined fixed singularities and thus does not enable control over the alignment of the resulting edges. In contrast, our method positions singularities dynamically during the vector-field optimization stage.

\cite{Schiftner_2010_Statics-Sensitive_Layout_of_Planar_Quadrilateral_Meshes} generate quad-dominant meshes by integrating two separable conjugate line field directions for structural purposes.
Similarly, in \cite{Zadravec_2010_Designing_Quad-dominant_Meshes_with_Planar_Faces}  quad-dominant meshes are created by integrating a pair of line fields aligned to principal curvature directions. In both cases, since the guiding line fields are separable and conjugate, the resulting quad meshes are strip-decomposable and quasi-planar. Their planarity is further improved with a perturbation optimization \cite{Liu_2006_Conical_meshes_and_developable_surfaces}. 
From our perspective these methods are the closest to our approach since we also rely on the integration of separable line fields. However, our work allows for more freedom and control of the directional constraints and location of singularities, and we consider a broader scope of application scenarios.

\subsection{Field-aligned parameterization and remeshing}
Our work uses a pair of line fields integrated into scalar functions from which the strips are extracted as their isolines, and are overlayed to create a field-guided SDQ mesh. We refer the reader to the surveys~\cite{goes2016_vector-field-processing-on-triangle-meshes,vaxman2016_directional-field-synthesis-design-and-processing} to get acquainted with field-aligned meshing methods. While our work also designs a quadrilateral mesh from directional-field-oriented parameterization (like eg.,~\cite{bommes2009_mixed-integer-quadrangulation,kalberer2007_quadcover---surface-parameterization-using}), it departs from such methods in that we create the quad mesh as two separable sets of \emph{strips} with embedded fabrication considerations. By that, our work is, in fact, more related to the concept of ``stripe patterns'' of ~\cite{knoppel2015_stripe-patterns-on-surfaces}, although those were not used for meshing or optimized for fabrication processes. 

Our design optimizes field integrability, which is essential for the fidelity of the parameterization and meshing. To this end, we aim to reduce curl, as considered in~\cite{diamanti2015_integrable-polyvector-fields,vekhter2019_weaving-geodesic-foliations}. However, our optimization follows an iterative smoothing and curl reduction process, adapted from methods such as ~\cite{sageman-furnas2019_chebyshev-nets-from-commuting,meekes2021_unconventional-patterns-on-surfaces,liu2020_practical-fabrication-of-discrete-chebyshev-nets}, and tailored to our specific requirements.

\subsection{Quadrilateral-mesh topology editing}
In the process of creating feasible, not entwined, and aesthetically pleasing strips, we introduce an interface for basic topology editing operations. These operations aim mainly to control the location and connectivity of the singularities of the mesh, which affect how the shape can be partitioned into a good patch layout. As such, these operations are similar in spirit to topology editing in quadrilateral meshes~\cite{peng2011_connectivity-editing-for-quadrilateral-meshes,takayama2013_sketch-based-generation-and-editing-of-quad-meshes,tarini2010_practical-quad-mesh-simplification,daniels-ii2009_localized-quadrilateral-coarsening}. Our operations, however, work on strips and not quadrilaterals, and as such have a more global effect, and guarantee that the strip-decomposability of the mesh is maintained. In~\cite{Noma2022FastEditing}, strip singularities are edited directly as scalar functions, but not as strip meshes. To the best of our knowledge, we are the first to perform topology editing on SDQ meshes.
\section{Overview}
\label{sec:Overview}

We aim to generate a fabrication-aware and editable SDQ mesh with edges aligned to user-defined directions. In this section, we describe in detail our objectives and pipeline. 

\subsection{Objectives}
We consider the following geometric objectives for the dual strip networks, commonly found in the fabrication scenarios mentioned in \secref{sec:intro}.

\begin{enumerate}

\item \emph{Alignment}. \label{obj:aligned}
The strip networks are aligned as well as possible with user-specified directions. 

\item \emph{Uniformity}. \label{obj:uniform} 
The strips have a width that is as uniform as possible, resulting in quads with as uniform as possible sizes.

\item \emph{Smoothness}. \label{obj:smooth}
The strips are as smooth as possible, meaning that they vary slowly in space without sharp turns or noisy boundaries.

\item \emph{Orthogonality}. \label{obj:orthogonal}
The two transversal strip networks are as orthogonal as possible to each other. This translates to an SDQ mesh where all regular vertices have four incident edges with angles close to $\frac{\pi}{2}$. 

\item \emph{Continuity}. \label{obj:continuous}
There are no interruptions or discontinuities of strips on the interior of the shape.  

\item \emph{Good strip topology}. \label{obj:strips_topology}
The strips have a good topology, meaning that they do not wind before they close or terminate on a boundary. Instead, they extend from boundary to boundary, or form closed loops without snake-like parts where they fold into themselves.

\item \emph{Good patch layout}. \label{obj:patch_layout}
Each strip network can be partitioned into a small number of patches of simply connected strips, i.e., it has a good patch layout.
\end{enumerate}

Objectives (\ref{obj:aligned})--(\ref{obj:strips_topology}) are optimized in the vector-field design stage (\secref{sec:Stripe_parametrizations}). In addition, objective (\ref{obj:continuous}) is further addressed in the subsequent seamless integration (\secref{subsec:integration-and-meshing}). Objectives (\ref{obj:strips_topology})--(\ref{obj:patch_layout}) are achieved by the topological operations that enable editing the mesh (\secref{sec:mesh_editing}).

\subsection{Pipeline}
\label{sec:pipeline}

Given a triangle mesh surface, possibly with open boundaries, our pipeline proceeds as follows.

\begin{enumerate}
\item	Collect user-specified directional constraints (\secref{par:alignment}).
\item	Compute two coupled integrable line fields that adhere to the objectives (\secref{subsec:guiding-fields}).
\item	Integrate the fields into two transversal parameterizations. Discretize those into two sets of strips, which, when overlaid, form an SDQ mesh (\secref{subsec:integration-and-meshing}).
\item	Carry out strip-based topological operations for eliminating winding strips and aligning singularities, with the help of user input, to create a good patch layout (\secref{sec:mesh_editing}).
\end{enumerate}

\section{Stripe parametrizations}
\label{sec:Stripe_parametrizations}

We describe steps (1)-(3) of the pipeline, where we first design a directional field and then integrate it into a parameterization suitable for our framework. 

\subsection{Transversal parametrizations}
\label{sec:stripe-parameterization}

We work with a triangular mesh $\MM = \left\{\VV,\EE,\FF\right\}$. 
We denote by $\mathcal{C}$ the set of \emph{corners} $(v,f)$, where $v \in \VV$ denotes a vertex and $\ f \in \FF$ a face adjacent to this vertex. We begin by constructing two transversal parameterizations $U$ (displayed throughout  in blue), and $V$ (displayed in green) on $\MM$ that represent the two strip networks.

Since strips are \emph{sign-invariant}, i.e., they can be traversed in both directions without any difference in the result, the parameterization must also be sign-invariant. To achieve that, we design a strip network as a branched $2$-function \cite{vaxman2016_directional-field-synthesis-design-and-processing}, which admits singularities of indices $\pm \frac{k}{2},\ k \in \ZZ$. Formally, a $2$-function $U:\mathcal{C} \rightarrow \RR$ is defined as the assignment of a scalar per corner, linearly interpolated inside each triangle. Furthermore, on every edge $e$ between vertices $i, j$ and adjacent to faces $f,g$, we have:
$$
U_{f,i}-U_{f,j} = s_e (U_{g,i}-U_{g,j}),\quad\text{for some } \, s_e = \pm 1.
$$
The specific choice of the sign $s_e$ is called the \emph{matching} on the edge $e$.
A regular patch $\mathcal{P} \subset \MM$ is a sub-mesh where a function $U$ can be ``combed'' into a single (1-)function without any sign ambiguity. Combing means re-signing all corner values on the patch so that they agree on all $e \in \mathcal{P}$ with $s_e=1$. Patches where this is not possible, are called \emph{singular}; they contain branching points, or \emph{singularities}, of the parameterization (\figref{fig:green_blue}, \ref{fig:vec_sings_to_mesh_sings}). 

The output of this stage is two $2$-functions $U$ and $V$, each representing one strip network. We design those functions by optimizing their gradients, which we denote with small case letters $u=\nabla U,v=\nabla V$. This is a common modus operandi to construct such \emph{seamless} parameterizations (eg.,~\cite{sageman-furnas2019_chebyshev-nets-from-commuting,meekes2021_unconventional-patterns-on-surfaces,verhoeven2022_dev2pq:-planar-quadrilateral-strip-remeshing}). $u$ and $v$ are the \emph{guiding fields} through which we can incorporate the objectives listed above.

\subsection{Guiding fields}
\label{subsec:guiding-fields}

The gradient of a $2$-function is a $2$-vector field (\figref{fig:vec_sings_to_mesh_sings}), which is also sign-invariant and has  the same singularities. Furthermore, the field is \emph{face-based} and \emph{piecewise-constant}, with a single $2$-vector in each face $f \in \FF$, and it is curl-free; that means on edge $e$ shared by faces $f$, $g$ we have:
 \begin{equation}
\langle u_f, e \rangle = s_e \langle u_g,e \rangle,\ \ \langle v_f, e \rangle = s_e \langle v_g,e \rangle.
\label{eq:curl_conditions}
 \end{equation}

Our main variables are two $2$-fields $X, Y$, piecewise constant per face,  that represent the \emph{candidate gradient fields}  $u, v$ of the parameterizations.
The $2$-fields are expressed in the \emph{power} representation~\cite{knoppel2013_globally-optimal-direction} to achieve sign invariance: we represent $u_f$ and $v_f$ as complex numbers in an arbitrary local basis in each face $f$, and then define:
$$
X_f = (u_f)^2,\ Y_f = (v_f)^2.
$$

\subsubsection{Energies}
\label{subsec:energies}

 A useful energy term is a generic \emph{closeness} energy for power fields with the structure:
 $$
 E_C(K,R,M) = (K - R)^H M (K - R). 
 $$
 This energy measures how close the field $K$ is to an input field $R$ in a metric defined via the mass matrix $M$. $(.)^H$ denotes the conjugate transpose of $(.)$. With this, we set the following optimization energy terms.

\paragraph{\textbf{Alignment} - objective \ref{obj:aligned}} 
\label{par:alignment}
This energy term aims to keep the two 2-vector fields aligned with the input directional constraints.
Each constrained face $f_S$ in the set of constrained faces $\FF_S$ has a constraint tangent alignment $2$-vector given in the power representation $\alpha_f$, and a confidence weight $\omega_f$. We collect the confidence weights in a confidence matrix $W = \text{diag}(\omega_f)$, and also compute an orthogonal tangent constraint $2$-vector $\beta_f = -\alpha_f$ for the transversal direction $Y$. Then the alignment energy is written as closeness energy to the directional constraints. 

 \begin{equation}
 E_A(X, Y) = E_C(X_S, \alpha, W\cdot M_{\FF|S}) + E_C(Y_S, \beta, W\cdot M_{\FF|S}).
 \label{eq:soft-alignment-term}
 \end{equation}
$M_{\FF|S}$ is the diagonal mass matrix of triangle areas restricted to the alignment faces $\FF_S$, as this energy is only applied to the constrained faces in $\FF_S$. 

\begin{figure}[h]
  \centering
  \includegraphics[width=0.7\linewidth]{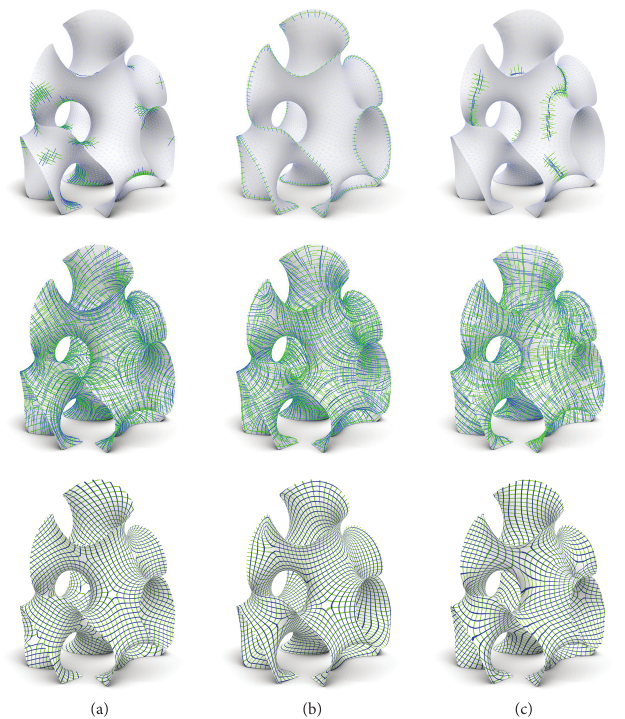}
  \caption{ Different input directional constraints: (a) curvature-aligned, (b) boundary-aligned, and (c)  aligned with user-drawn curves. Top to bottom: directional constraints on a sparse set of faces, optimized coupled 2-vector fields, and the resulting parameterization.}
  \label{fig:different_input_constraints}
\end{figure}

We employ three presets of input alignment constraints, each beneficial in a different context: 
\begin{itemize}
    \item \emph{Curvature alignment.} We use principal curvature directions for alignment, with confidence value (Eq.~\ref{eq:soft-alignment-term}) for each face $\omega = \kappa_{\text{max}}-\kappa_{\text{min}}$, where $\kappa_{(\cdot)}$ are the principal curvatures.  We then choose as the  constrained faces $\FF_S$ those with the highest confidence values (by default: top $\% 10$). We set $\alpha$ and to $\beta$ be the square of the minimum and maximum principal curvature vectors respectively (\figref{fig:different_input_constraints}a).
    \item \emph{Boundary constraints.} We add each boundary face with one boundary edge $e$ to the constrained faces and set $\alpha$ and $\beta$ to be exactly orthogonal and parallel to $e$ respectively (\figref{fig:different_input_constraints}b), with confidence values $\omega = 1$ on all constrained faces.
    \item \emph{User-drawn directions.} The user draws a curve on the surface from which we extract constrained faces. We then set $\alpha$ to be orthogonal and to the curve and $\beta$ parallel (\figref{fig:different_input_constraints}c), with confidence values $\omega = 1$ on all constrained faces.
\end{itemize}

\paragraph{\textbf{Unit length} - objective \ref{obj:uniform}} This energy term aims to keep the 2-vector fields as close as possible to unit norm, and can be written as a closeness energy to 
$\hat{X}=\frac{X}{|X|}$ and $\hat{Y}=\frac{Y}{|Y|}$ (normalized $X$ and $Y$). 
    \begin{equation}
    E_U(X, Y) = E_C(X,\hat{X},M_\FF) + E_C(Y,\hat{Y},M_\FF)
    \end{equation}
where $M_{\FF}$ is the diagonal mass matrix of triangle areas. We note the use of the auxiliary variables $\hat{X}$ and $\hat{Y}$ that are ``snapshot'' (to a previous iteration in our algorithm), keeps this energy quadratic.

 \paragraph{\textbf{Smoothness} - objective \ref{obj:smooth}} 
 This energy term aims to keep the 2-vector fields as smooth as possible.
 For smoothness, we consider the $|\EE_I|\times|\FF|$ discrete \emph{covariant derivative} matrix $D$ for power $2$-fields (with the weights of ~\cite{brandt2018_modeling-in/i-symmetry-vector-fields-using}), which for every internal (non-boundary) edge $e \in \EE_I$ on adjacent faces $f$ and $g$ computes:
 $$
 (DX)_{|e} = \left(X_f\cdot (\overline{e}_f)^2 - X_g\cdot (\overline{e}_g)^2\right),
 $$
 where $\overline{e}_f$ (resp. $\overline{e}_g$) is the \emph{normalized} conjugate edge vector $e$ in the basis of $f$ (resp. $g$). We also need the $|\EE_I| \times |\EE_I|$ diagonal \emph{mass matrix} $M_D$, where $M_D(e,e) = \frac{3|e|^2}{A(f) + A(g)}$, and $A(\cdot)$ denotes the area operator. Then, the smoothness energy is:
 \begin{equation}
     E_S(X,Y) = X^H (D^H M_D D) X + Y^H (D^H M_D D) Y 
 \end{equation}

 \paragraph{\textbf{Orthogonality} - objective \ref{obj:orthogonal}} This energy is the only term that links $X$ with $Y$. It aims to keep the two 2-fields as orthogonal as possible.
 \begin{equation}
     E_O(X,Y) = (X+Y)^H M_\FF (X+Y) 
 \end{equation}
 When $X=-Y$ we get $u = \pm i\cdot v$, which implies that the vectors are perfectly orthogonal (and with equal magnitudes).

 \paragraph{Normalization of energies}
 We normalize all energy terms to make them of comparable magnitudes with each other. The energies have the from $E = P^H \mathcal{A}P$. To normalize them, we divide by $\mathbf{1}^T \mathcal{A} \mathbf{1}$, where $\mathbf{1}$ is the vector of all $1$'s.

\subsubsection{Constraints}

 \paragraph{\textbf{Integrability}} 
 As described above, the guiding fields must be curl-free to represent the gradients of the parametrization functions. The curl reduction strategy aims to keep the curl of the fields $u$ and $v$ to zero during the optimization. Note that, unlike all previous energy terms, the curl of a field is not sign-invariant. Therefore we need to compute an explicit matching $s_e$ on every inner edge $e \in \EE_I$, as per \eqref{eq:curl_conditions}.

\paragraph{Matching}
\label{par:matching}
To find the matching (i.e. the sign of each vector) we first compute $u_f = \sqrt{X_f}$ and $v_f = \sqrt{Y_f}$ in each face. Note that the solution to this has arbitrary signs. 
We then find the sign for one of the 2-fields, say $u$; namely for each edge $e$ between faces $f$, $g$, we find the sign $s_e \in \left\{-1,0,1\right\}$ per edge $e \in \EE_I$ so that it minimizes rotation effort between $\left\{u,-u\right\}_f$ and $\left\{u,-u\right\}_g$ (also known as the \emph{principal} matching~\cite{diamanti2014_designing-n-polyvector-fields-with}). We then update the signs of the other field to have the same matching. This results in two positively-oriented fields with the same topology by design.
 
\paragraph{Curl elimination} 
Equipped with the matching, we build the $|\EE_I|\times |\FF|$ \emph{curl matrix} $C_s$ which, on every edge $e$ between faces $f$ and $g$, computes, for the $u$ field:
 \begin{equation}
     (C_s\cdot u)_{|e} = \langle s_e u_g - u_f, e\rangle,
 \end{equation}
 and similarly for $v$. To eliminate curl, we solve a projection problem to the nearest curl-free field with the given matching $\mathbf{s}$, namely:
\begin{align*}
    (u,v)^* &= \text{argmin}|u^*-u|^2+|v^*-v|^2\ s.t.\\ C_s\begin{pmatrix}u^*\\v^*\end{pmatrix}&=0
\end{align*}
via the minimum $2$-norm projection on the null space of $C_s$ \cite{verhoeven2022_dev2pq:-planar-quadrilateral-strip-remeshing}.

\subsubsection{Optimization problem}
With these, our full optimization problem for the fields is:

\begin{align}
    (X, Y,u, v,s) = &\text{argmin}( \lambda_S E_S+ \lambda_OE_O + \lambda_U E_U+ \lambda_A E_A)        \label{equ:optimizationproblem}\\
    \text{subject to \quad} & C_s\cdot u_f = 0,\ C_s\cdot v_f = 0   ,\quad \forall f \in \FF
    \tag{\small curl-free} \\
    &X_f = u_f^2,\ Y_f = v_f^2, \quad \forall f \in \FF \tag{\small compatibility}
\end{align}

\subsection{Optimization}
\label{subsec:optimization}
\subsubsection*{Optimization algorithm}
The directional-field optimization problem in \eqref{equ:optimizationproblem} is nonlinear, due to the normalized compatibility and unit-length term, and discrete, because of the matching variable $s$. We use an alternating optimization scheme, summarized in Alg.~\ref{alg:full-optimization-directional-field}, where we alternate between reducing a quadratic energy $E^k$ with an implicit step, principal matching (in closed-form),  curl-elimination (another quadratic energy), and closed-form projection to the compatiblity constraints. To encourage convergence, we introduce a factor $\lambda_E$ that attenuates the objective terms in every iteration.

\begin{algorithm}[h!]
\SetAlgoLined
 Initialize $k\leftarrow 0$, \\
 $\left(X^0,Y^0, E^0\right)\leftarrow \text{argmin} (\lambda_SE_S+\lambda_AE_A+\lambda_OE_O)$\hfill \emph{(variables initialization)}\\
$\forall f\in \FF,\ X_f^0\leftarrow \frac{X_f^0}{|X_f^0|},\ Y_f^0\leftarrow \frac{Y_f^0}{|Y_f^0|}\  $\hfill \emph{(normalization of initial solution)} \\
$\lambda_E^0 = 1$\\
 $\mu = \text{smallestEigenValue}(E^0,M_\FF),\ dt = \nicefrac{1}{\mu}$\\
 \Repeat{$ \|E^k-E^{k-1}\| < 10^{-4}$ \text{or} $k=100$}{
 	
    $(X^{k+1}, Y^{k+1}, E^{k+1}) \rightarrow \text{ImplicitStep}\left(X^k, Y^k,  E^k, M_\FF, dt\right)$\\

    $\hat{X}_f^{k+1}\leftarrow \frac{X_f^{k+1}}{|X_f^{k+1}|},\ \hat{Y}_f^{k+1}\leftarrow \frac{Y_f^{k+1}}{|Y_f^{k+1}|}$\hfill \emph{(re-normalization)} \\
    
    $(u^{k+1},v^{k+1},s)\leftarrow\text{PrincipalMatching}(X^{k+1},Y^{k+1})$\\

    $(u^{k+1},v^{k+1})\leftarrow\text{CurlElimination}(u^{k+1},v^{k+1},s)$ \\
    $X^{k+1}=(u^{k+1})^2$, $Y^{k+1}=(v^{k+1})^2$\hfill \emph{(compatibility)}\\
    $\lambda_E^{k+1} \leftarrow 0.8\lambda_E^k$\hfill \emph{(dampening)}\\
    $k \leftarrow k+1$\\
 }
 \caption{Directional-Field Optimization}
 \label{alg:full-optimization-directional-field}
\end{algorithm}

\paragraph{Implicit Step} The generic function $$y=\text{ImplicitStep}(x, A, M, dt)$$ performs an implicit step to minimize an objective of the type $E = x^TAx$ with the a $M$. It returns $y$ (and the energy) as the solution to the linear system:
$$
(M + dt\cdot A)y = Mx.
$$
Our time step $dt$ is inversely proportional to the scaling of the energy by the Fiedler value $\mu$, as explained in~\cite{sageman-furnas2019_chebyshev-nets-from-commuting}.

\subsubsection*{Choice of parameters.}
\label{sec:directional_constraints}

We use the following energy weights: 
     smoothness $\lambda_S = 10.0$,
     orthogonality $\lambda_O = 2.0$,
     alignment  $\lambda_A = 0.1$,
     and unit $\lambda_U = 1.0$.
$\lambda_E$ has an initial value of 1.0, which is multiplied by 0.8 after every iteration to encourage convergence to the constraints.

\subsection{Integration and Meshing}
\label{subsec:integration-and-meshing}

\begin{figure}[h]
  \centering
  \includegraphics[width=\linewidth]{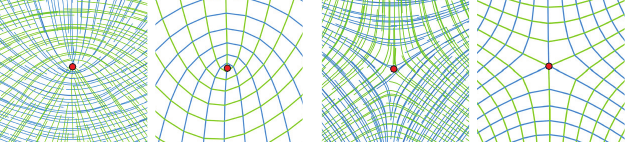}
  \caption{The singularities of the vector field become singular vertices of the quad mesh. Left: a $\nicefrac{1}{2}$ singularity becomes a vertex with valence 2. Right: a $-\nicefrac{1}{2}$ singularity becomes a vertex with valence 6.}
  \label{fig:vec_sings_to_mesh_sings}
\end{figure}

Having optimized the two coupled fields $u$ and $v$, we integrate them into parameterization functions $U$ and $V$. 
To do that, we combine $u$ and $v$ in one 4-field so that the vectors in each face are arranged in the order $\left\{u,v,-u,-v\right\}$ in CCW fashion around the normal. The matching of the 4-field on every edge $\mathbf{s_e}$ equals $2 s_e$, i.e. it has values in $\left\{-2,0,2\right\}$ so that $u$ and $v$ never intermix in the 4-field. We can then use a standard frame-field integration and meshing algorithm to integrate the 4-field into the  functions $U$ and $V$. These functions result in two coupled transversal strip networks (Fig. \ref{fig:green_blue}) that are overlaid into a single quad mesh $\MM_\QQQ = \left\{\VV_\QQQ, \EE_\QQQ, \FF_\QQQ\right\}$ that is strip-decomposable by design. The singularities of the field are directly translated into singular vertices of $\MM_\QQQ$ (~\figref{fig:vec_sings_to_mesh_sings}). Integration and meshing are done using the library \texttt{Directional}~\cite{amir_vaxman_and_others_2021_5746726}.

Note that we use the same notation ($U$, $V$) for the parameterization functions and for resulting strips, as they essentially refer to the same geometry. $U$-strips (blue) are bounded by $U$-edges which are isolines of the integration of the $u$ vectors, i.e., always orthogonal to the $u$-field. Similarly, $V$-strips (green) are bounded by $V$-edges orthogonal to the $v$-field. 
$\MM_\QQQ$ has non-quad faces only on the boundaries, where the strips are cut by the boundary.

\begin{figure}[h]
  \centering
  \includegraphics[width=0.6\linewidth]{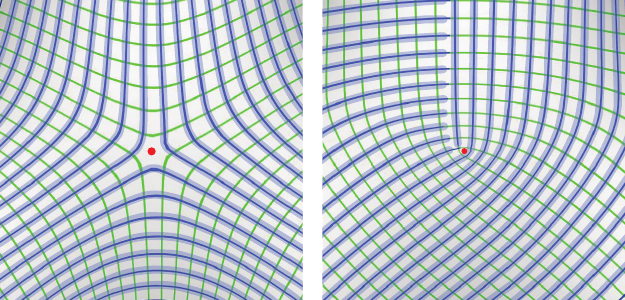}
  \caption{A pair of coupled, but non-interchangeable $2$-parameterizations (left) vs. a general $4$-parameterization (right), from identical directional constraints. A $4$-parameterization (produced via a cross-field, right) allows switching between $U$ and $V$; this produces strips that flip direction around the singularity, which create discontinuities and are unsuitable for our purposes. In contrast a pair of $2$-parameterizations (left) creates two strip networks that never intermix.}
  \label{fig:comparison-with-cross-field}
\end{figure}

 \paragraph{Discussion} Our directional-field framework is inspired by  similar frameworks for integrable aligned parameterization. Specifically, there is considerable similarity with integrable $4$-fields (``frame fields'') since we create the combined field $\left\{u,v,-u,-v\right\}_f$ in each face. The overlay of our resulting parameterizations is technically a $4$-function (of the kind used for quad-meshing). However, existing algorithms for integrable frame fields also allow both directions to interchange in a way that permits $\pm \frac{1}{4}$ singularities, leading to a general quad mesh that does not necessarily separate into transversal strips without cuts or discontinuities (see \figref{fig:comparison-with-cross-field}). To the best of our knowledge, no existing algorithm specifically treats our case of SDQ meshes with directional fields.

\section{Mesh editing}
\label{sec:mesh_editing}

We next describe step (4) of our pipeline (\secref{sec:pipeline}), namely, editing the topology of the mesh to simplify it in order to achieve good strips topology (objective \ref{obj:strips_topology}) and a good patch layout of both strips networks (objective \ref{obj:patch_layout}). This consists of aligning singularities and correcting winding strips (eg. the blue strips in \figref{fig:topological-defect}b and \figref{fig:routes}e). 
This step includes user interaction which we simplified to a small number of intuitive steps.
For our editing interface, we use the \emph{libigl} \cite{jacobson2018_libigl:-a-simple-c-geometry-processing-library} viewer.

\subsection{Topology Editing} 
\label{subsec:topology-editing}


\subsubsection*{Dealing with topological defects} 
\begin{figure}
    \centering
    \includegraphics[width=0.85\textwidth]{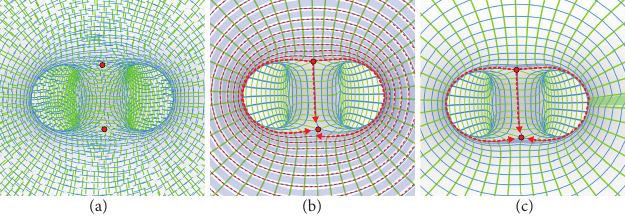}
    \caption{Even well-aligned fields (a) may create misaligned singular vertices leading to winding strips (b). A simple topological edit (c) of rewiring the highlighted green strip fixes the topological defect.}
    \label{fig:topological-defect}
\end{figure}

Our directional field, and consequent seamless parameterization and mesh, are optimized for smoothness and good alignment with directional constraints, which incurs well-placed singularities for the most part. However, the mixed-integer seamless integration can potentially cause \emph{misalignment} between two close-by singularities that should have been naturally paired. For example, in \figref{fig:topological-defect}b, the singularities are slightly misaligned along the route that goes around the right handle, while in \figref{fig:topological-defect}c they are aligned along all highlighted routes. Even such slight misalignment can cause long undesirable winding strips with self-folding geometry or very thin patches. We offer a paradigm for topological editing that can easily fix most defects with simple and minimal user control. 

We note that there are paradigms for canonically partitioning quad meshes into singularity-free sub-meshes without altering the underlying topology of the mesh, such as motorcycle graphs
\cite{Eppstein-2008-Motorcycle_Graphs, Schertler_2018_Generalized_Motorcycle_Graphs_for_Imperfect_Quad-Dominant_Meshes}. However, such approaches would not work for our case since we are, in essence, partitioning strip networks instead of quad meshes. This means that we consider cuts along edges of the same color for the most part, thus few or no intersections are created where cuts can terminate. Therefore, we opt for a new topological editing workflow described below.

\begin{figure}[h]
  \centering
  \includegraphics[width=\linewidth]{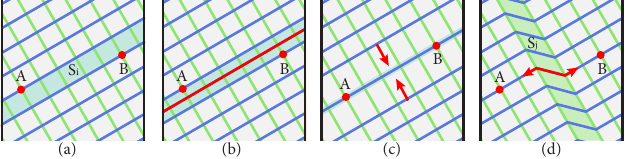}
  \caption{Atomic strip-editing operations affect  alignment between quad vertices. (a) Initially, $A$ and $B$ are on different sides of the highlighted strip $S_i$. (b) Subdividing $S_i$ moves them one strip further apart. (c) Collapsing $S_i$ or (d) rewiring the transversal strip $S_j$   aligns $A$ and $B$.}
  \label{fig:editing_operations}
\end{figure}

\paragraph{Atomic operations}
Editing a strip-decomposable mesh is a more constrained problem than editing typical quad meshes~\cite{tarini2010_practical-quad-mesh-simplification,daniels-ii2009_localized-quadrilateral-coarsening,peng2011_connectivity-editing-for-quadrilateral-meshes,takayama2013_sketch-based-generation-and-editing-of-quad-meshes}, as we have a strict requirement that $U$ and $V$ strips don't intermix, so that the strip decomposition remains unambiguous.  To this end, we devised two novel strip-based atomic operations, whose purpose is to rearrange the connectivity of adjacent strips and align vertices that lie on different strips: 

\begin{enumerate}
    \item \emph{Collapsing and splitting.} Given a strip $S_i$, described as a sequence of quadrilaterals, we can either subdivide it by splitting it into two adjacent topologically identical strips (\figref{fig:editing_operations}b) or collapse it so that its immediate neighbors are now adjacent (\figref{fig:editing_operations}c). 
    \item \emph{Rewiring.} Given a strip $S_j$, we rewire its quads by switching the edges of all quads, either one vertex forward or backward on the strip (\figref{fig:editing_operations}d). As a result, all transversal strips that intersect $S_j$ are reconfigured.
\end{enumerate}

\paragraph{Alignment of a pair of vertices}
Assume we want to align the vertices $A$ and $B$ that lie on the two sides of a strip $S_i$ (\figref{fig:editing_operations}a) along the $U$ (blue) direction. This can be achieved in two ways: either by collapsing strip $S_i$  (\figref{fig:editing_operations}c) or by rewiring a \emph{transversal} (green) strip along the $V$ direction that runs between $A$ and $B$, such as strip $S_j$ (\figref{fig:editing_operations}d). For vertices further apart, there might be an entire decision graph of possible choices. To make a reasonable choice for a sequence of operations for alignment, it is important to consider:

\begin{itemize}
    \item Collapsing removes mesh faces. If it is applied on many strips, or on a strip that covers a large part of the mesh, for example, a winding strip, it can lead to considerable geometric distortion. 
    \item Rewiring can misalign other pairs of vertices that were already aligned, because it changes the connectivity of all strips intersected by the transversal rewired strip.
\end{itemize}

The alignment of vertices can be achieved along either the $U$ or the $V$ direction using the presented atomic operations. For brevity, in the following, we consider alignment along the $U$ (blue) direction, which we call \emph{dominant}, and we call the $V$ (green) direction \emph{subdominant}.

\subsubsection*{User-controlled topology editing}
Determining what edits must be carried out is a difficult combinatorial problem that is hard to automate fully. In addition, editing decisions have a large impact on the final aesthetics of the produced object. Therefore, it makes sense to delegate some of the combinatorial choices to the user. On the other hand, reasoning about the desired topology of a mesh geometry is a challenge, as the connectivity of strips can be too complex for a novice user to grasp or change manually. As a practical compromise, we abstract full control over mesh connectivity away from the user, and instead have them follow a short pipeline where they choose pairs of vertices they wish to align. Our algorithm then automatically generates the sequence of atomic operations that realizes the chosen alignment. 
The accompanying video shows the full process of aligning singularities on various shapes.

\begin{figure*}[h!]
  \centering
  \includegraphics[width=\linewidth]{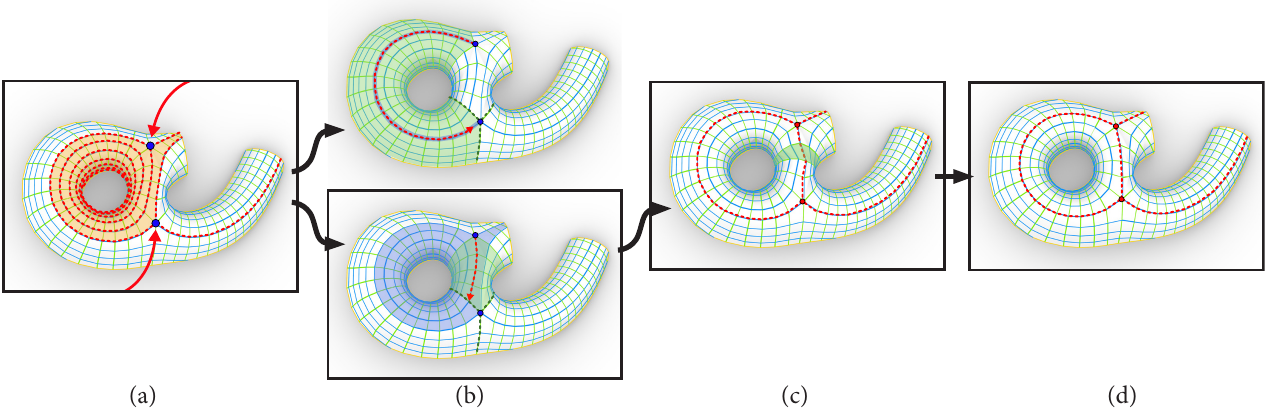}
  \caption{The control pipeline for topology editing. (a) The separatrices and winding strips of the $U$ strip network are displayed, and the user selects two vertices to align. (b) All existing routes between the two vertices are displayed, and the user selects along which route to align. (c) The algorithm proposes the next best step and the user approves it. (d) The result is smoothed iteratively.}
  \label{fig:control_pipeline}
\end{figure*}

\paragraph{Control pipeline}

Every alignment of a pair of vertices entails the following steps.

\begin{figure}[h]
  \centering
  \includegraphics[width=0.8\linewidth]{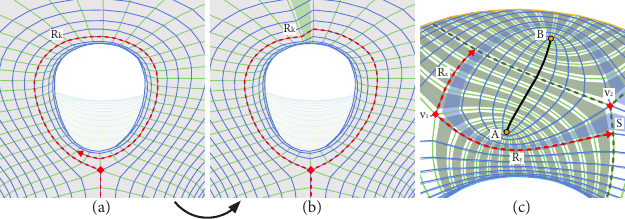}
  \caption{(a) Winding strip around the handle caused by the misalignment of a singular vertex with itself along the route $R_k$. (b) Rewiring the subdominant strip aligns the vertex along $R_k$ and removes the winding strip. (c) The fixed alignment of vertices $A$ and $B$ locks the rewiring of all subdominant strips highlighted in dark green.}
  \label{fig:strips_editing}
\end{figure}

\begin{enumerate}[label=(op\arabic*)] 

\item 
\label{editop:choose_vertices}
The user selects two vertices $v_1$, $v_2$ to be aligned along the dominant direction (\figref{fig:control_pipeline}a). {Note that we allow $v_1=v_2$, for example, the vertex $v$ in \figref{fig:strips_editing}a,b can be aligned with itself around the handle.}
\item 
\label{editop:choose_route}
The algorithm enumerates all routes from $v_1$ to $v_ 2$, and the user selects one route along which to align (\figref{fig:control_pipeline}b). The default choice is the route that has the smallest non-zero length.
\item 
\label{editop:choose_ops}
The algorithm points out the ``best'' next step, and the user confirms its application (\figref{fig:control_pipeline}c). This is repeated until the two vertices are aligned along the selected route.
\end{enumerate}

\begin{wrapfigure}{r}{0.2\textwidth}
    \vspace{-15pt}
    \includegraphics[width=0.2\textwidth]{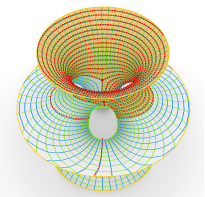}
    \vspace{-40pt}
\end{wrapfigure}
\paragraph{Choosing vertices to be aligned \ref{editop:choose_vertices}}
The singularity separatrices and winding strips are highlighted as guides (see \figref{fig:control_pipeline}a and inset) to help the user choose which vertices it makes sense to align.

\begin{figure}[h]
  \centering
  \includegraphics[width=0.8\linewidth]{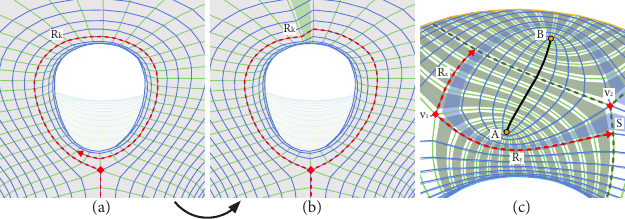}
  \caption{All routes from $v_1$ to $v_2$. Top: sequence of subdominant strips. Bottom: sequence of dominant strips. (a) Route $R_a$ terminates on vertex $v_2$. (b) Route $R_b$ terminates on a green edge directly connected with vertex $v_2$. (c) Route $R_c$ terminates on a boundary (irrelevant route). (d) The vertices are already aligned along $R_a$, ie. they are 0 strips apart ($d_{R_a} = 0$). (e) The vertices are one strip apart along $R_b$ (the blue strip that covers the handle), $d_{R_b} = 1$. (f) One rewiring operation aligns $v_1$ and $v_2$ along $R_b$.}
  \label{fig:routes}
\end{figure}

\paragraph{Choosing routes between vertices \ref{editop:choose_route}}
Having chosen two vertices, $v_1$ and $v_2$, we automatically enumerate all relevant routes  between them, the set of which we denote by $R$.  Each route $R_i \in R$ consists of a sequence of $U_{R_i}$ (dominant) and a sequence of $V_{R_i}$ (subdominant) strips, which we collect as follows:
\begin{enumerate}
    \item \emph{Subdominant sequences $\{V_{R_i}\}$:}  We start from the vertex  ($v_1$ or $v_2$) that has the higher valence, 
    and trace every incident dominant edge until we intersect a subdominant edge directly connected to the other vertex (\figref{fig:routes} top). This defines the route $R_i$, and the crossed subdominant strips comprise its subdominant sequence $V_{R_i}$. If another singularity or a boundary is reached instead (\figref{fig:routes}c), the route is discarded as irrelevant. Therefore, the number of possible routes is bounded by the dominant-edge valence of the first vertex. In \figref{fig:routes}, we have two relevant routes $R_a$ and $R_b$.
 
    \item \emph{Dominant sequences $\{U_{R_i}\}$:} For each relevant route $R_i$ found in the previous step, we find the shortest among the paths of dominant (blue) strips between $v_1, v_2$ that intersects the sequence of subdominant strips $V_{R_i}$ (\figref{fig:routes}d,e). This is then the dominant strip sequence $U_{R_i}$. We define the \emph{route distance} $d_{R_i}$ as the length of $U_{R_i}$; it specifies the number of operations that must be performed for the two vertices to  align along $R_i$. In \figref{fig:routes} we have $d_{R_a} = 0$, i.e. the vertices are already aligned, and $d_{R_b} = 1$, i.e. the vertices are one strip apart, and therefore require one operation to be aligned.
\end{enumerate}

In \figref{fig:routes}b,e the depicted route $R_b$ is the smallest non-zero length route chosen by default.  

\paragraph{Iterative next best step \ref{editop:choose_ops}}
Having chosen a route $R_i$ along which $v_1$ and $v_2$ should be aligned, the algorithm iteratively brings them towards alignment, one strip at a time, by  either (a) rewiring a subdominant strip $V_{R_i} \in V_R$, or (b) collapsing a dominant strip $U_{R_i} \in U_R$. To choose the strip to be modified, we consider the following  ``fitness'' criteria applied to all strips $S \in \{U_{R_i}, V_{R_i}\}$:
\begin{itemize}
    \item The total area $A$ of the strip 
    $ A(S) = \sum_{q \in S} A(q)$
    measured as the sum of the areas of its quadrilaterals $\{q\}$.
    \item The total deviation from perfect ``squareness'' $D= \sum_{q \in S} D(q)$. We measure the squareness of a quad $q$ as
    $$D(q) = \sum_{i=1}^4{\left|a_i - \frac{\pi}{2}\right|} + \sum_{i=1}^4{\left|l_i - \bar l\right|}$$

    where $a_i$ is the $i$-th inner angle of the quad, $l_i$ is the $i$-th edge length and $\bar l$ is the average edge length of the quad.
    
\end{itemize}
The fitness score for $S$ is then $1/(A(S) + D(S))$, and we choose the strip of highest fitness. As a result, we end up editing small and well-shaped strips, which helps avoid geometric artifacts.

In the example of  \figref{fig:routes}(b,e), $v_1$ and $v_2$ are one dominant strip apart along route $R_b$, thus one operation is required. All strips of $R_b$ are nicely shaped with almost square quads. Therefore, their area is the deciding factor for picking a strip.
Since the blue strip is winding (\figref{fig:routes}e), it has a large area sum, therefore, its fitness is very low. As a result, the green strip with the smallest area is selected for rewiring, resulting in the alignment of $v_1$ and $v_2$ along that route (\figref{fig:routes}f). 

While the algorithm iteratively offers the next atomic operation in this step, we allow the user to approve every such iteration before it is applied. This is useful because the proposed next step might still not be a preferable operation to perform (e.g. if all strips of a route are winding). Then the user can abort and attempt to align another pair of vertices, which can, for example, unwind these strips before returning to the current route. 

\begin{figure}[h]
  \centering
  \includegraphics[width=\linewidth]{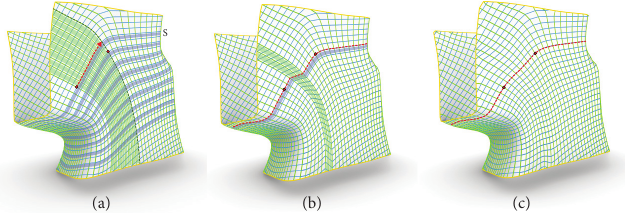}
  \caption{Unwinding strips through the alignment of a pair of vertices. (a) Initial mesh with winding strips (for example, the highlighted blue strip $S$ is winding, i.e., it has a snake-like shape folding upon itself), and two selected vertices along the displayed route. (b) After two rewiring operations, the two vertices are aligned, and the winding strip is removed. (c) Final smoothing.}
  \label{fig:unwinding_strips}
\end{figure}

In both examples \figref{fig:routes}f and \figref{fig:strips_editing}b, the alignment of singularities removes the winding strips. In \figref{fig:unwinding_strips}, we show how aligning a non-singular pair of vertices can remove winding strips. Aligning any two vertices that lie on the same side of a winding strip in two consecutive turns unwinds the strip.

\paragraph{Locking alignments}
Rewiring operations may misalign already aligned pairs of vertices. To avoid this, the user may select vertex pairs whose alignment should remain locked. In that case, we lock the rewiring of any subdominant strips that intersect a dominant alignment path between the two locked vertices, by setting their fitness to $0$. Collapse operations are never blocked, as they can only bring vertices together but not apart.  For example, on \figref{fig:strips_editing}c, the alignment between vertices $A$ and $B$ has been locked, so none of the subdominant strips (in dark green) that intersect the path between $A$ and $B$ can be rewired. As a result, to align $v_1$ and $v_2$ along either one of the two available routes ($R_x$ or $R_y$), only the collapse of the highlighted dominant (blue) strip  $S$ is possible.  

\subsubsection*{Post-processing}
Once all operations are completed, the algorithm selectively subdivides all elongated strips, smooths local deformations, and projects the resulting mesh to the original mesh boundary and surface, as follows.

\begin{wrapfigure}{r}{0.25\textwidth}
    \vspace{-15pt}
    \includegraphics[width=0.25\textwidth]{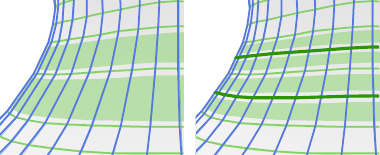}
    \vspace{-30pt}
\end{wrapfigure}
 \paragraph{Selective Subdivision} We define the \emph{elongation} of a strip by the ratio of lengths of edges,
$r = max(l_U,l_V) / min(l_U,l_V)$, where $l_U$ and $l_V$ are the total lengths of the strip's $U$ and $V$ edges respectively. If $r > 2$, we subdivide the strip $\text{round}(r)$ times to reduce the elongated direction (see inset). This is done iteratively until no strip is elongated.

\begin{figure}[h]
  \centering
  \includegraphics[width=0.7\linewidth]{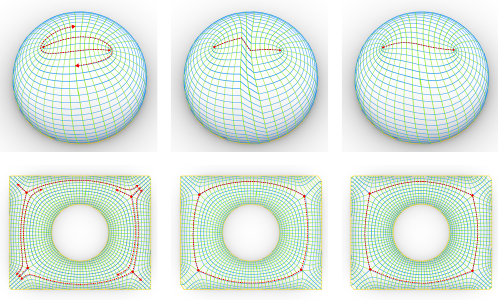}
  \caption{
  Left: Misaligned singularities create winding strips (top) or thin patches (bottom). Middle: Editing operations align the singularities. Right: Smoothing (with 40 iterations) alleviates the deformation caused by editing. 
  }
  \label{fig:smoothing-boundaries_editing}
\end{figure}

\paragraph{Smoothing}
We further reduce the geometric distortion by applying iterative implicit Laplacian smoothing on the vertices $v \in \VV_\QQQ$ by
$(I + dt L_\QQQ) v^{t+1} = v^t$
(\figref{fig:smoothing-boundaries_editing}). Each smoothing iteration consists of the following two steps:
\begin{enumerate}
    \item Smoothen the boundary of the quad mesh $\partial \MM_\QQQ$ as an independent curve and project it onto the original boundary $\partial \MM$.
    \item Smoothen $\MM_\QQQ$ keeping boundary vertices fixed and project it onto the original surface $\MM$.
\end{enumerate}
For the boundary smoothing we use $dt = 0.5$, and for the mesh smoothing, we use $dt = 0.001/\lambda$, where $\lambda$ is the smallest non-zero eigenvalue of the uniform quad-mesh Laplacian $L_\QQQ$.

\subsection{Topological Partitioning}
\label{subsection:patitioning}

The topological partitioning of the resulting SDQ mesh into singularity-free strip sub-networks is useful for all sequential fabrication processes, as it allows to determine an ordering of the resulting strips up to orientation. For example this is useful for deciding the assembly sequence of a surface made of flat sheets, or a print sequence of a series of print paths for 3D printing. 

Once a good patch layout has been achieved with desired alignments of singularities, separating the strip network into good patches is trivial. As we have two transversal strip networks, the user must select along which network the cuts should be aligned. In the following we assume that partitioning happens along the $U$ direction, meaning that cuts run along the $U$ edges (except close to singularities with valence 2) and cut transversally the $V$ strips.   

\begin{figure}[h]
  \centering
  \includegraphics[width=\linewidth]{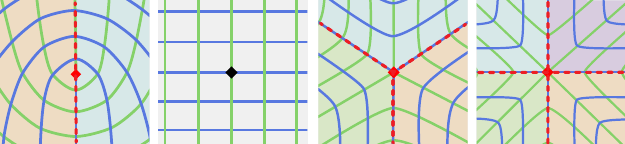}
  \caption{Vertices with valence 2, 4, 6, 8 from left to right. When we walk around a vertex, we always encounter a $U$ (blue) edge after a $V$ (green) edge. For singular vertices (a,c,d) we cut along the separatrices of singularities.}
  \label{fig:singularities}
\end{figure}

To create such a partitioning, we make cuts along the $U$ separatrices of singularities (\figref{fig:singularities}). 
However, for singularities with valence 2 (\figref{fig:singularities} left), only cutting the separatrix in the $U$ direction does not separate the surface. In this case, we cut ``through'' the singularity along both the $U$ and $V$ directions.

\begin{figure}[h!]
  \centering
  \includegraphics[width=0.95\linewidth]{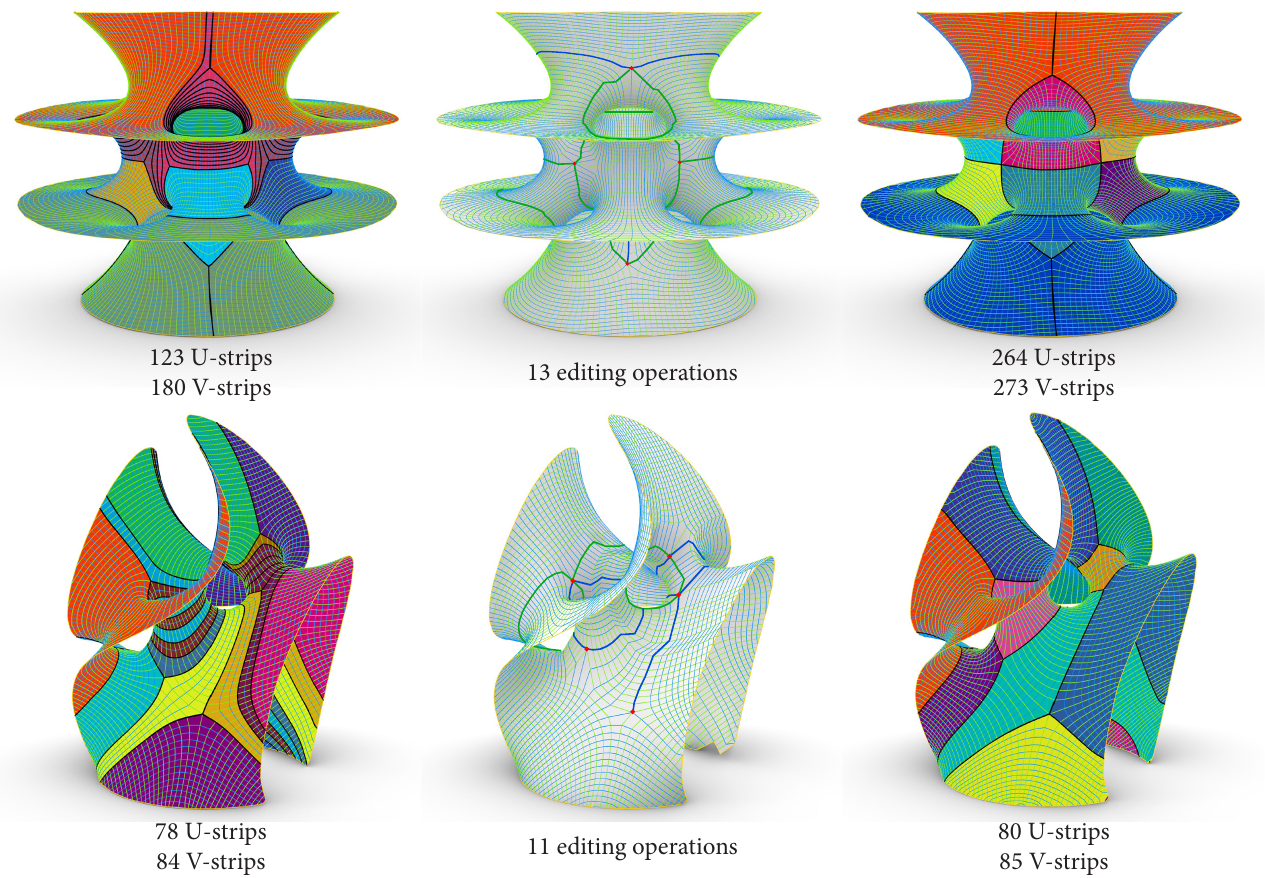}
  \caption{
      Topological editing and partitioning on path networks generated using curvature alignment constraints for the Costa minimal surface (top) and a custom-modeled surface (bottom). 
      Left: original partitioning from singularities. Various small thin and winding patches have formed. 
      Middle: simplified topology achieved by aligning the marked pairs of vertices through the highlighted routes. 
      Right: the resulting network after selective subdivision, 40 iterations of smoothing, and topological partitioning. 
      Note that after the editing operations the number of strips often increases because each winding strip is untangled into several strips.}
  \label{fig:full_editing_operations_pipeline_2}
\end{figure}

Having cut away all singularities, we also cut along short topological handles to obtain simply-connected networks where no singularity or separatrice lies in the interior of the patch. 
\figref{fig:full_editing_operations_pipeline_2} shows results of editing and topological partitioning on more complex shapes.

\section{Results}
\label{sec:Results}
We present the results of our methodology for generating SDQ meshes on various inputs. As mentioned, we focus on relatively smooth geometries with open boundaries and without significant surface detail features. In addition, we apply SDQ meshes for the fabrication scenario of robotic non-planar 3D printing of shell surfaces. 

We use a set of metrics to quantify to what extent the requirements are met on the resulting meshes. Given a sufficiently high-resolution parametrization, these metrics are approximately invariant to the global scaling of the discretization, and are as follows:

\begin{itemize}
    \item \emph{Edge length uniformity error} ($\mathcal{L}$) is the standard deviation of the lengths of all non-boundary edges  divided by the average edge length. It evaluates the deviation from uniform length on the strip networks (objective \ref{obj:uniform}).
    
    \item \emph{Alignment error} ($\mathcal{A}$) with the user-specified directions  measures the difference between the input directional constraints and the orientation of the non-boundary edges of $\MM_\QQQ$ (objective \ref{obj:aligned}). To compare the orientation of an edge $e$ of $\MM_\QQQ$ with directional constraints on the faces of $\MM$, we take the projections of the start ($P_1$) and end ($P_2$) points of $e$ on the closest triangle faces $f_1$ and $f_2$ of $\MM$ that have directional constraints $a_1$, $a_2$ and confidence weights $\omega_1$, $\omega_2$. Note that unconstrained faces have $a, \omega = 0$, and we discard edges with both $f_1$ and $f_2$ unconstrained. After matching the signs of $a_1$ and $a_2$ we find the average constraint direction $a_e = (d_2 a_1 + d_1 a_2) / (d_1 + d_2)$, and the average confidence weight $\omega_e = (d_2 \omega_1 + d_1 \omega_2) / (d_1 + d_2)$, where $d_1$, $d_2$ are the distances of the projected points from the barycenters of $f_1$ and $f_2$. Then we set $\mathcal{A} = \frac{\sum \omega_e * \lvert \widehat{a_e} \cdot \widehat{(P_1 - P_2)} \lvert}{N_e}$, where $N_e$ is the number of edges projected on at least one constrained face. Note that since the directions $a$ constrain the gradient of the parametrization, perfect alignment means that the dominant edges of $\MM_\QQQ$ are orthogonal to the original constraints. 

    \item \emph{Orthogonality error} ($\mathcal{O}$) of the intersections of the strips around each vertex of $\MM_\QQQ$ measures how orthogonal the two strip networks are (objective \ref{obj:orthogonal}). $\mathcal{O} = \frac{\sum{\lvert \frac{\pi}{2} - \theta \lvert}}{N_c}$ for every non-boundary corner of $\MM_\QQQ$ with angle $\theta$, where $N_c$ is the total number of non-boundary corners.
    
\end{itemize}

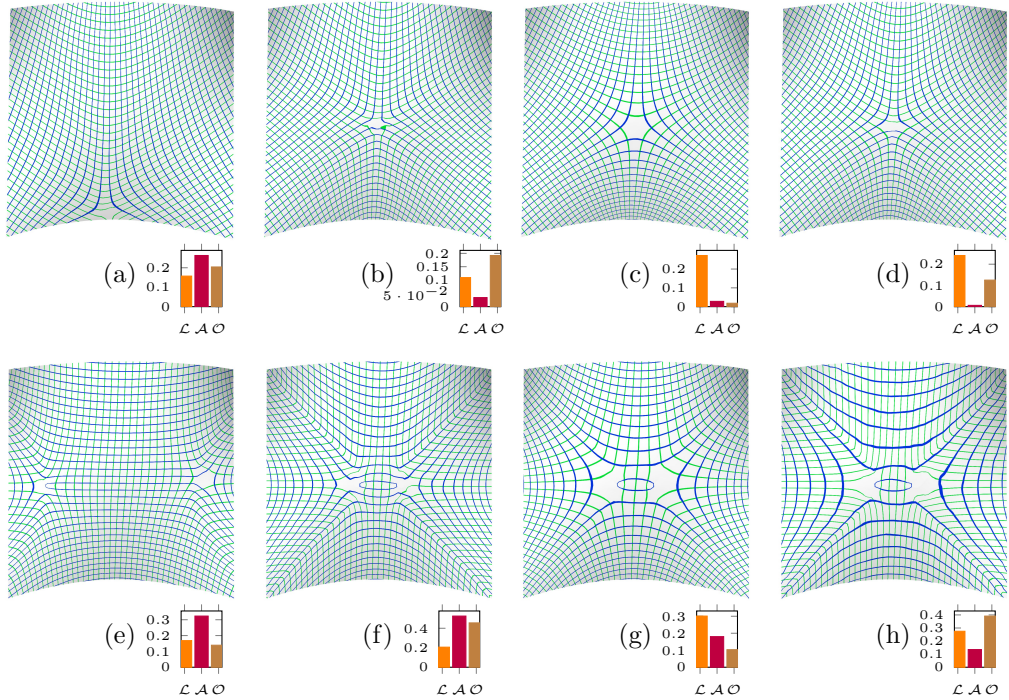
\begin{figure}
    \centering
    \begin{tikzpicture}
        
        \node[inner sep=0pt, anchor=south west] (a) at (0, 0)  {\includegraphics[trim=400 10 240 40, clip, width=0.24\textwidth]{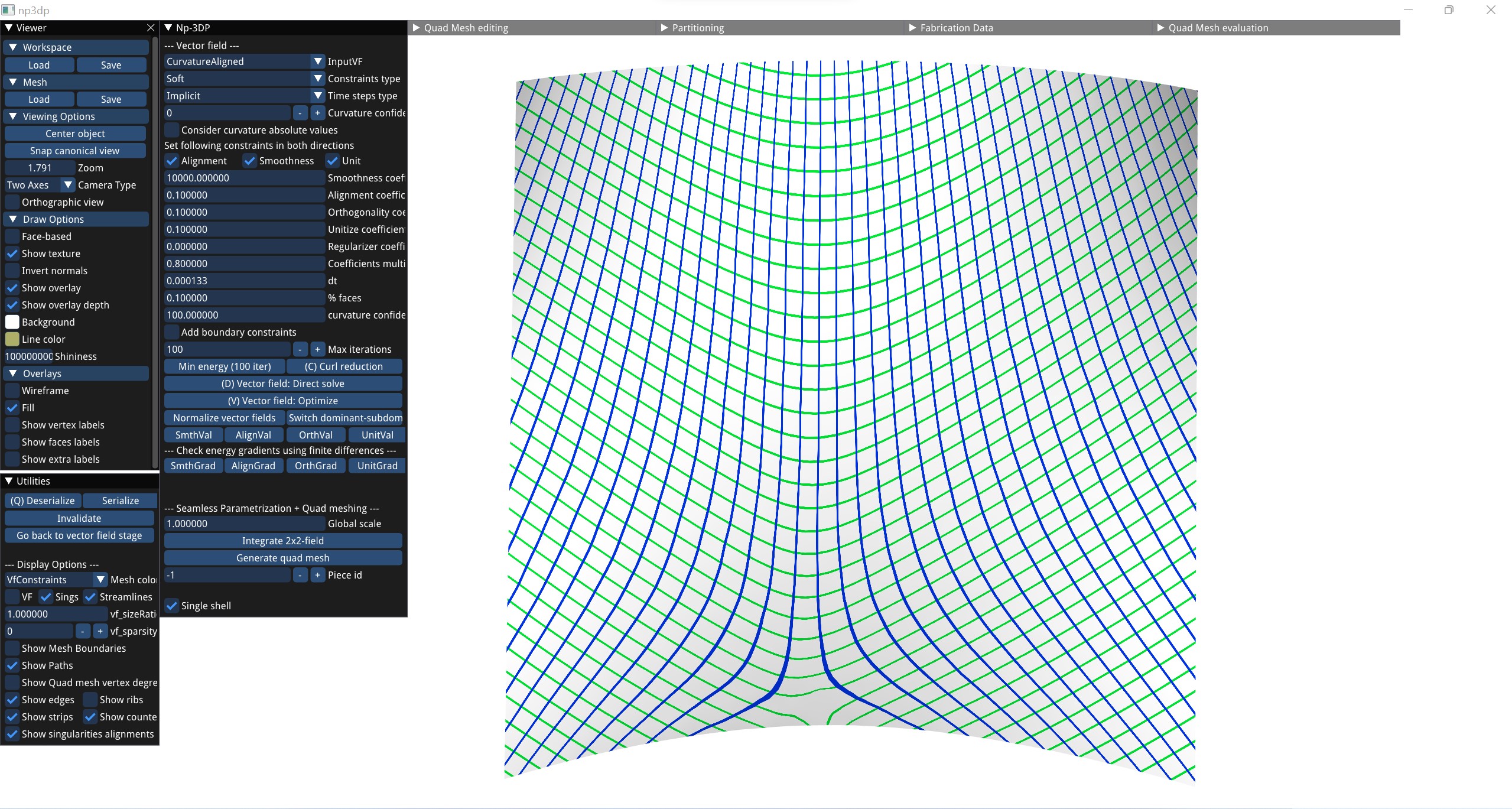}};
        \node[inner sep=0pt, anchor=south west] (b) at (0.25\textwidth, 0) {\includegraphics[trim=400 10 240 40, clip, width=0.24\textwidth]{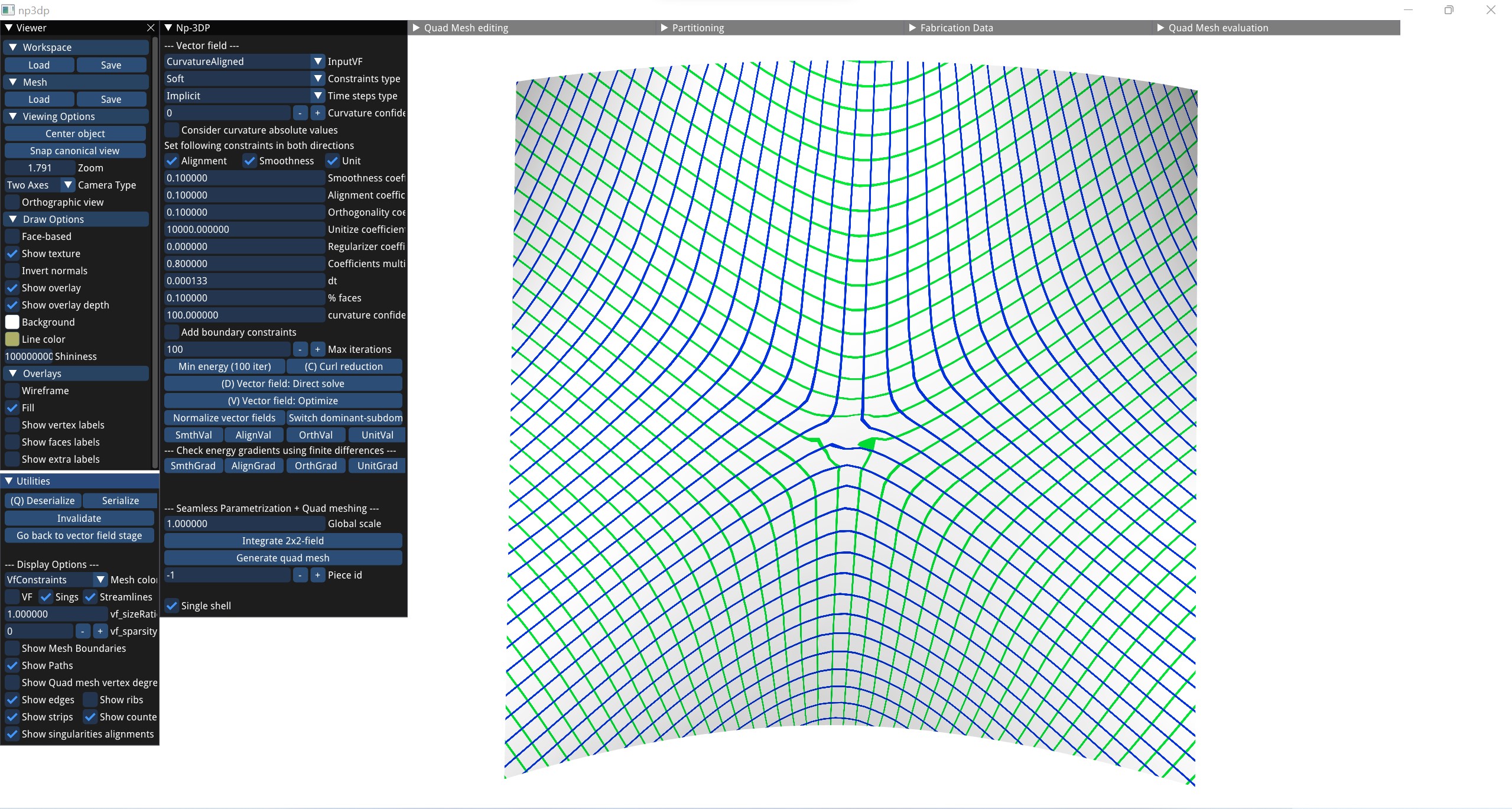}};
        \node[inner sep=0pt, anchor=south west] (c) at (0.50\textwidth, 0) {\includegraphics[trim=400 10 240 40, clip, width=0.24\textwidth]{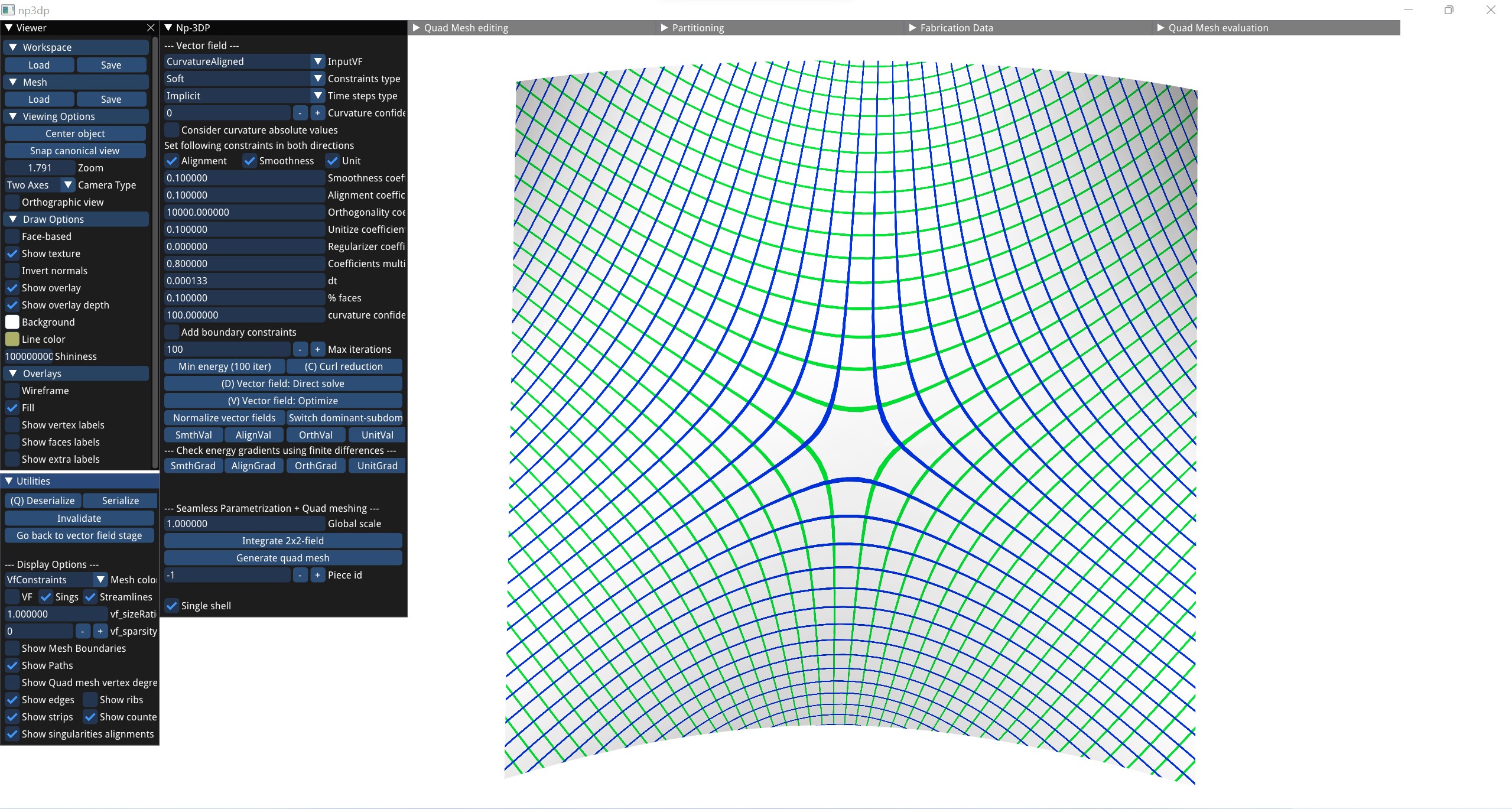}};
        \node[inner sep=0pt, anchor=south west] (d) at (0.75\textwidth, 0) {\includegraphics[trim=400 10 240 40, clip, width=0.24\textwidth]{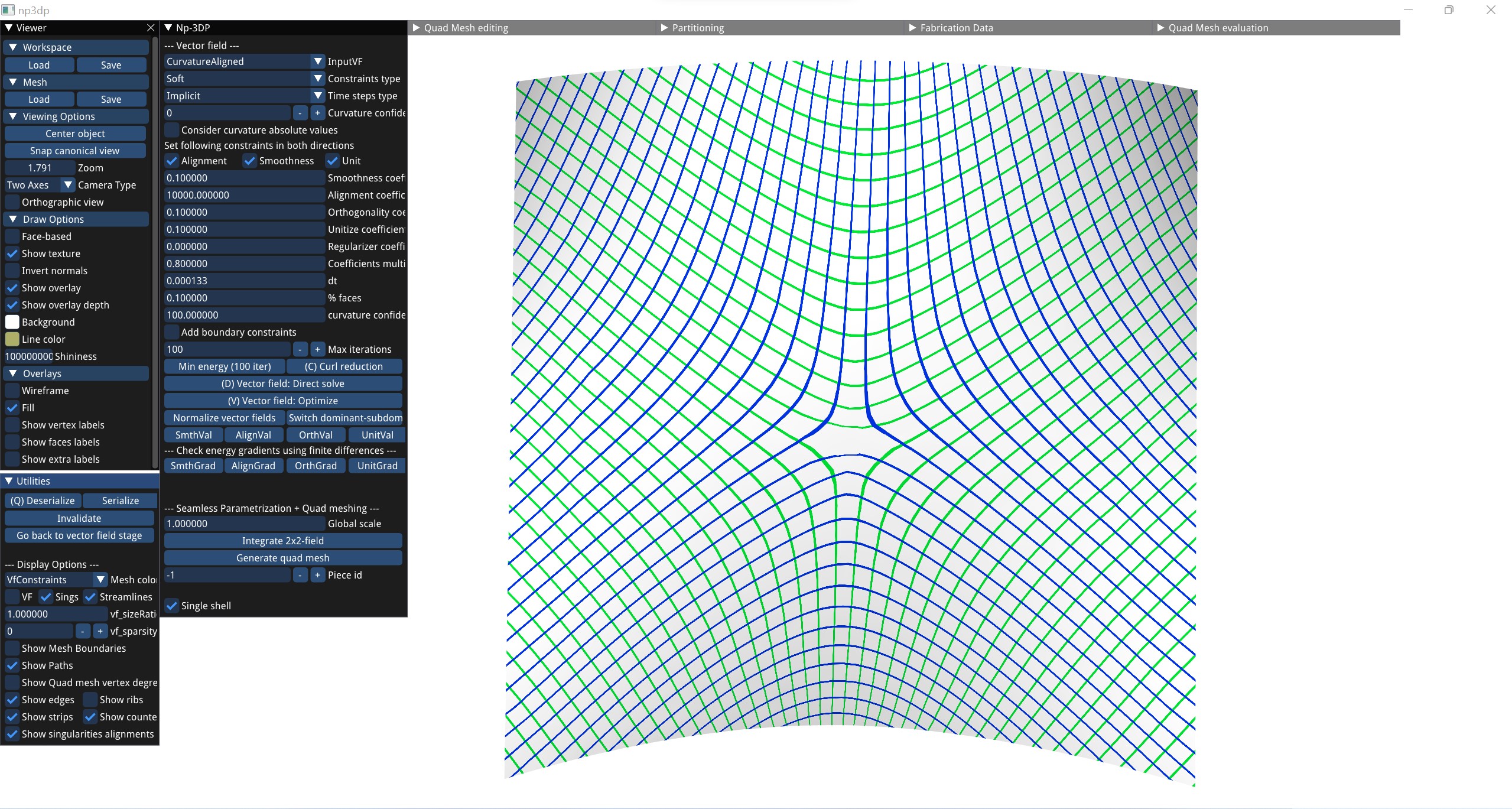}};

        \node[anchor=north, yshift=-3pt] at (a.south) {\footnotesize{(a)}};
        \node[anchor=north, yshift=-3pt] at (b.south) {\footnotesize{(b)}};
        \node[anchor=north, yshift=-3pt] at (c.south) {\footnotesize{(c)}};
        \node[anchor=north, yshift=-3pt] at (d.south) {\footnotesize{(d)}};

        \node[inner sep=0pt, anchor=south west] (e) at (0, -0.35\textwidth) {\includegraphics[trim=400 10 240 40, clip, width=0.24\textwidth]{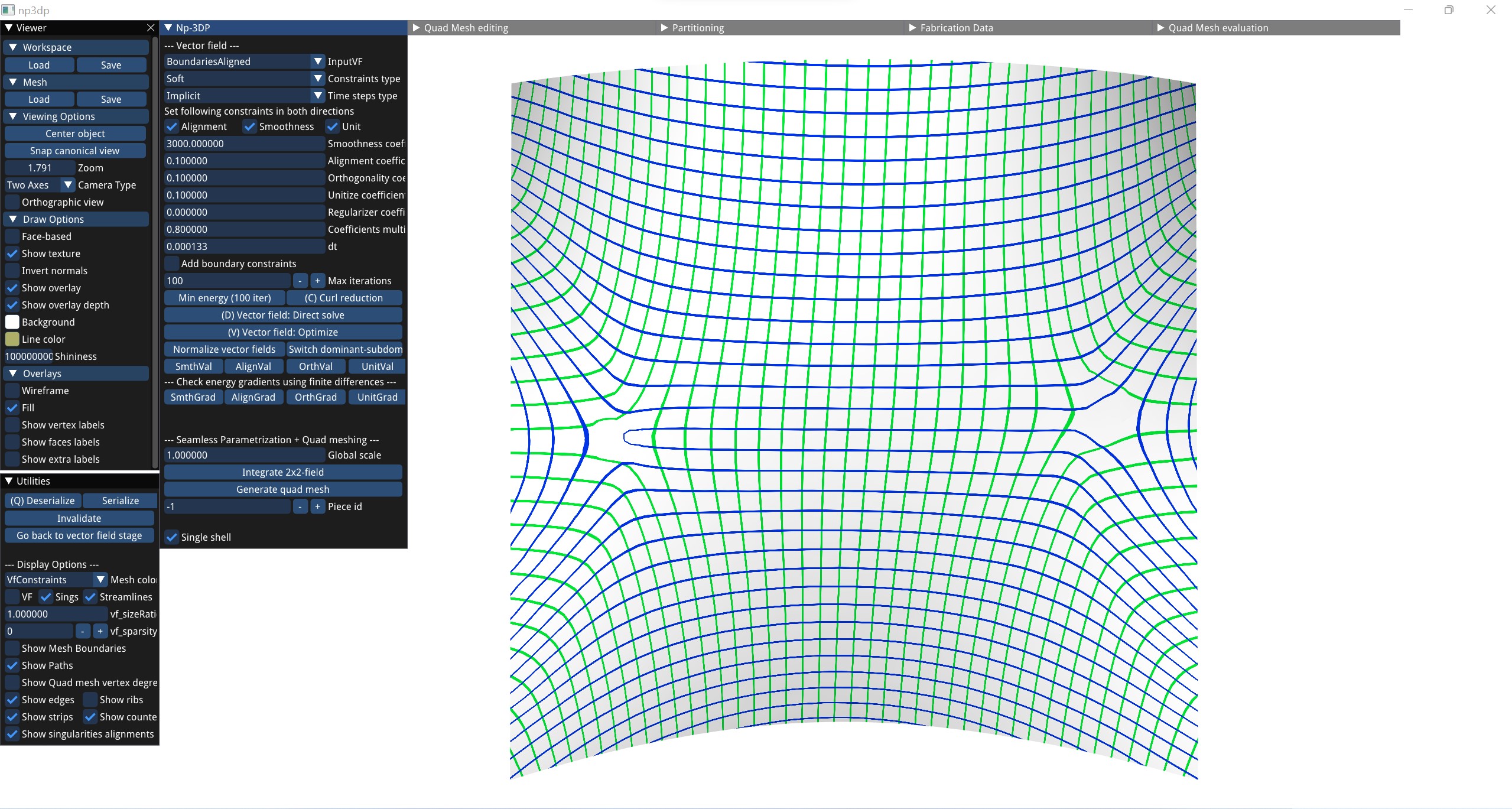}};
        \node[inner sep=0pt, anchor=south west] (f) at (0.25\textwidth, -0.35\textwidth) {\includegraphics[trim=400 10 240 40, clip, width=0.24\textwidth]{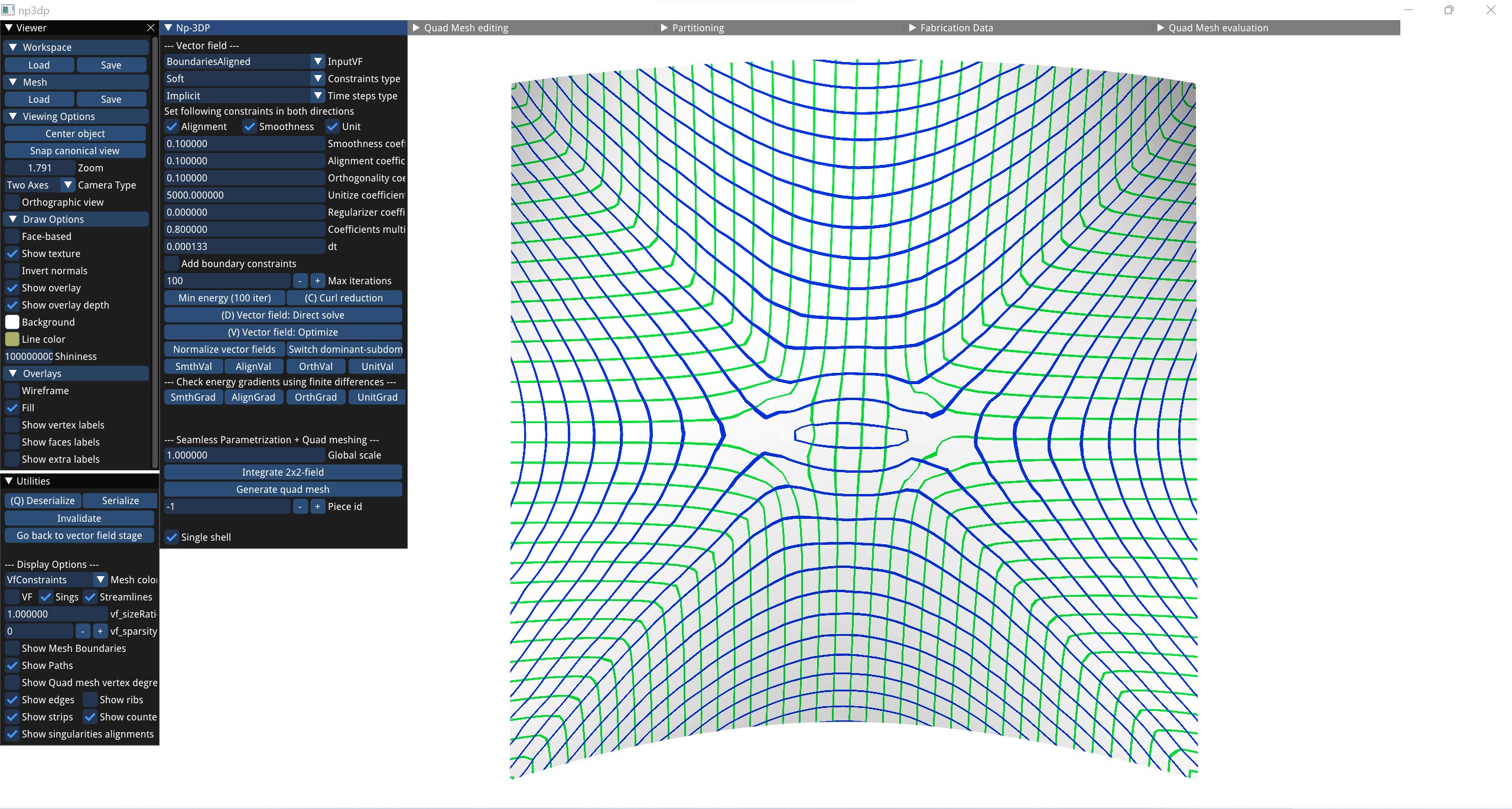}};
        \node[inner sep=0pt, anchor=south west] (g) at (0.50\textwidth, -0.35\textwidth) {\includegraphics[trim=400 10 240 40, clip, width=0.24\textwidth]{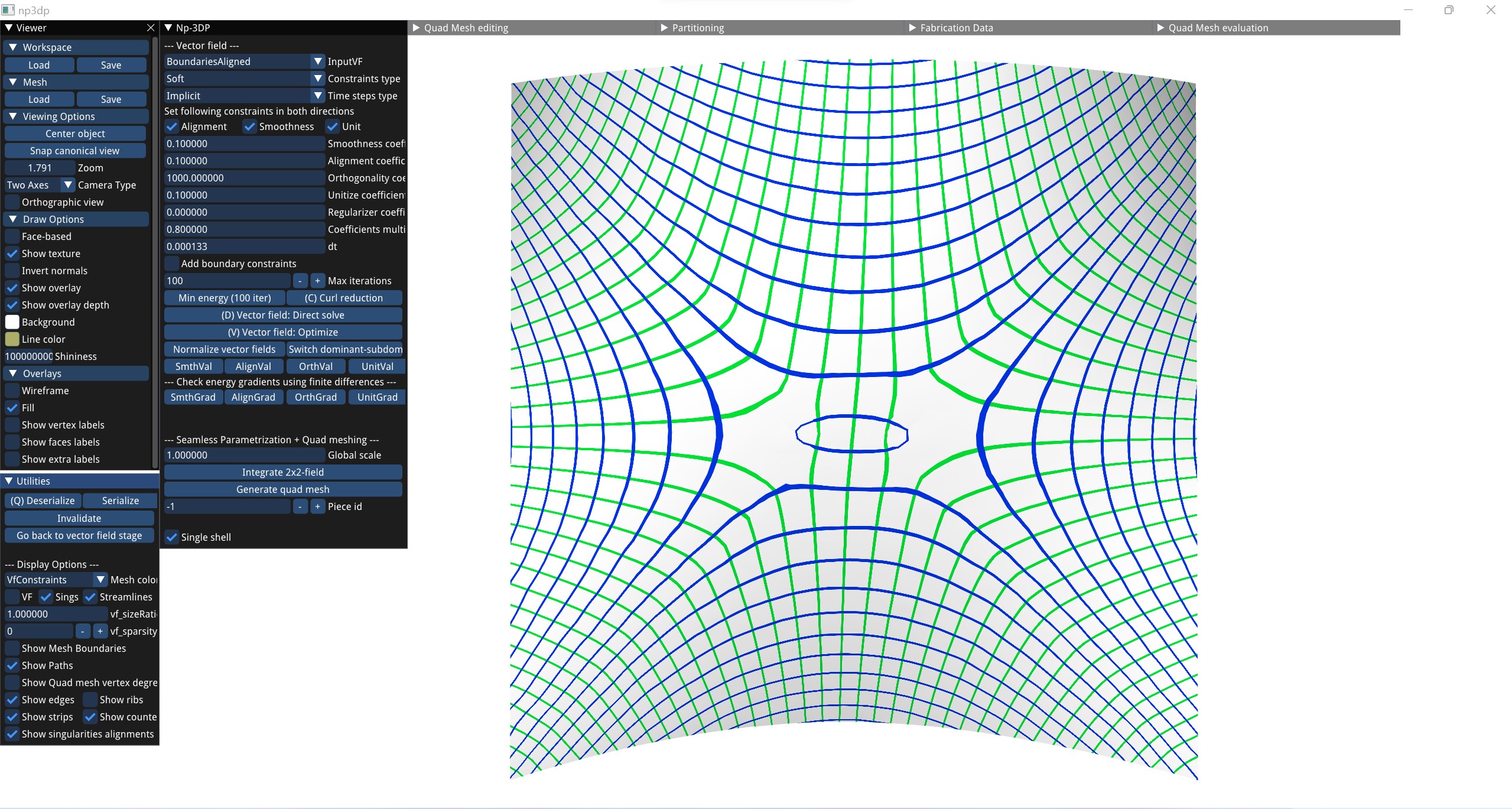}};
        \node[inner sep=0pt, anchor=south west] (h) at (0.75\textwidth, -0.35\textwidth) {\includegraphics[trim=400 10 240 40, clip, width=0.24\textwidth]{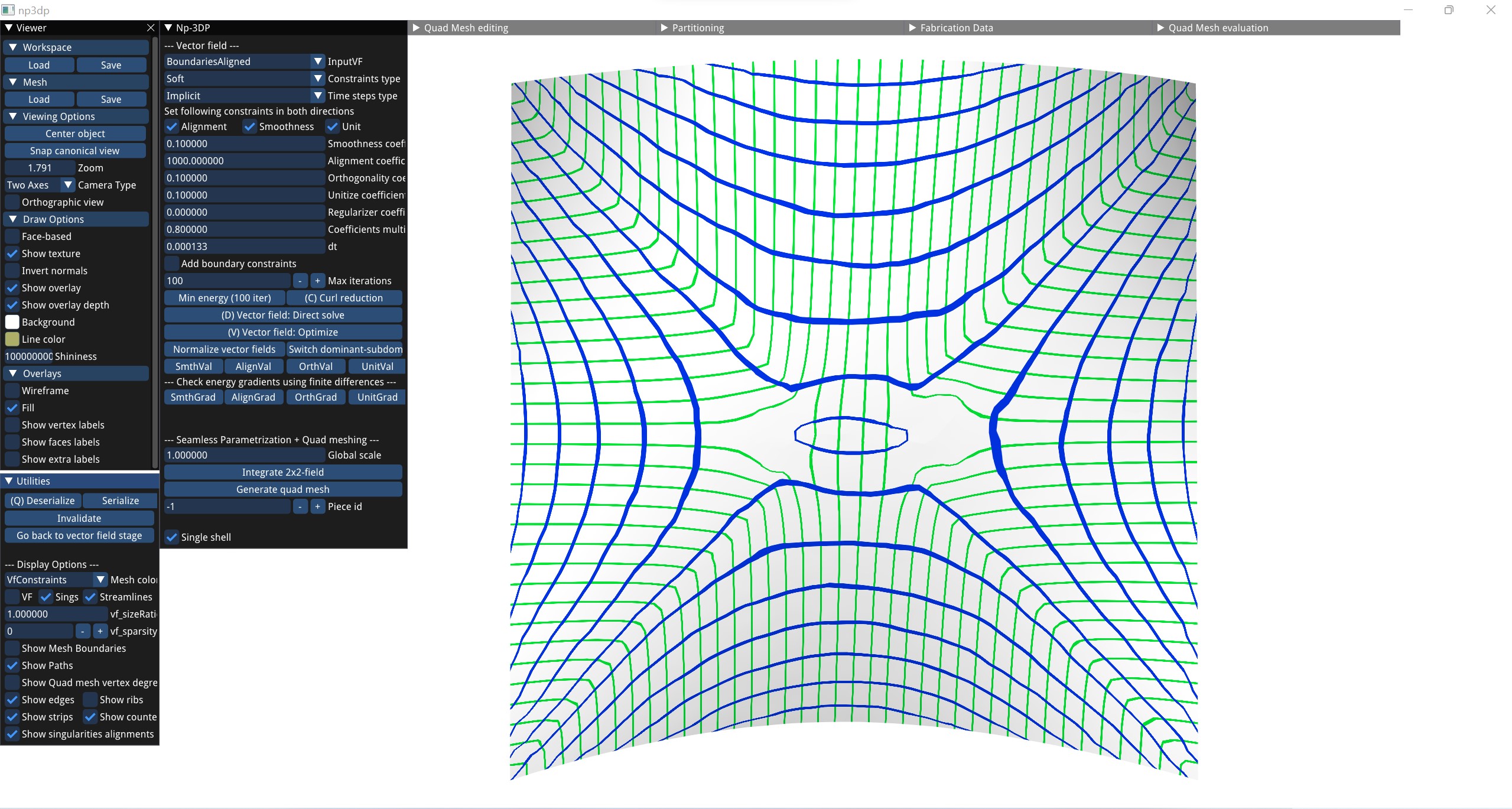}}; 

        \node[anchor=north, yshift=-3pt] at (e.south) {\footnotesize{(e)}};
        \node[anchor=north, yshift=-3pt] at (f.south) {\footnotesize{(f)}};
        \node[anchor=north, yshift=-3pt] at (g.south) {\footnotesize{(g)}};
        \node[anchor=north, yshift=-3pt] at (h.south) {\footnotesize{(h)}};

        \newcommand{\customGraph}[3]{
            \begin{tikzpicture}
                \begin{axis}[  width=0.155\textwidth, height=0.17\textwidth,
                    ybar=0pt,
                    ymin=0,
                    symbolic x coords={$\mathcal{L}$, $\mathcal{A}$, $\mathcal{O}$},
                    tick label style={font=\fontsize{4}{1}}, 
                    label style={font=\fontsize{4}{1}, inner sep=10pt }, 
                    tickwidth=2.5pt
                    ]
                    \addplot[fill=orange, draw=orange, ybar, bar width=6pt, bar shift=-0.05*\pgfplotbarwidth] coordinates {($\mathcal{L}$, #1)};
                    \addplot[fill=purple, draw=purple, ybar, bar width=5pt, bar shift=0.0*\pgfplotbarwidth] coordinates {($\mathcal{A}$, #2)};
                    \addplot[fill=brown, draw=brown, ybar, bar width=6pt, bar shift=0.05*\pgfplotbarwidth] coordinates {($\mathcal{O}$, #3)};
            
                \end{axis}
            \end{tikzpicture}
        }

        \newcommand{\placeCustomGraph}[4]{
            \begin{scope}[shift={(#1.south)}]
                \node[anchor=south] at (27pt, -42pt) { \customGraph{#2}{#3}{#4} };
            \end{scope}
        }

        \placeCustomGraph{a}{0.157}{0.263}{0.204} 
        \placeCustomGraph{b}{0.109}{0.034}{0.192} 
        \placeCustomGraph{c}{0.272}{0.028}{0.018} 
        \placeCustomGraph{d}{0.241}{0.006}{0.125} 
        
        \placeCustomGraph{e}{0.168}{0.323}{0.139} 
        \placeCustomGraph{f}{0.206}{0.525}{0.455} 
        \placeCustomGraph{g}{0.301}{0.180}{0.103} 
        \placeCustomGraph{h}{0.275}{0.135}{0.391} 

    \end{tikzpicture}

    \caption{
    Resulting parametrization using different energy parameters, strongly prioritizing one energy every time. Top: curvature-aligned directional constraints, Bottom: boundary-aligned directional constraints. (a,e) Smoothness coefficient = 1000 results in low lengths uniformity and orthogonality error, but high alignment error. (b,f) Unit coefficient = 1000 results in strips with more uniform widts, i.e., low lengths uniformity error. (c,g) Orthogonality coefficient = 1000 results in low orthogonality error. (d,h) Alignment coefficient = 1000 results in low alignment error. All other coefficients are set to 0.1. The type of constraints also significantly affects the resulting evaluation metrics.}
    
  \label{fig:energy_params}
\end{figure}

In \figref{fig:energy_params}, we ablate the prioritizing of each energy term presented in \secref{subsec:energies}. 

The bar graphs show how the metrics $\mathcal{L}$,  $\mathcal{A}$ and $\mathcal{O}$ vary with the different optimization parameters. 

\begin{figure}
    \centering
    \begin{tikzpicture}
        
        \node[inner sep=0pt, draw=white, anchor=south west] (a) at (0, 0)  {\includegraphics[trim=480 30 610 190, clip, width=0.19\textwidth]{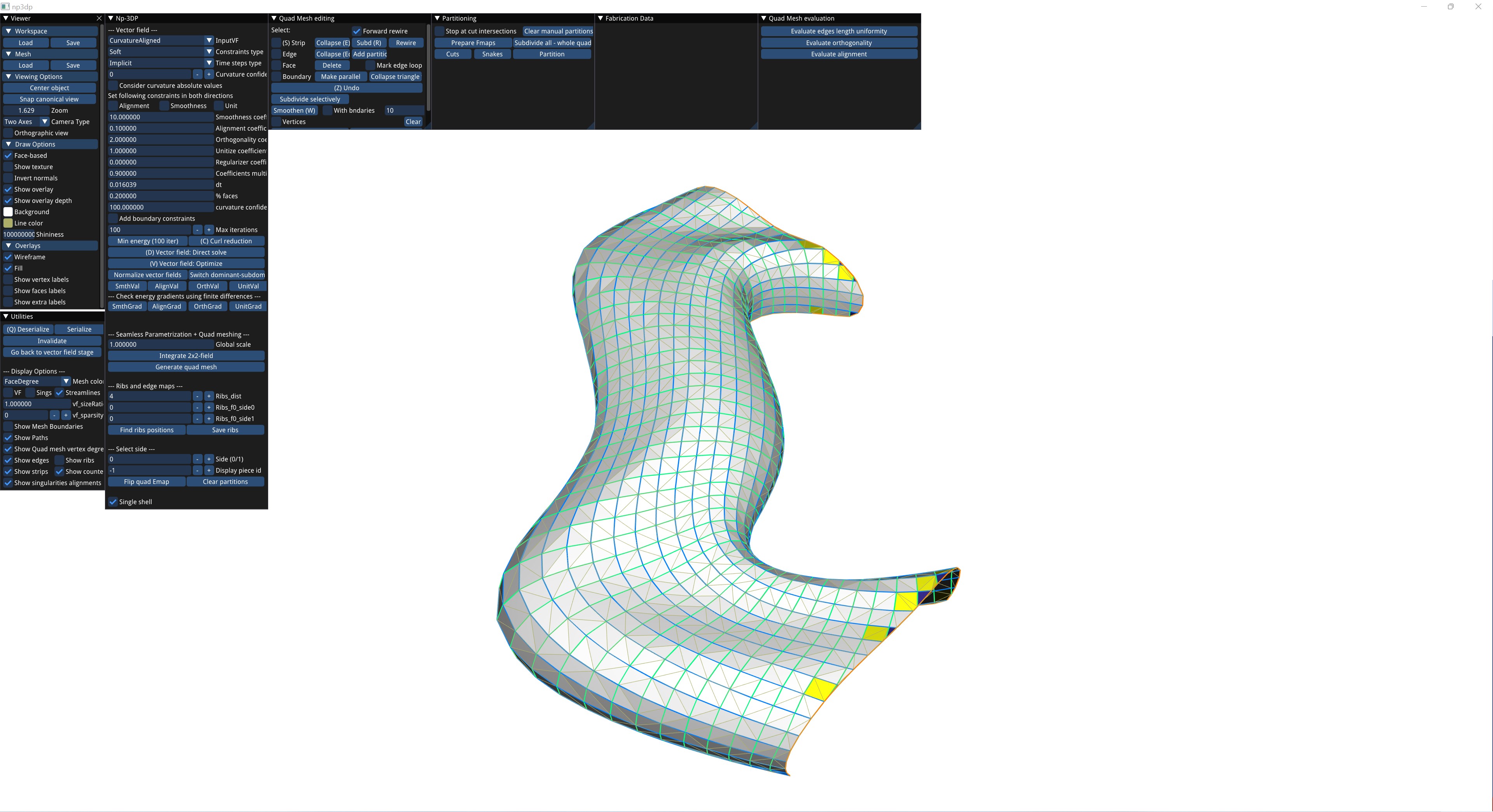}};
        \node[inner sep=0pt,  draw=white, anchor=south west] (b) at (0.20\textwidth, 0) {\includegraphics[trim=510 00 520 170, clip, width=0.19\textwidth]{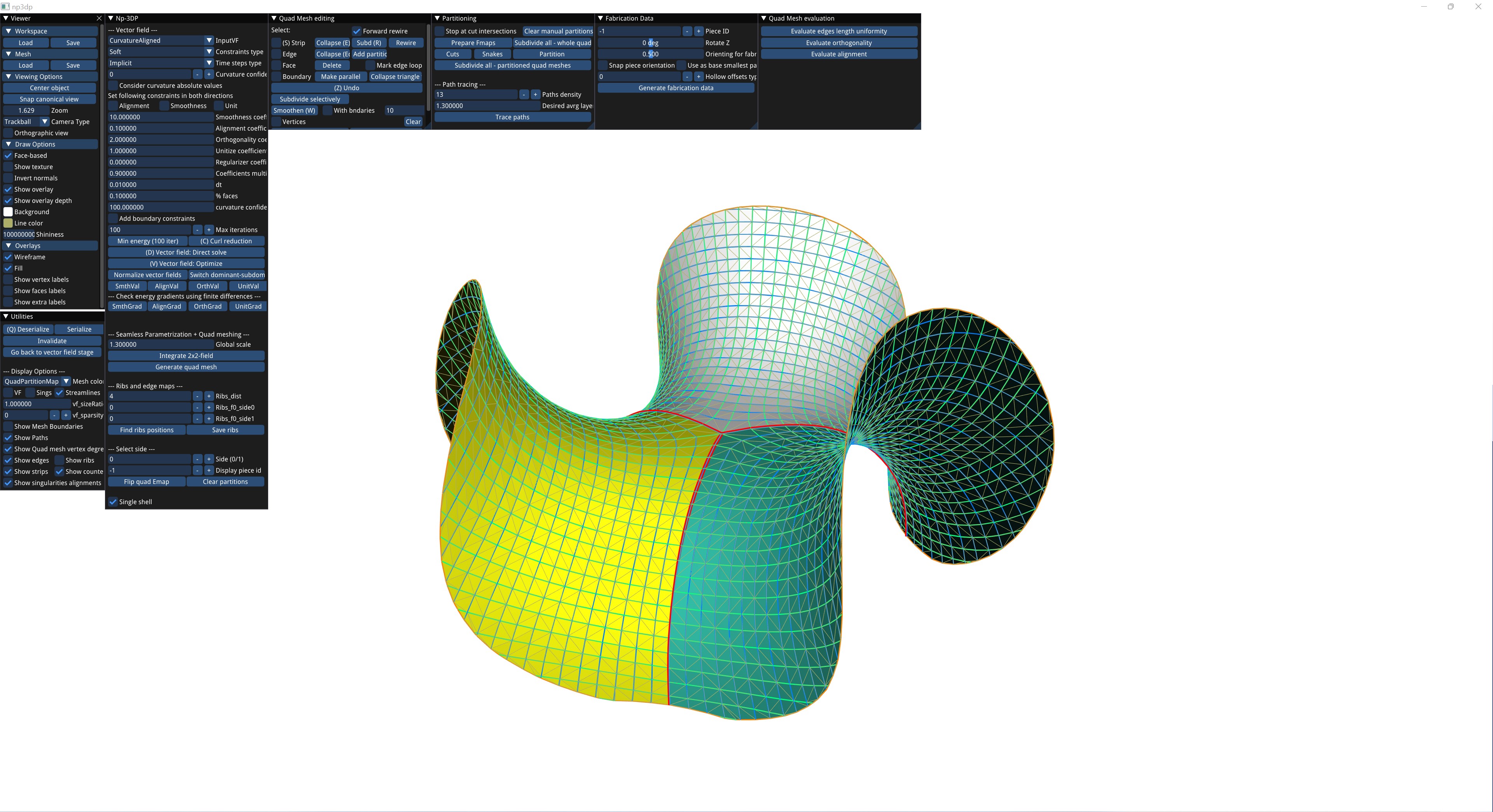}};
        \node[inner sep=0pt,  draw=white, anchor=south west] (c) at (0.40\textwidth, 0) {\includegraphics[trim=440 0 470 170, clip, width=0.19\textwidth]{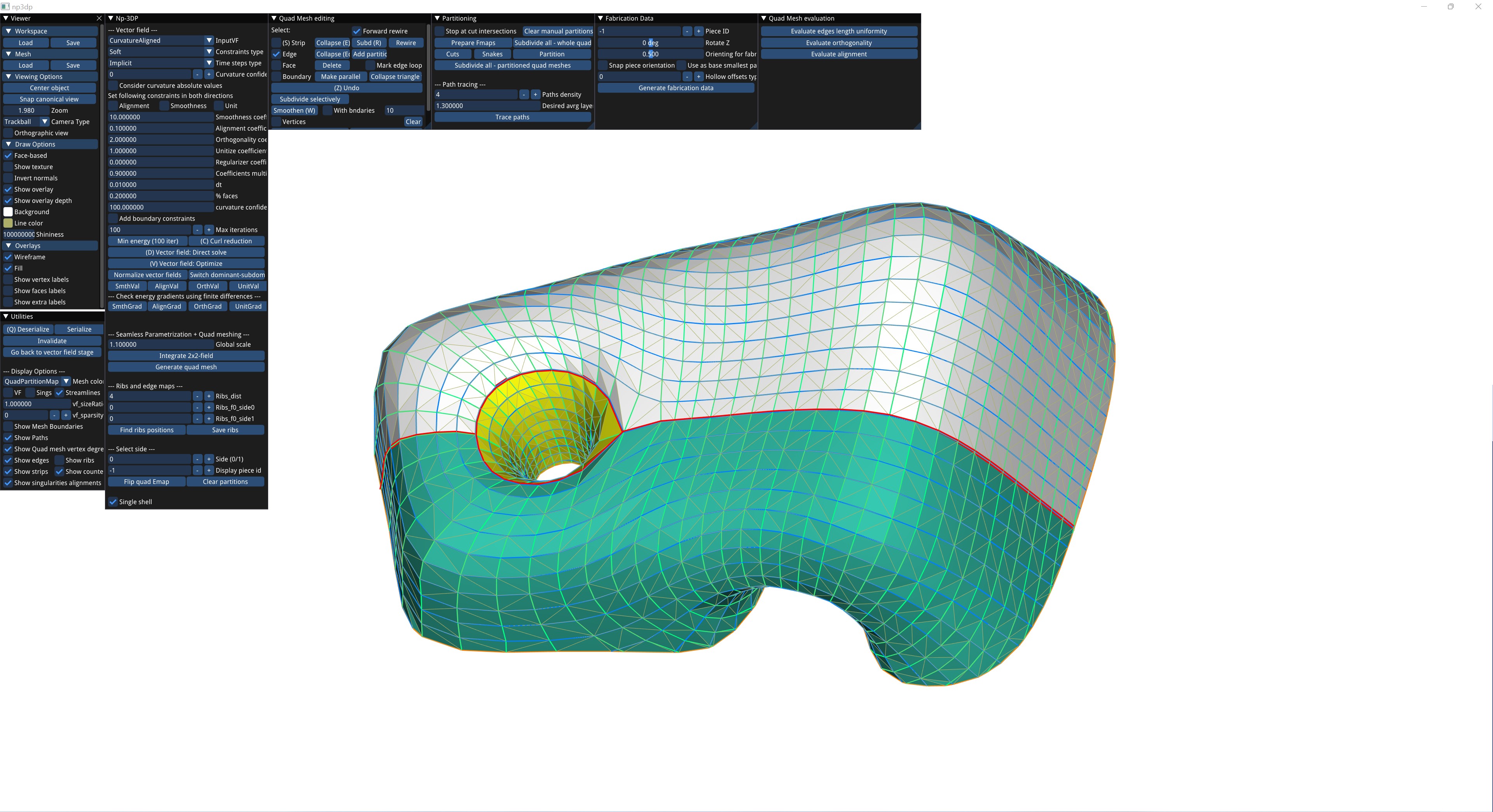}};
        \node[inner sep=0pt,  draw=white, anchor=south west] (d) at (0.60\textwidth, 0) {\includegraphics[trim=540 40 520 170, clip, width=0.19\textwidth]{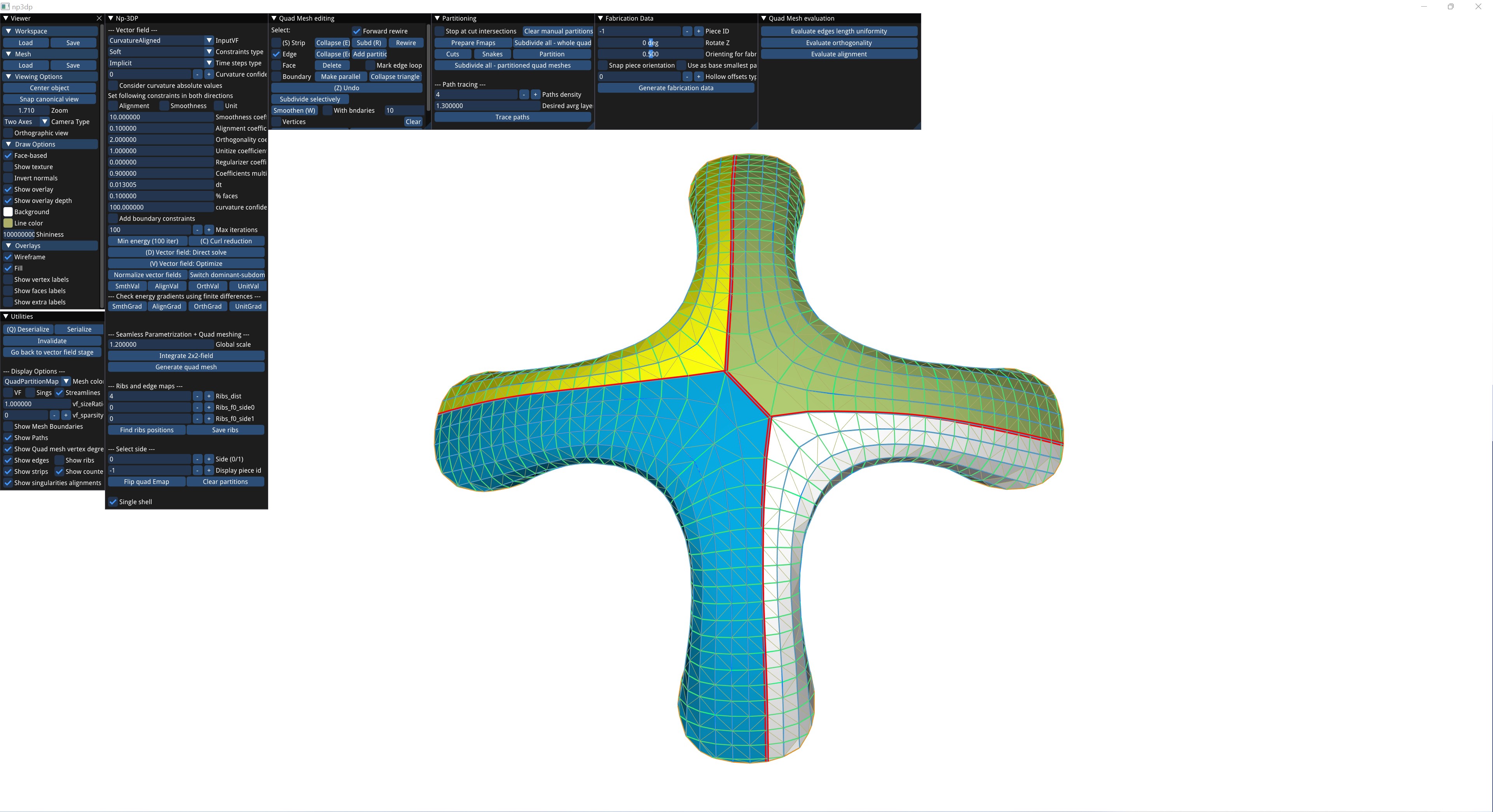}};
        \node[inner sep=0pt,  draw=white, anchor=south west] (e) at (0.80\textwidth, 0) {\includegraphics[trim=550 50 640 230, clip, width=0.19\textwidth]{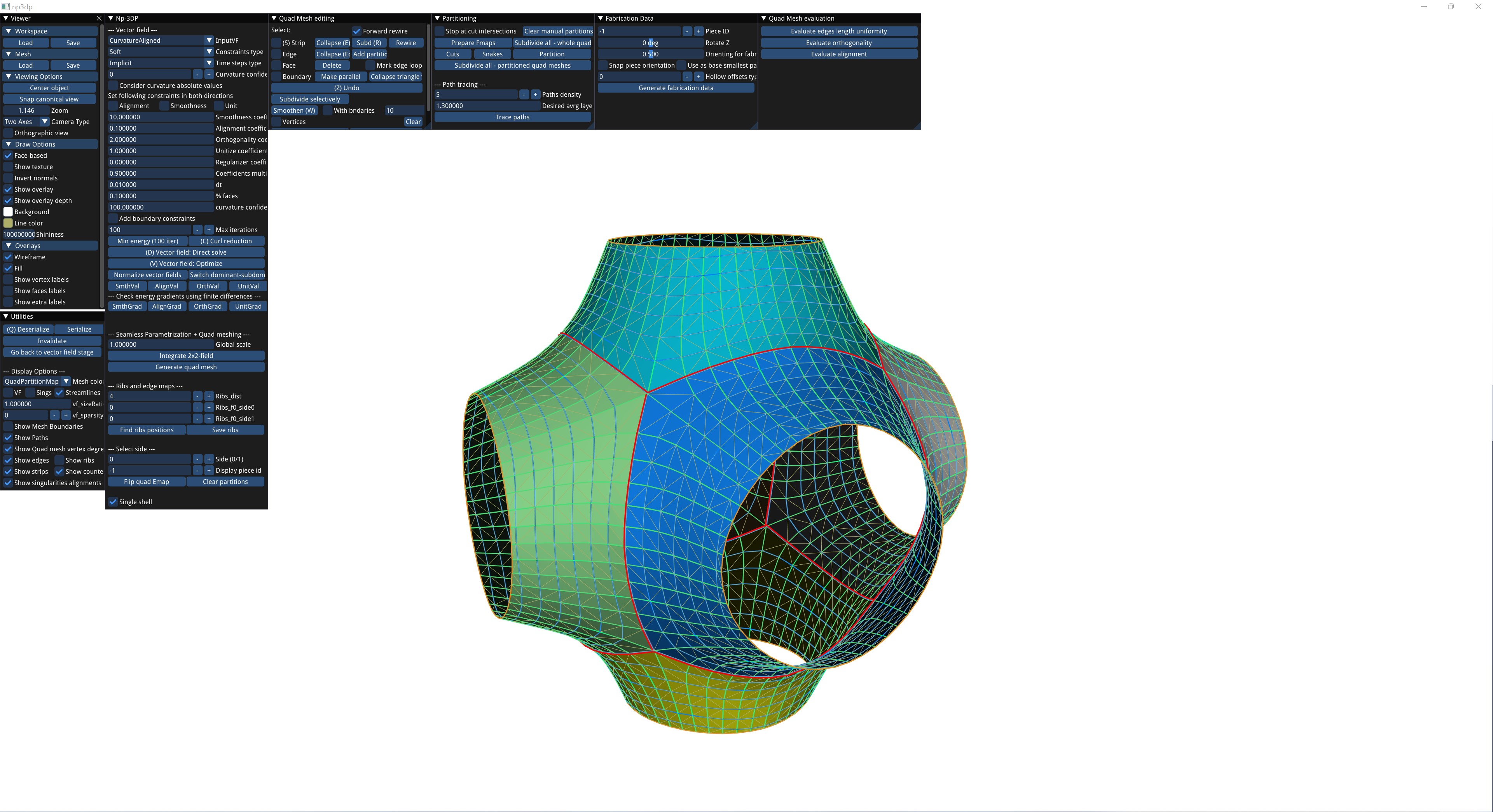}};
        
        \node[anchor=north, yshift=-3pt] at (a.south) {\footnotesize{(a)}};
        \node[anchor=north, yshift=-3pt] at (b.south) {\footnotesize{(b)}};
        \node[anchor=north, yshift=-3pt] at (c.south) {\footnotesize{(c)}};
        \node[anchor=north, yshift=-3pt] at (d.south) {\footnotesize{(d)}};
        \node[anchor=north, yshift=-3pt] at (e.south) {\footnotesize{(e)}};

        \node[inner sep=0pt, draw=white, anchor=south west] (f) at (0, -0.30\textwidth)  {\includegraphics[trim=550 30 550 170, clip, width=0.19\textwidth]{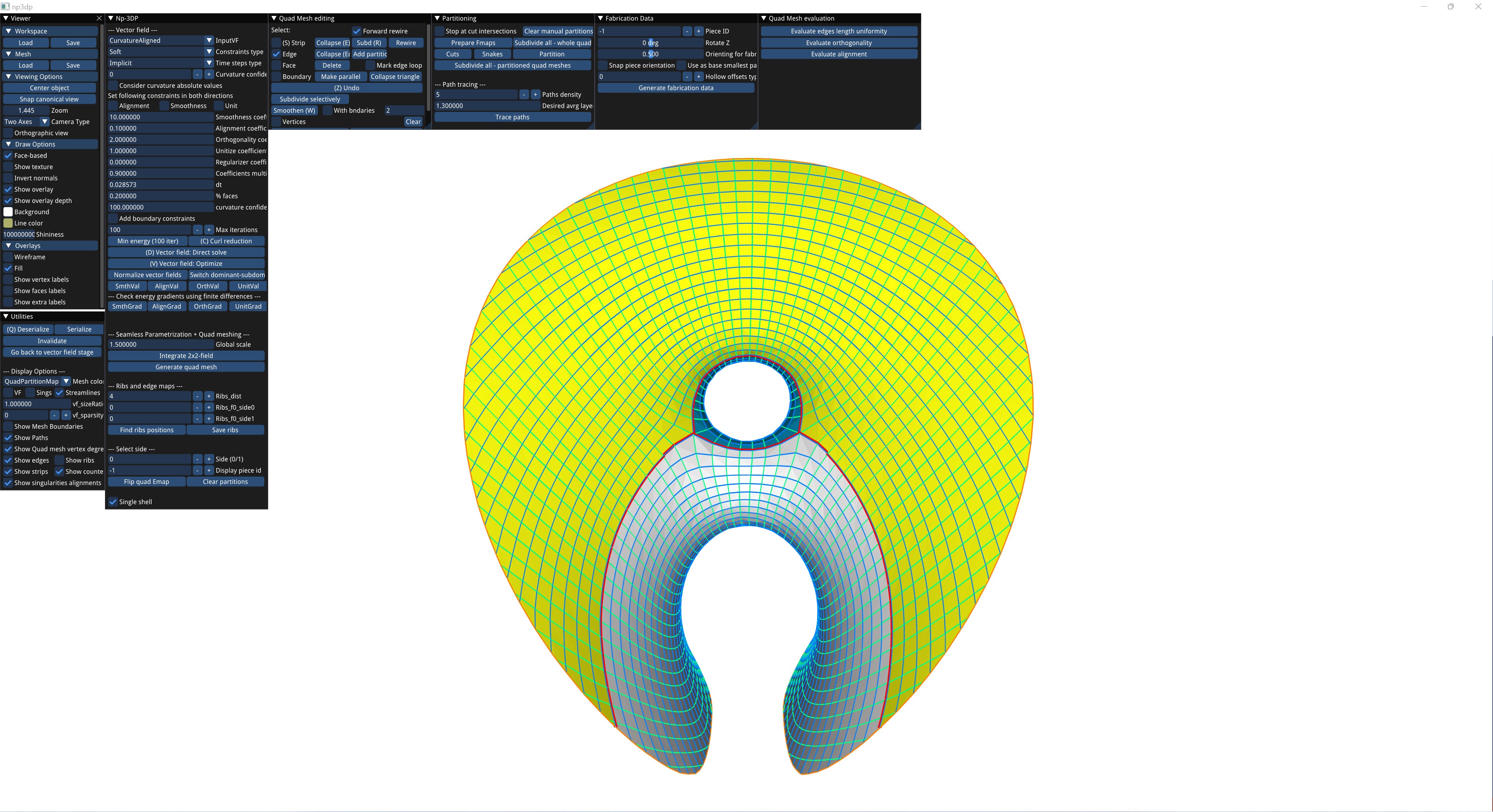}};
        \node[inner sep=0pt,  draw=white, anchor=south west] (g) at (0.20\textwidth, -0.30\textwidth) {\includegraphics[trim=0 0 0 0, clip, width=0.19\textwidth]{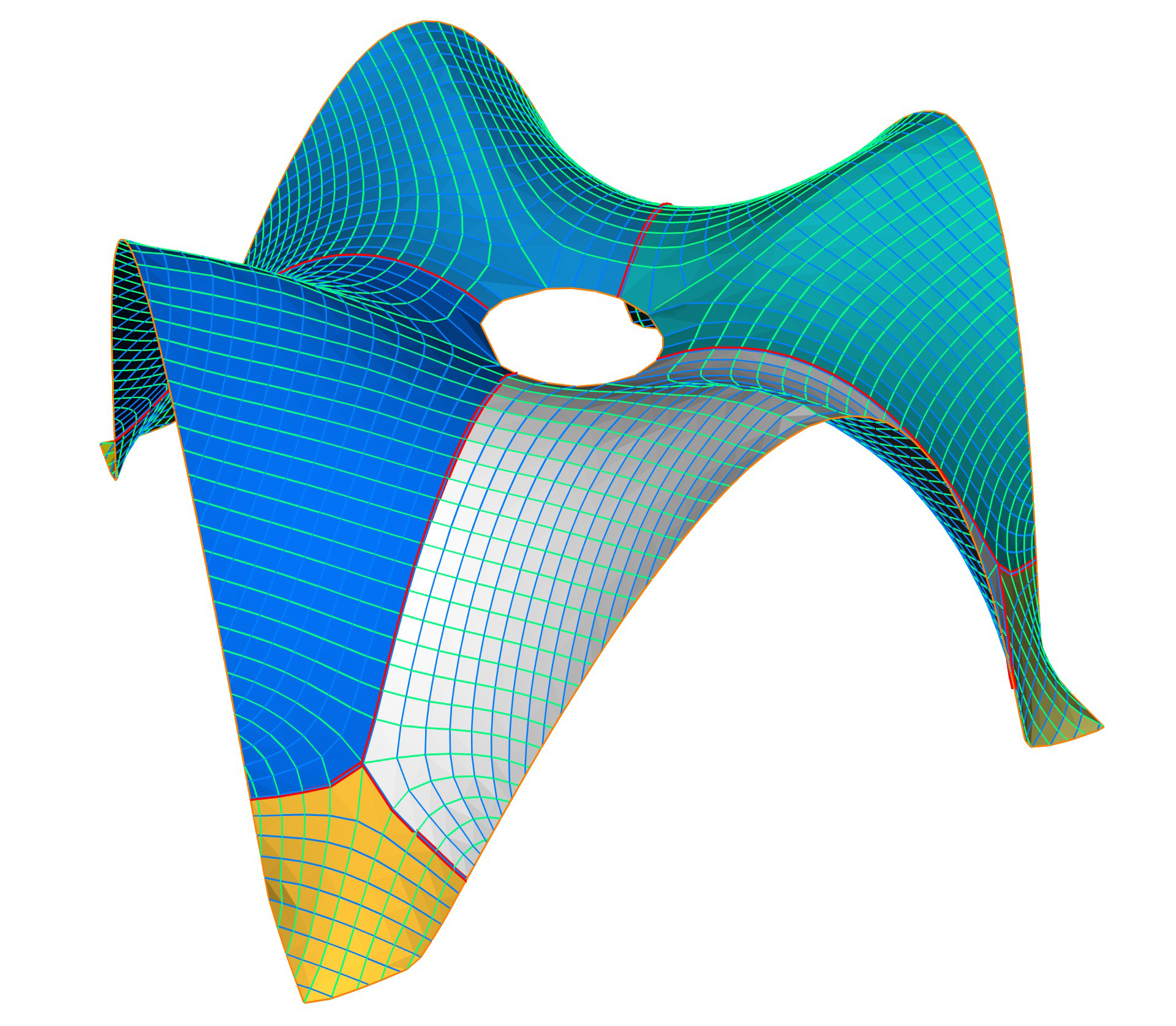}};
        \node[inner sep=0pt,  draw=white, anchor=south west] (h) at (0.40\textwidth, -0.30\textwidth) {\includegraphics[trim=510 0 470 170, clip, width=0.19\textwidth]{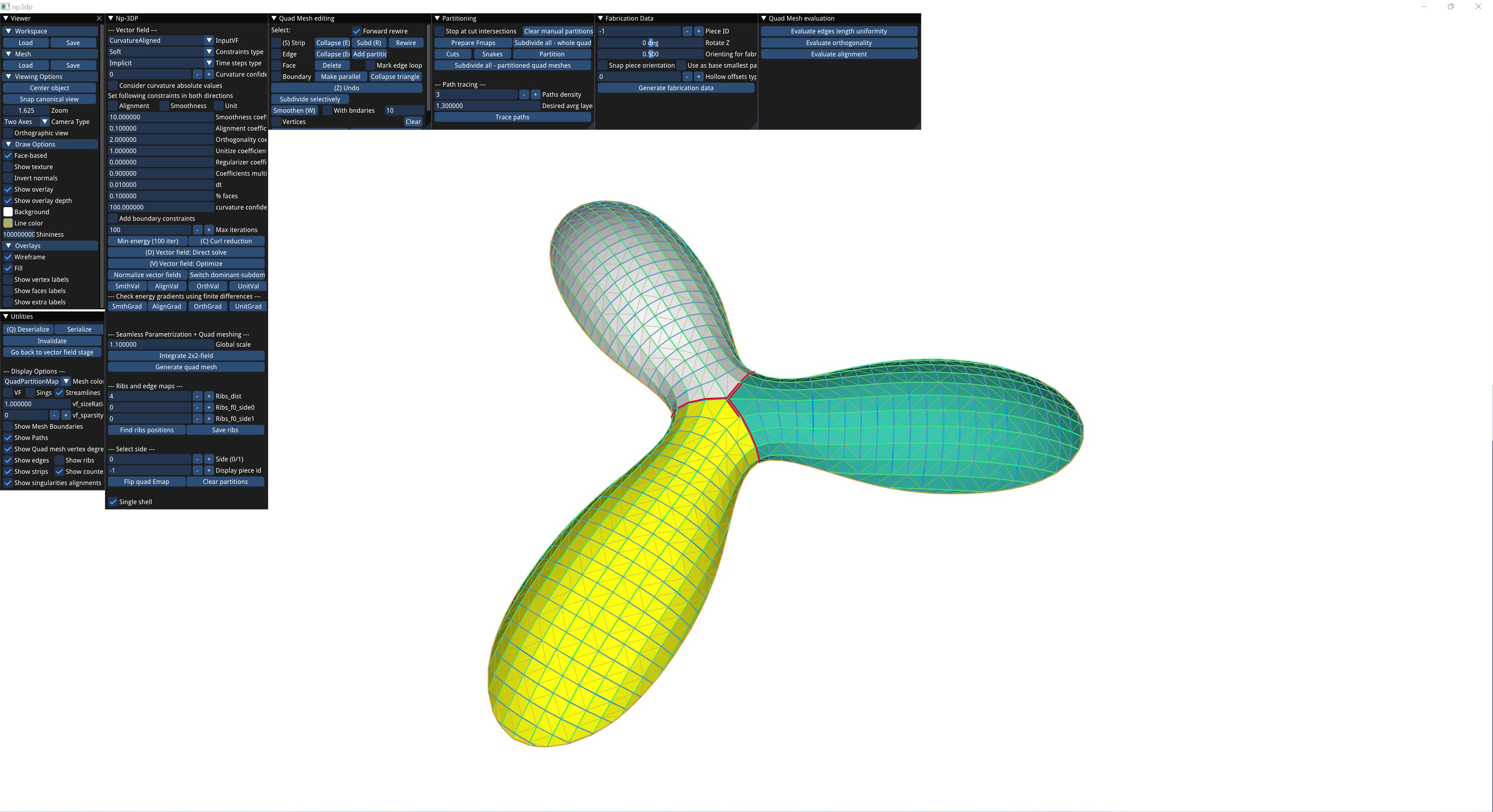}};
        \node[inner sep=0pt,  draw=white, anchor=south west] (i) at (0.60\textwidth, -0.30\textwidth) {\includegraphics[trim=500 00 660 200, clip, width=0.19\textwidth]{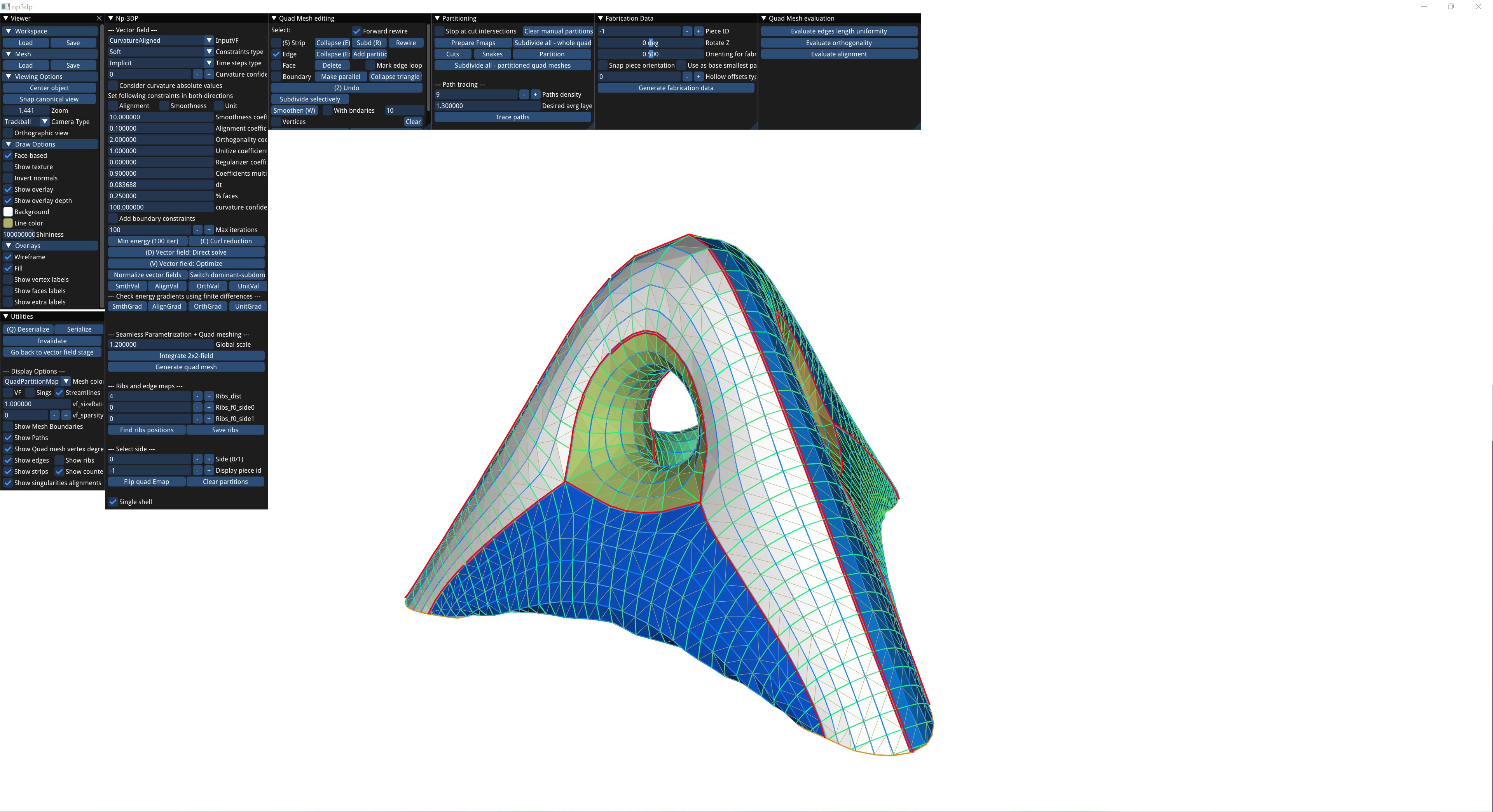}};
        \node[inner sep=0pt,  draw=white, anchor=south west] (j) at (0.80\textwidth, -0.30\textwidth) {\includegraphics[trim=550 50 640 230, clip, width=0.19\textwidth]{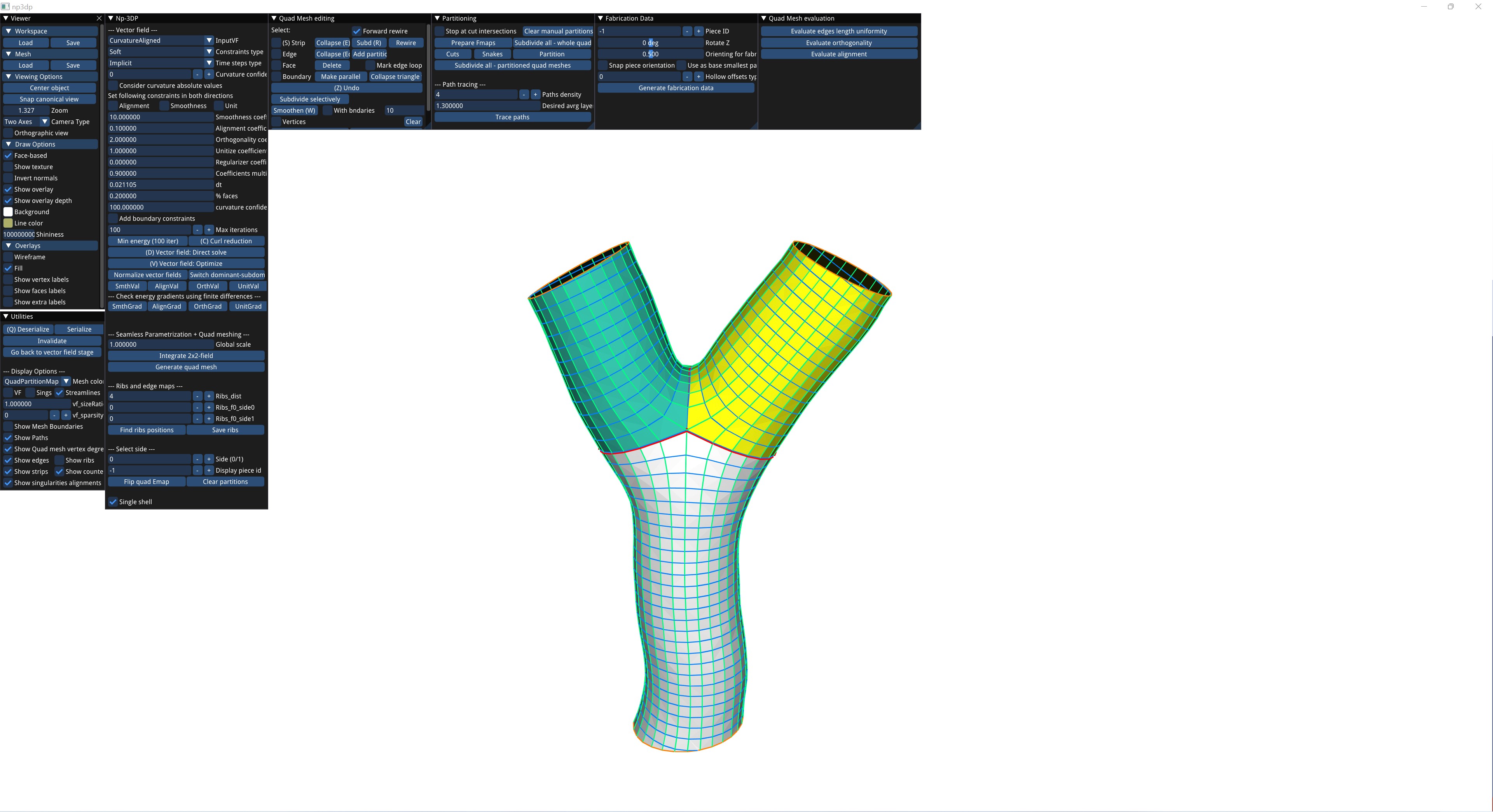}};
        
        \node[anchor=north, yshift=-3pt] at (f.south) {\footnotesize{(f)}};
        \node[anchor=north, yshift=-3pt] at (g.south) {\footnotesize{(g)}};
        \node[anchor=north, yshift=-3pt] at (h.south) {\footnotesize{(h)}};
        \node[anchor=north, yshift=-3pt] at (i.south) {\footnotesize{(i)}};
        \node[anchor=north, yshift=-3pt] at (j.south) {\footnotesize{(j)}};

        \node[inner sep=0pt, draw=white, anchor=south west] (k) at (0, -0.60\textwidth)  {\includegraphics[trim=410 20 470 190, clip, width=0.19\textwidth]{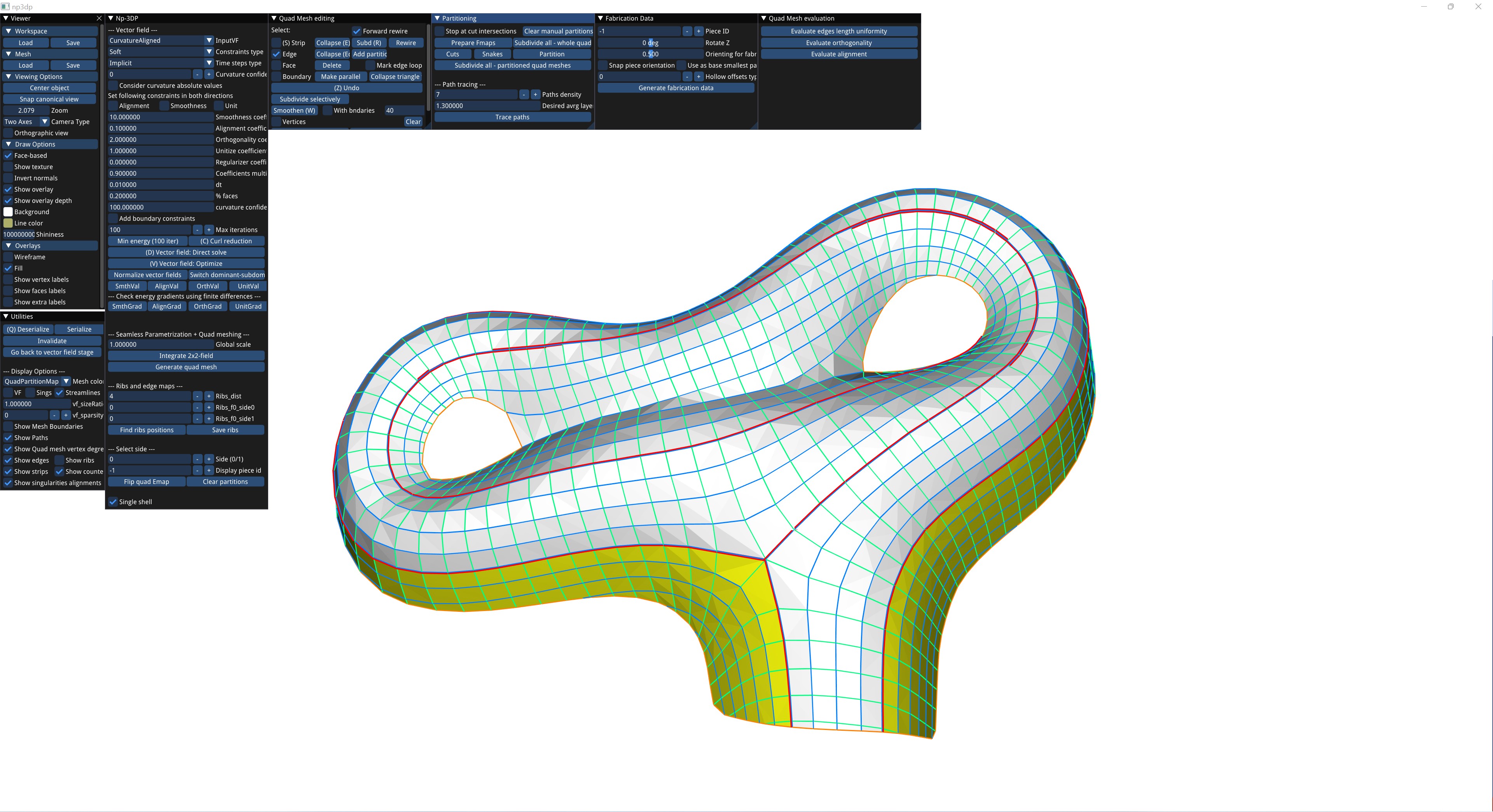}};
        \draw[->, ultra thick, black] ($(k.south)-(0.0,0.25)$) -- ($(k.south)-(0.0,1.0)$);
        \node[anchor=north, xshift=-16pt, yshift = -10pt, align=center] at (k.south) {\tiny{\#4  ops}};
        \node[inner sep=0pt, draw=white, anchor=south west] (k2) at (0, -0.86\textwidth)  {\includegraphics[trim=410 20 470 190, clip, width=0.19\textwidth]{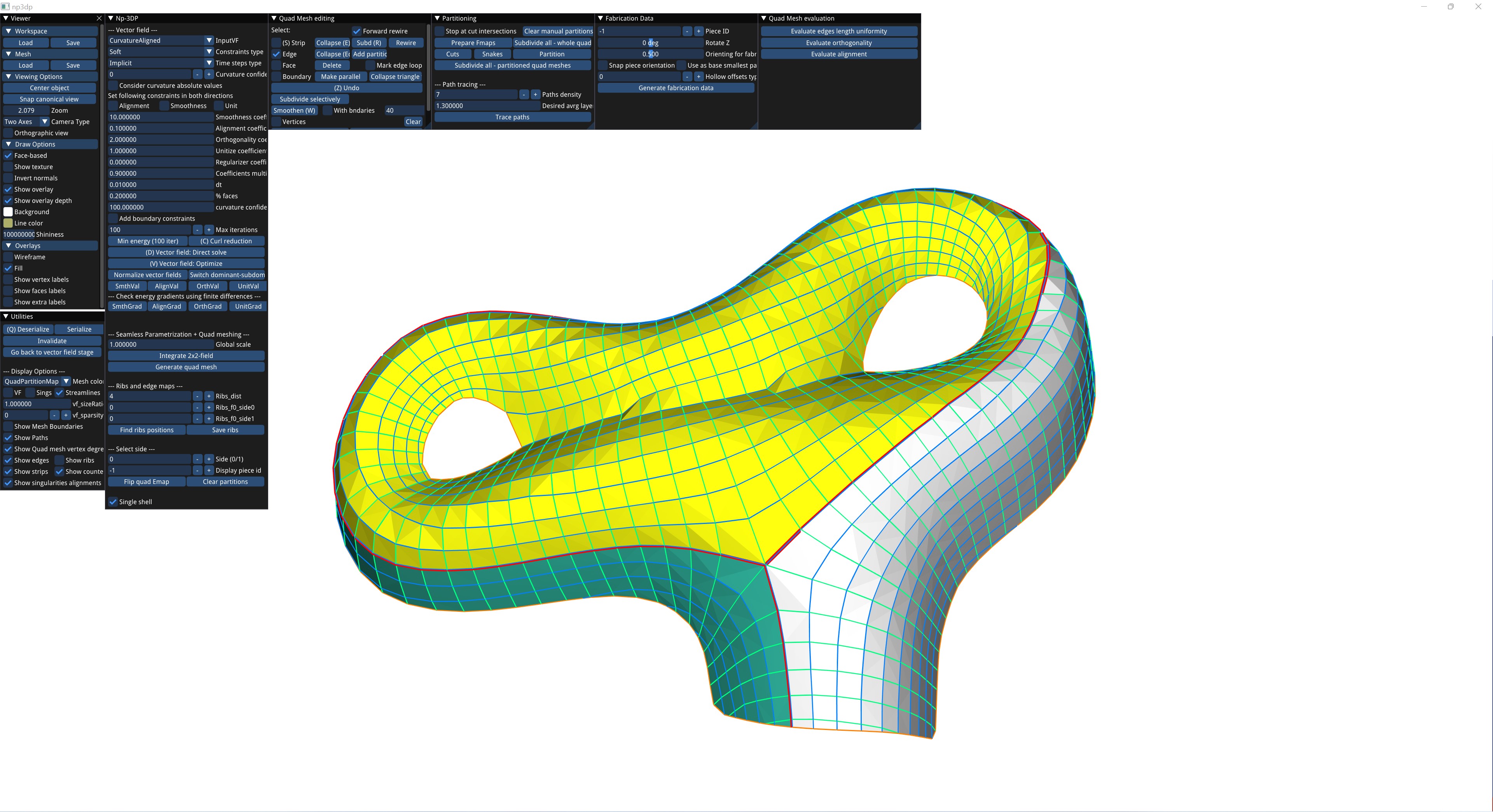}};
        
        \node[inner sep=0pt, draw=white, anchor=south west] (l) at (0.20\textwidth, -0.60\textwidth)  {\includegraphics[trim=490 70 480 190, clip, width=0.19\textwidth]{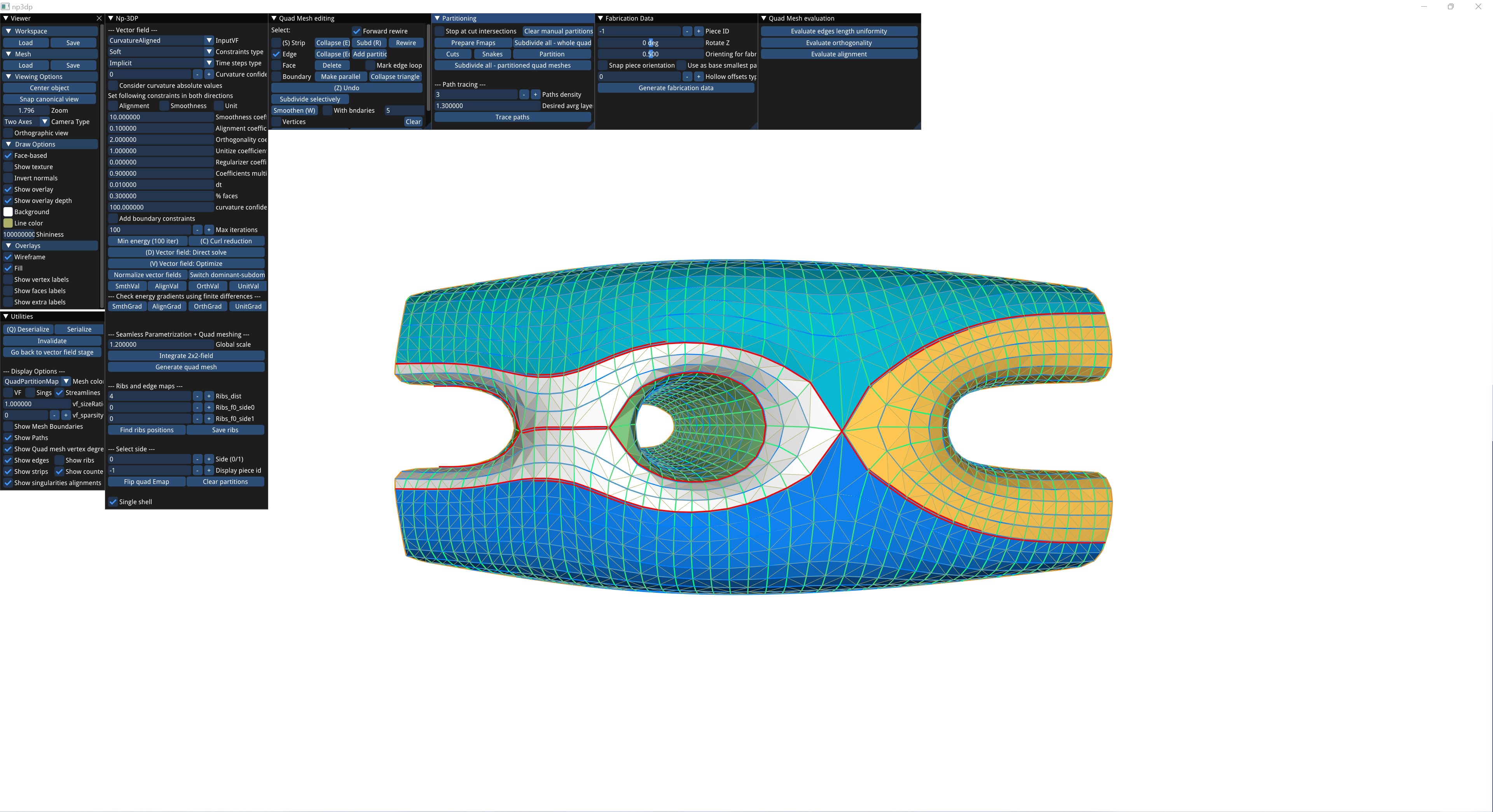}};
        \draw[->, ultra thick, black] ($(l.south)-(0.0,0.25)$) -- ($(l.south)-(0.0,1.0)$);
        \node[anchor=north, xshift=-16pt, yshift = -10pt, align=center] at (l.south) {\tiny{\#3  ops}};
        \node[inner sep=0pt, draw=white, anchor=south west] (l2) at (0.20\textwidth, -0.86\textwidth)  {\includegraphics[trim=490 70 480 190, clip, width=0.19\textwidth]{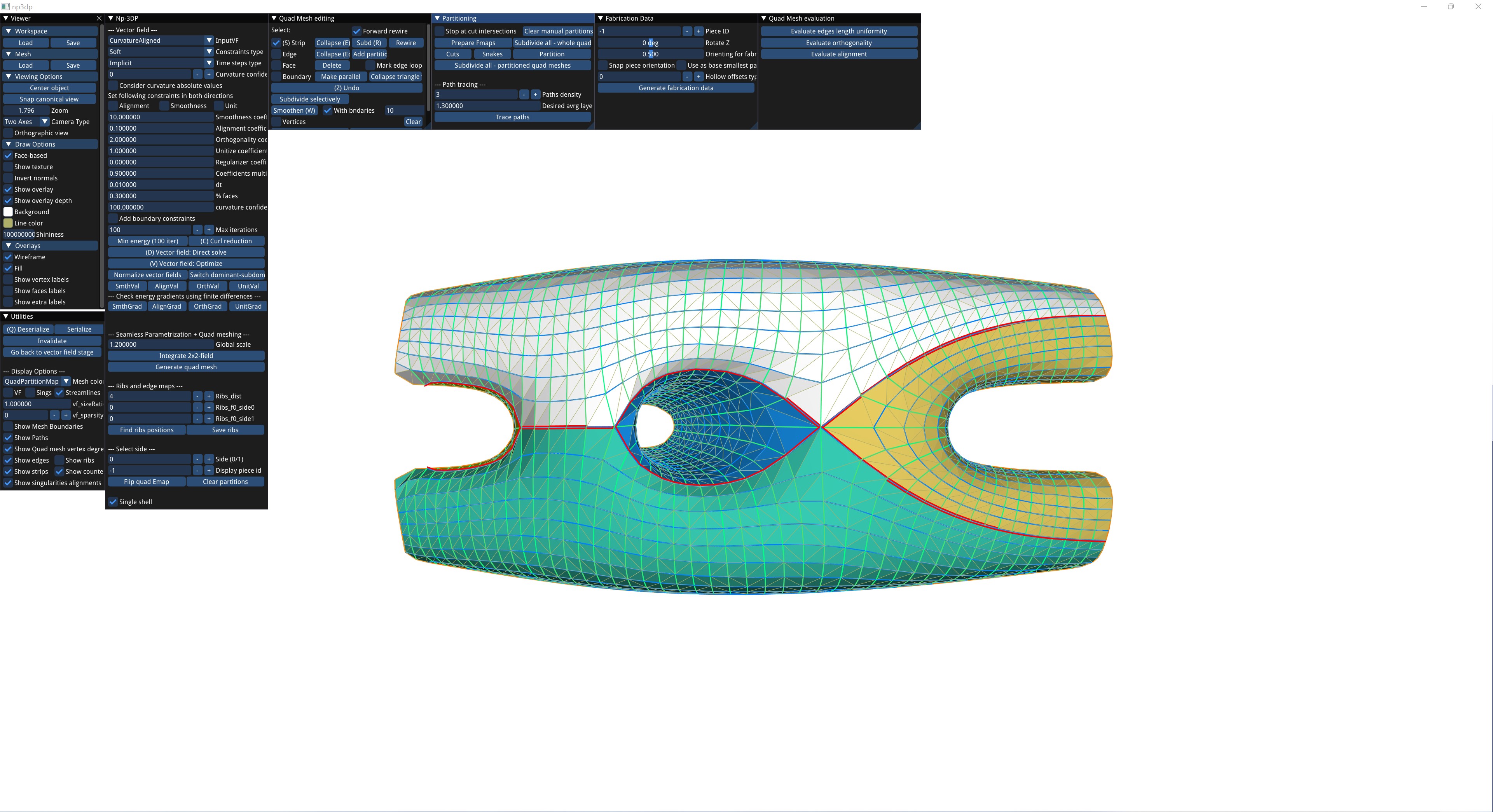}};

        \node[inner sep=0pt, draw=white, anchor=south west] (m) at (0.40\textwidth, -0.60\textwidth)  {\includegraphics[trim=440 10 550 170, clip, width=0.19\textwidth]{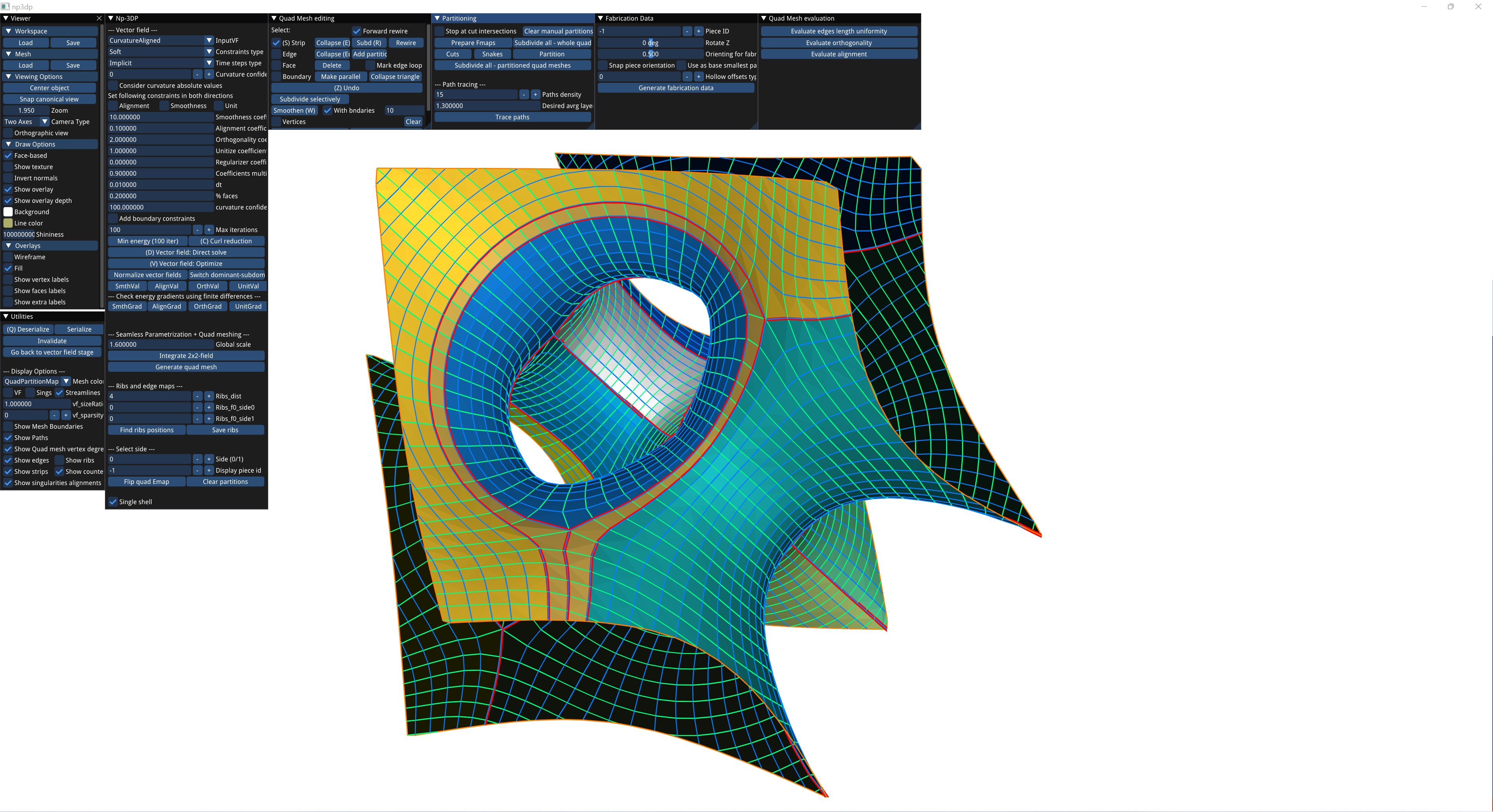}};
        \draw[->, ultra thick, black] ($(m.south)-(0.0,0.25)$) -- ($(m.south)-(0.0,1.0)$); 
        \node[anchor=north, xshift=-14pt, yshift = -10pt, align=center] at (m.south) {\tiny{\#1  op}};
        \node[inner sep=0pt, draw=white, anchor=south west] (m2) at (0.40\textwidth, -0.86\textwidth)  {\includegraphics[trim=440 10 550 170, clip, width=0.19\textwidth]{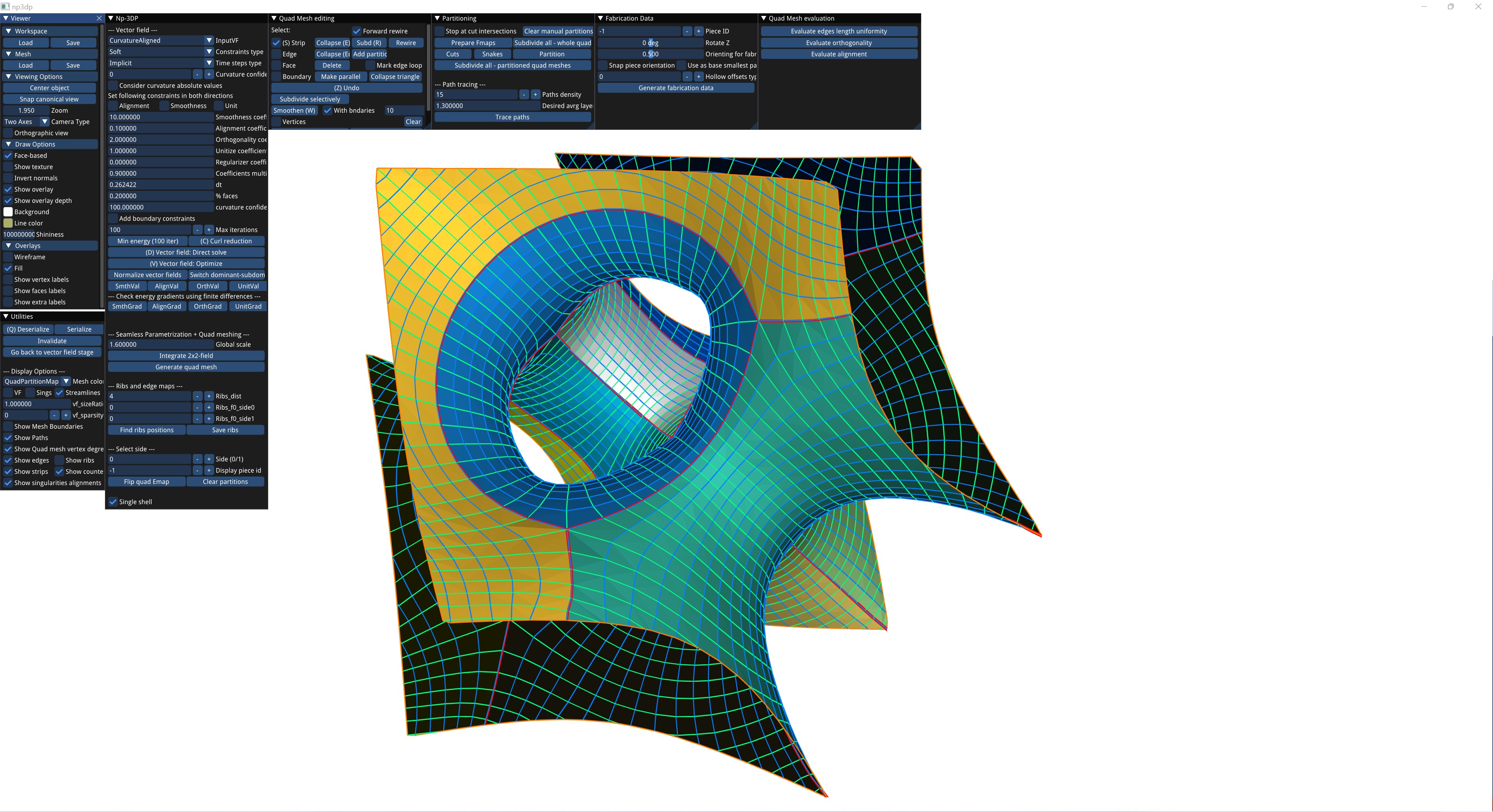}};

        \node[inner sep=0pt, draw=white, anchor=south west] (n) at (0.60\textwidth, -0.60\textwidth)  {\includegraphics[trim=570 20 390 100, clip, width=0.19\textwidth]{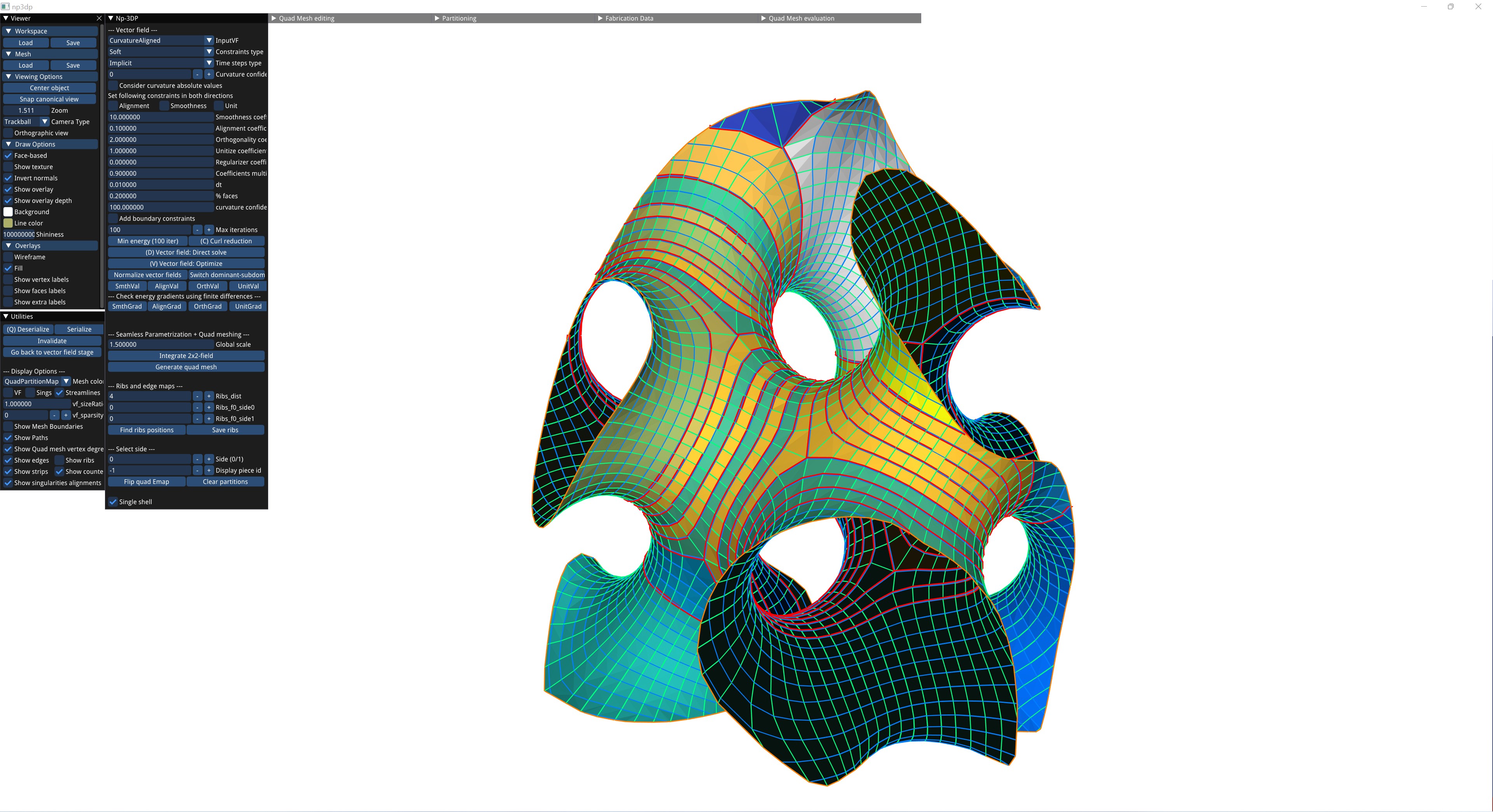}};
        \draw[->, ultra thick, black] ($(n.south)-(0.0,0.25)$) -- ($(n.south)-(0.0,1.0)$); 
        \node[anchor=north, xshift=-16pt, yshift = -10pt, align=center] at (n.south) {\tiny{\#12  ops}};
        \node[inner sep=0pt, draw=white, anchor=south west] (n2) at (0.60\textwidth, -0.86\textwidth)  {\includegraphics[trim=570 20 390 100, clip, width=0.19\textwidth]{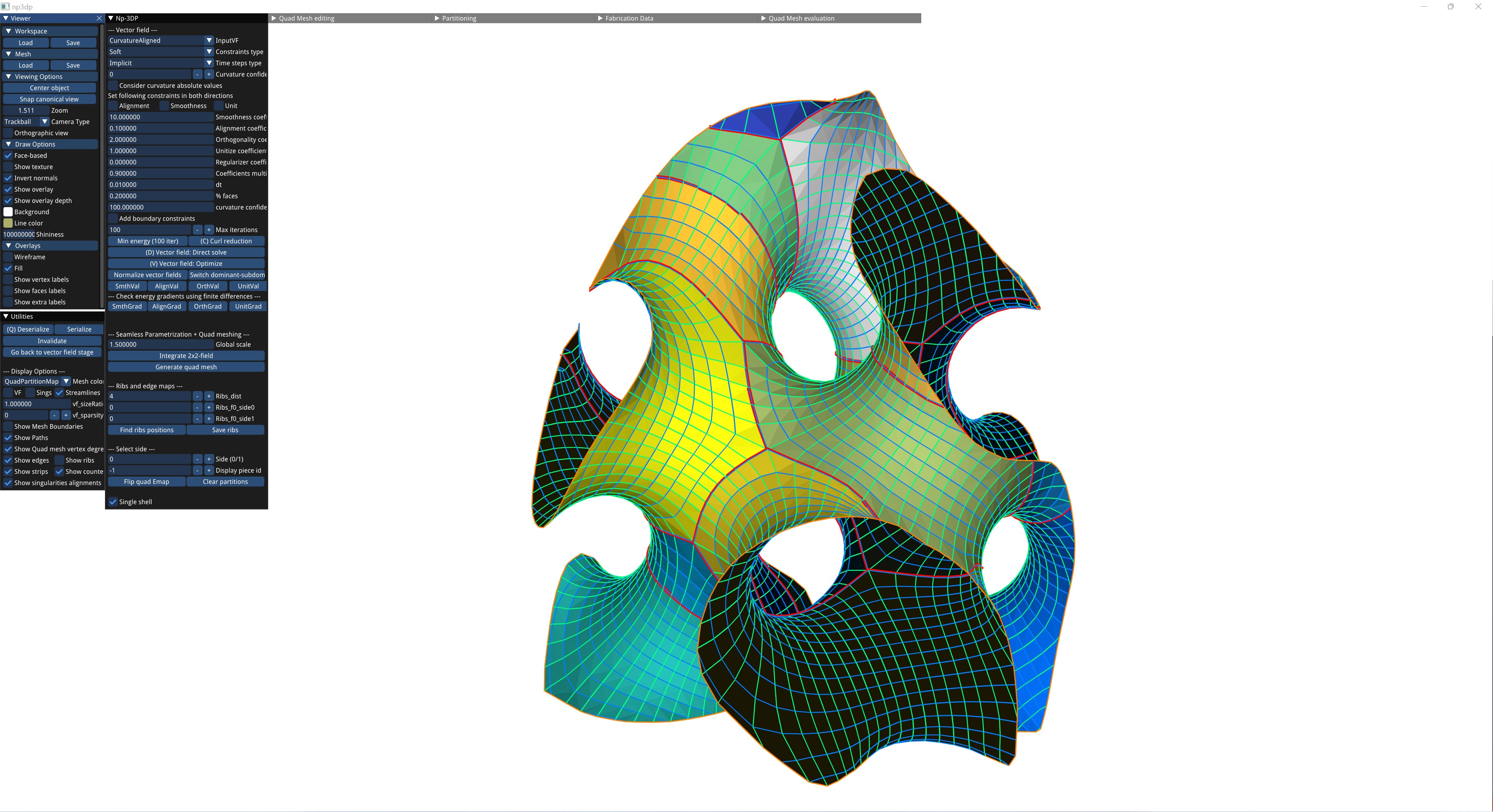}};

        \node[inner sep=0pt, draw=white, anchor=south west] (o) at (0.80\textwidth, -0.60\textwidth)  {\includegraphics[trim=500 20 490 175, clip, width=0.19\textwidth]{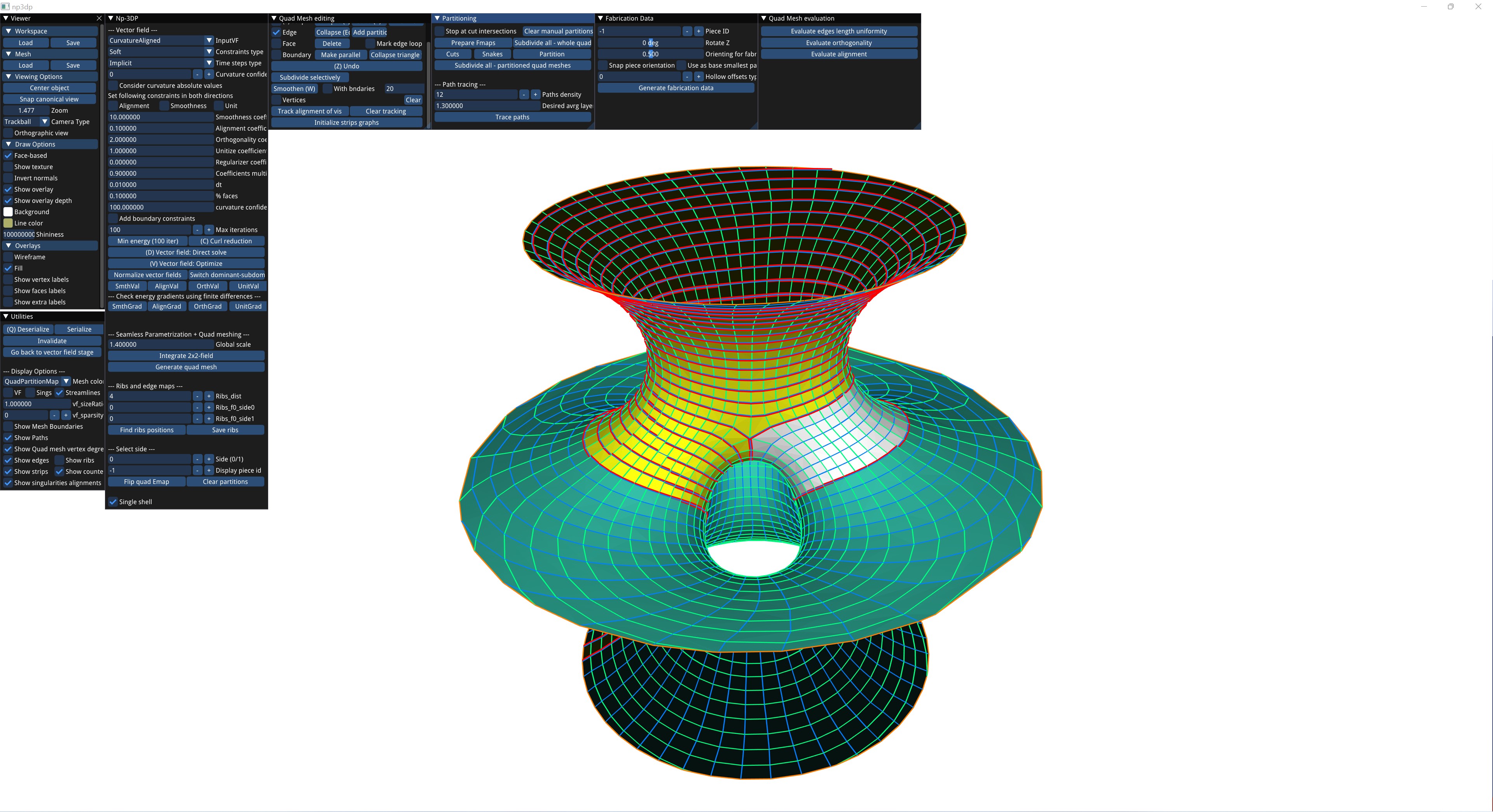}};
        \draw[->, ultra thick, black] ($(o.south)-(0.0,0.25)$) -- ($(o.south)-(0.0,1.0)$); 
        \node[anchor=north, xshift=-14pt, yshift = -10pt, align=center] at (o.south) {\tiny{\#1  op}};
        \node[inner sep=0pt, draw=white, anchor=south west] (o2) at (0.80\textwidth, -0.86\textwidth)  {\includegraphics[trim=500 20 490 175, clip, width=0.19\textwidth]{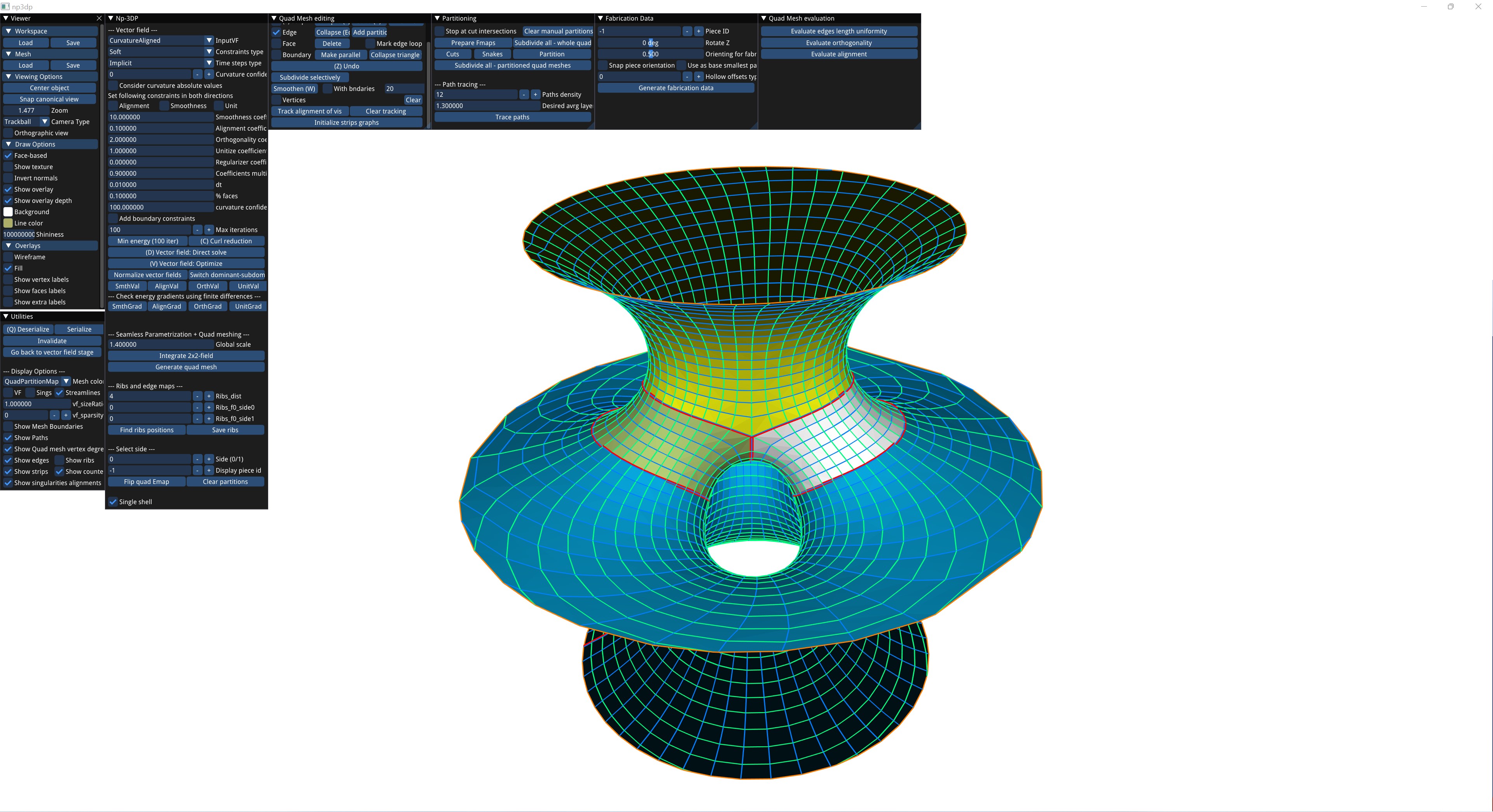}};

        \node[anchor=north, yshift=-3pt] at (k2.south) {\footnotesize{(k)}};
        \node[anchor=north, yshift=-3pt] at (l2.south) {\footnotesize{(l)}};
        \node[anchor=north, yshift=-3pt] at (m2.south) {\footnotesize{(m)}};
        \node[anchor=north, yshift=-3pt] at (n2.south) {\footnotesize{(n)}};
        \node[anchor=north, yshift=-3pt] at (o2.south) {\footnotesize{(o)}};

        \newcommand{\customGraph}[3]{
            \begin{tikzpicture}
                \begin{axis}[  width=0.155\textwidth, height=0.17\textwidth,
                    ybar=0pt,
                    ymin=0,
                    symbolic x coords={$\mathcal{L}$, $\mathcal{A}$, $\mathcal{O}$},
                    tick label style={font=\fontsize{4}{1}},
                    label style={font=\fontsize{4}{1}},
                    tickwidth=2.5pt
                    ]
                    \addplot[fill=orange, draw=orange, ybar, bar width=6pt, bar shift=-0.05*\pgfplotbarwidth] coordinates {($\mathcal{L}$, #1)};
                    \addplot[fill=purple, draw=purple, ybar, bar width=5pt, bar shift=0.0*\pgfplotbarwidth] coordinates {($\mathcal{A}$, #2)};
                    \addplot[fill=brown, draw=brown, ybar, bar width=6pt, bar shift=0.05*\pgfplotbarwidth] coordinates {($\mathcal{O}$, #3)};
            
                \end{axis}
            \end{tikzpicture}
        }

        \newcommand{\placeCustomGraph}[4]{
            \begin{scope}[shift={(#1.south)}]
                \node[anchor=south] at (26pt, -37pt) { \customGraph{#2}{#3}{#4} };
            \end{scope}
        }

        \placeCustomGraph{a}{0.268}{0.454}{0.089} 
        \placeCustomGraph{b}{0.157}{0.043}{0.053} 
        \placeCustomGraph{c}{0.209}{5.331}{0.120} 
        \placeCustomGraph{d}{0.251}{0.505}{0.059} 
        \placeCustomGraph{e}{0.178}{1.623}{0.077} 

        \placeCustomGraph{f}{0.233}{1.454}{0.118} 
        \placeCustomGraph{g}{0.238}{0.140}{0.084} 
        \placeCustomGraph{h}{0.191}{0.766}{0.042} 
        \placeCustomGraph{i}{0.278}{0.111}{0.106} 
        \placeCustomGraph{j}{0.119}{3.409}{0.036} 

        \placeCustomGraph{k}{0.230}{0.451}{0.074} 
        \placeCustomGraph{l}{0.298}{0.806}{0.068} 
        \placeCustomGraph{m}{0.216}{0.078}{0.103} 
        \placeCustomGraph{n}{0.221}{0.177}{0.080} 
        \placeCustomGraph{o}{0.372}{9.484}{0.127} 

        \placeCustomGraph{k2}{0.225}{1.659}{0.091} 
        \placeCustomGraph{l2}{0.284}{0.935}{0.093} 
        \placeCustomGraph{m2}{0.230}{0.050}{0.061} 
        \placeCustomGraph{n2}{0.243}{0.352}{0.096} 
        \placeCustomGraph{o2}{0.377}{9.540}{0.106} 

    \end{tikzpicture}
    
    \caption{Resulting parametrization and patch layout for different input surfaces using curvature-aligned directional constraints. The evaluation metrics $\mathcal{L}$, $\mathcal{A}$, $\mathcal{O}$ are displayed in the bar graphs for each shape. (a-j) The resulting patch layout did not require any topological modifications. (k-o) Topological modifications are used to simplify the patch layout and remove winding strips. The editing operations have the following effect on the evaluation measures; $\mathcal{L}$ and $\mathcal{O}$ remain constant or are reduced, while $\mathcal{A}$ increases, i.e., the edge length uniformity and the orthogonality are improved, while the alignment error increases. }
  \label{fig:gallery}
\end{figure}

We present a gallery of the resulting SDQ meshes and their patch layout along the $U$ direction for a series of input meshes using curvature-aligned directional constraints (\figref{fig:gallery}). The bar graphs indicate the evaluation of the metrics on the gallery shapes, and table \ref{table:performance} shows the details of the models and the algorithm's performance. The editing operations have the following effect on the evaluation measures; $\mathcal{L}$ and $\mathcal{O}$ remain constant or are reduced, while $\mathcal{A}$ increases. This means that the edge length uniformity and the orthogonality are improved, an effect caused by the iterative smoothing operations, while the alignment error increases, since the editing operations and smoothing slightly alter the orientations of the mesh edges.  

\paragraph{Alignment to principal curvature directions}
The alignment of the guiding fields to principal curvature directions is probably the most common scenario from the alignment options we present. However, in exceptional cases, it can produce undesirable results. This happens for meshes with large flat surfaces and undulating details, where the principal curvatures flip signs.
\begin{wrapfigure}{r}{0.5\textwidth}
    \includegraphics[width=0.5\textwidth]{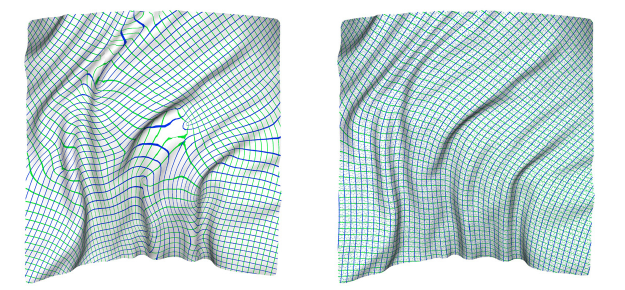}
    \vspace{-25pt}
\end{wrapfigure}
The cloth surface (see inset) is one such example; our alignment with signed principal curvatures produces the result on the left. To correct this, we need to align instead with absolute curvature directions, where it is not the sign but the magnitude that defines the minimum and the maximum direction (inset, right).

\begin{table}[h!]
\centering
\vspace{20pt}
\begin{tiny}
\setlength{\tabcolsep}{3pt} 
\begin{tabular}{c | c c c c c c c c c}
  \toprule
{model}  &  
\vtop{\hbox{\strut input tri mesh}\hbox{\strut  \#V , \#F}}  & 
\vtop{\hbox{\strut field}\hbox{\strut  iter.}}  & 
\vtop{\hbox{\strut field}\hbox{\strut  time } \hbox{\strut (s)}} & 
\#sing. &  
\vtop{\hbox{\strut quad mesh}\hbox{\strut  \#V , \#F}} & 
\vtop{\hbox{\strut \#patches}\hbox{\strut  (initial)}} & 
\#ops &
\vtop{\hbox{\strut \#patches}\hbox{\strut  (final)}} & 
\vtop{\hbox{\strut tracing}\hbox{\strut  8 paths/quad}\hbox{\strut  time (ms)}} \\
\midrule
(a) & 1653 , 3136     & 20       &  3.34    & 0   & 1055 , 980      & 1  & 0  & -  & 28 \\ 
(b) & 4716 , 9186     & 18       &  8.42    & 1   & 2271 , 2147     & 3  & 0  & -  & 262 \\ 
(c) & 2887 , 5592     & 26       &  7.23    & 1   & 1659 , 1579     & 3  & 0  & -  & 134 \\ 
(d) & 2105 , 4092     & 25       &  4.99    & 3   & 1293 , 1228     & 4  & 0  & -  & 95 \\ 
(e) & 3260 , 6144     & 79       &  18.48   & 8   & 2156 , 2016     & 6  & 0  & -  & 22 \\ 
(f) & 2623 , 5014     & 30       &  7.03    & 4   & 3949 , 3804     & 4  & 0  & -  & 402 \\ 
(g) & 2495 , 4725     & 27       &  5.97    & 4   & 2752 , 2534     & 8  & 0  & -  & 123 \\ 
(h) & 3954 , 7606     & 15       &  5.34    & 1   & 886  , 793      & 3  & 0  & -  & 35 \\ 
(i) & 4144 , 8221     & 29       &  13.81   & 14  & 2568 , 2510     & 9  & 0  & -  & 130 \\ 
(j) & 1935 , 3791     & 22       &  3.85    & 2   & 1525 , 1479     & 3  & 0  & -  & 14 \\ 

(k) & 2072 , 3963     & 21       &  3.81    & 1   & 1214 , 1129     & 7  & 4  & 3  & 26 \\ 
(l) & 1753 , 3330     & 26       &  4.14    & 3   & 1783 , 1660     & 7  & 3  & 5  & 89 \\ 
(m) & 4759 , 9150     & 28       &  20.84   & 9   & 3470 , 3282     & 11 & 1  & 9  & 425 \\ 
(n) & 1653 , 3136     & 26       &  8.43    & 17  & 4931 , 4655     & 62 & 12 & 16 & 21 \\ 
(o) & 3714 , 7240     & 26       &  9.16    & 4   & 2688 , 2624     & 45 & 1  & 5  & 142 \\ 

\end{tabular}
\end{tiny}
\bigskip
\caption{ Details for all the presented results in \figref{fig:gallery}. The examples were run on a Windows 11 laptop with an AMD Ryzen 9 6900HS CPU, an NVIDIA GeForce RTX 3080 GPU and 32GB of RAM. The performance of the integration and meshing processes are not included as they are done with an external library \cite{vaxman2016_directional-field-synthesis-design-and-processing}.}
\label{table:performance}
\end{table}

In addition, we present two examples of applying our meshing
workflow to more typical test geometries within graphics using
alignment to principal curvature directions. These geometries have no open boundaries, therefore, they are not within the scope of geometries we investigate. In such geometries, the field optimization and meshing step lead to good 
results (see inset). However, it is usually hard or impossible to reach a good patch layout using editing operations. Due to the large number of singularities and lack of boundaries where strips can terminate, most strips are winding around the mesh, making the application of editing operations difficult as each edit alters a large part of the mesh with results that are hard to control. The presented examples display the strips layout directly after integration without any editing operations applied.

\begin{figure}[h!]
  \centering
  \includegraphics[width=0.75\linewidth]{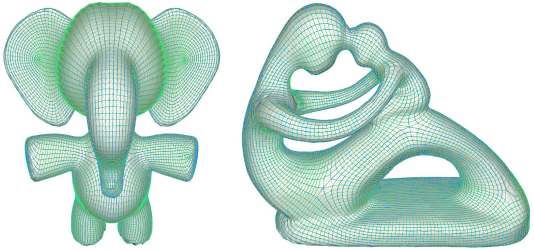}
  \caption{Application of meshing workflow on graphics meshes without open boundaries. Several winding strips are created due to the lack of boundaries where strips can terminate, making the use of editing operations difficult. }
  \label{fig:full_editing_operations_pipeline_1}
\end{figure}


\subsection{Application of SDQ meshes for non-planar 3D printing}

We present a case study for applying the presented SDQ meshes to design non-planar print paths for robotic 3D printing of shell surfaces. 

\subsubsection{Characteristics of printing method}
\begin{wrapfigure}{r}{0.4\textwidth}
    \vspace{-35pt}
    \includegraphics[width=0.4\textwidth]{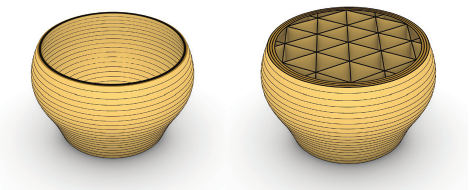}
    \vspace{-35pt}
\end{wrapfigure}
We focus on printing \emph{shells} (inset, left), where only the surface of the object is printed, as opposed to printing volumes (inset, right), where multiple offsets and infill structures are created in the interior. While a closed, watertight input shape is required for printing a volume, shell printing is more commonly applied to shapes with open boundaries.

In particular, we investigate the 3D printing of \emph{standing} shells (inset, left), where only the first path of the print lies on external support, while all 
\begin{wrapfigure}{r}{0.4\textwidth}
    \vspace{-15pt}
    \includegraphics[width=0.4\textwidth]{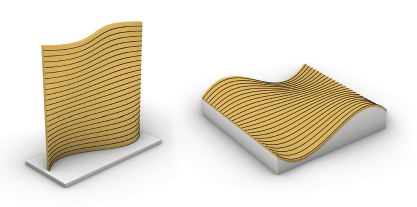}
    \vspace{-25pt}
\end{wrapfigure}
subsequent paths are supported by their preceding paths. This contrasts with the common ``horizontal'' printing of shells (inset, right), where the entire surface lies on existing support \cite{tam2017_additive-manufacturing-along-principal-stress, Bi2021}. This technique can save on material, time, and energy required for printing, as it requires considerably less external support. However, the printing of standing shells \cite{mitropoulou2020_print-paths-key-framing:-design-for-non-planar, zhong2020_ceramic-3d-printed-sweeping-surfaces} introduces an additional requirement for producing a feasible sequencing of the paths so that each path being printed is both accessible by the robotic arm, and fully supported by previously completed paths. 

\begin{wrapfigure}{r}{0.4\textwidth}
    \vspace{-15pt}
    \includegraphics[width=0.4\textwidth]{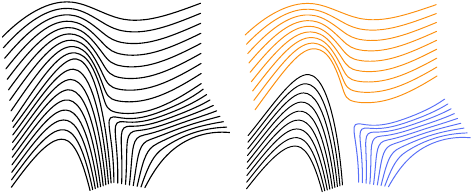}
    \vspace{-25pt}
\end{wrapfigure}
In general, strip networks with branched connectivity (inset-left) do not have a feasible print sequence for the entire shape, meaning that they cannot be printed in one go as a standing shell.  
The partitioning process described in \secref{subsection:patitioning} decomposes the strip network into simply-connected patches (inset-right), where each has a feasible sequence. These are then printed separately and assembled afterwards.

In addition to the topological cuts that separate the shape into simply-connected patches, in certain cases geometric cuts must also be introduced to split partitions into smaller pieces; This might be necessary for the following reasons, either the patches are too large to fit within the space of the fabrication setup, or they have an angle variation that exceeds what can be fabricated with the extruder. We document the number of geometric cuts for the presented prototypes in table \ref{table:prints_details}.

\subsubsection{Use of SDQ meshes}

\begin{wrapfigure}{r}{0.5\textwidth}
    \vspace{-5pt}
    \includegraphics[width=0.5\textwidth]{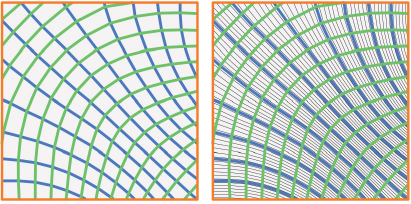}
    \vspace{-30pt}
\end{wrapfigure}

An SDQ mesh consists of the overlay of two transversal networks of strips. We chose to align the paths along the $U$ (blue) direction and use the $V$ (green) direction to design rigidifying ribs on the printed surface.  

\begin{wrapfigure}{r}{0.5\textwidth}
    \vspace{-5pt}
    \includegraphics[width=0.5\textwidth]{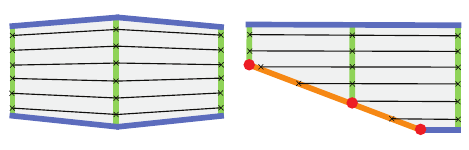}
    \vspace{-30pt}
\end{wrapfigure}

To trace the print paths along the dominant direction we subdivide each strip's subdominant edges $N$ times and connecting the subdivision points (see inset). On the boundaries of the mesh (shown in orange), we consider natural boundary conditions.

\subsubsection{Prototypes}

\begin{figure}[h!]
  \centering
  \includegraphics[width=\linewidth]{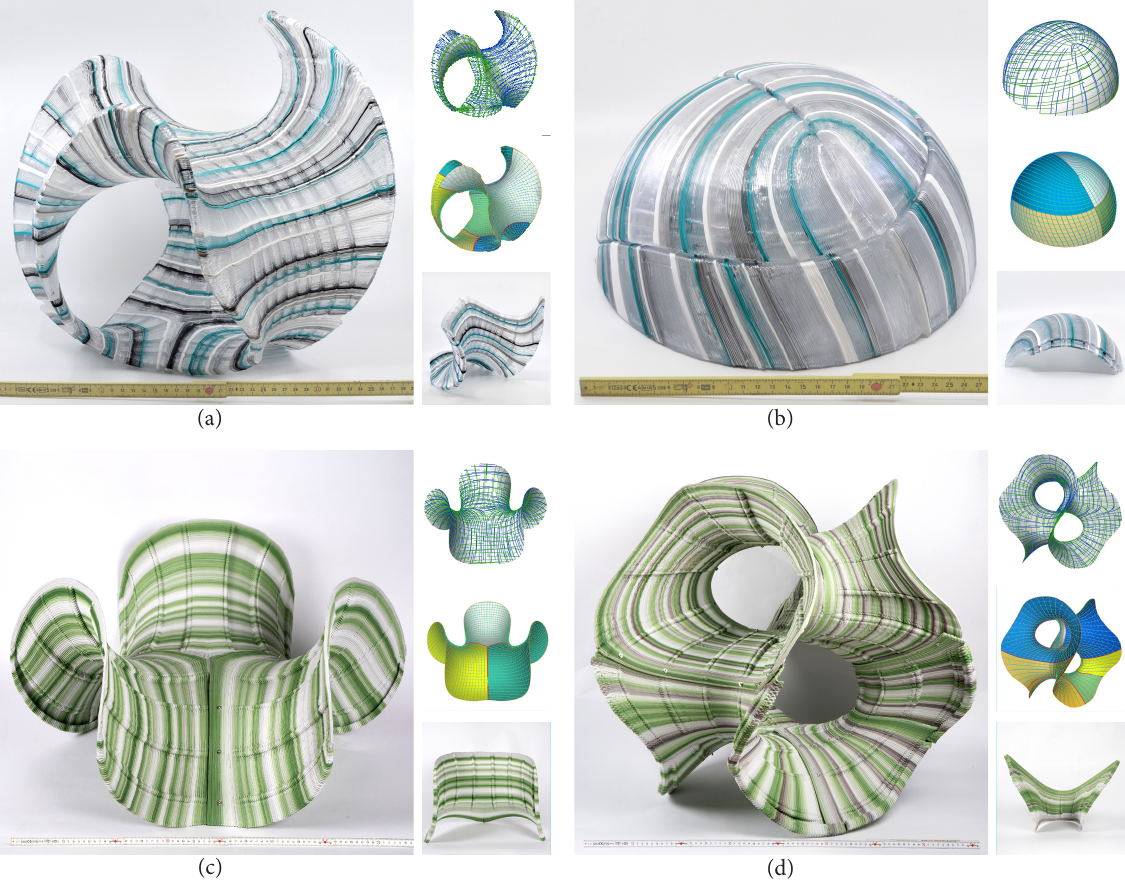}
  \caption{Fabricated prototypes. (a) TPMS using boundary-aligned directional constraints. (b) Half-sphere using user-drawn directional constraints. (c) Enepper surface, and (d) Batwing using curvature aligned directional constraints. On the right of each prototype, top: vector field streamlines, middle: partitioned SDQ mesh, bottom: one piece before assembly. 
  Photo credits: (a,b) Ioanna Mitropoulou, (c,d) Dominik Vogel
  }
  \label{fig:prints}
\end{figure}

\begin{figure}[h!]
  \centering
  \includegraphics[width=\linewidth]{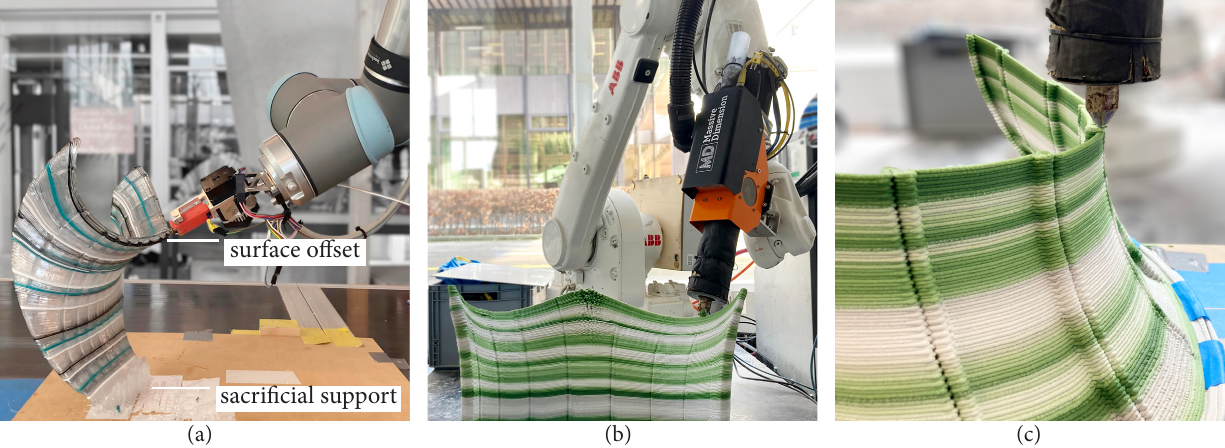}
  \caption{Print process. (a) Printing with a UR10 robot, extruding PETG filament. (b,c) Printing with an IRB1600 ABB robot, extruding PIPG with glass fibers using the MDPH2 pellet extruder.}
  \label{fig:print_process}
\end{figure}

We present a series of fabricated prototypes that showcase the physical results that can be achieved with the presented methods. 
The TPMS (\figref{fig:prints}a) and half-sphere (\figref{fig:prints}b) models were printed with a UR10 robot, using a plastic extruder tool that prints filament of polyethylene terephthalate glycol (PETG) with diameter 2.75m, using a 2.5 mm aperture nozzle (\figref{fig:print_process}a). The Enepper (\figref{fig:prints}c) and Batwing (\figref{fig:prints}d) surfaces were printed with an IRB1600 ABB robot, using an MDPH2 pellet extruder \cite{Massive_Dimension_pellet_extruder} with a 3mm aperture nozzle, that prints recycled PIPG pellets reinforced with glass fiber (\figref{fig:print_process}b,c). Table \ref{table:prints_details} provides more information on the logistics of the printing process for each prototype.

The first path of each patch is printed on sacrificial support, and all subsequent paths are printed support-free. The surfaces are printed with an offset of size negligible to the surface's scale to increase the final print's stability (\figref{fig:print_process}a). 
The change of printing color is carried out manually without stopping the printing process, and its purpose is to differentiate the paths visually.

\begin{table}[h!]
\centering
\begin{tiny}
\setlength{\tabcolsep}{3pt} 
\begin{tabular}{c | c c c c c c}
\toprule
{model}  &  \vtop{\hbox{\strut dimensions}\hbox{\strut  (cm)}} & \#sings  &  \vtop{\hbox{\strut \#geometric}\hbox{\strut cuts}} &\#pieces  &  \vtop{\hbox{\strut total print}\hbox{\strut time (hrs)}}  &  \vtop{\hbox{\strut \% sacrificial}\hbox{\strut support}}  \\
\midrule
\figref{fig:prints}a   & 38 x 38 x 38  & 2  & 3  & 6   & 18.3   & 16\% \\
\figref{fig:prints}b   & 30 x 30 x 15  & 2  & 0  & 4   & 10.3   & 11\% \\
\figref{fig:prints}c   & 70 x 70 x 45  & 1  & 0  & 3   & 12.5   & 29\% \\
\figref{fig:prints}d   & 76 x 76 x 76  & 3  & 6  & 9   & 27.0   & 39\% \\
\end{tabular}
\end{tiny}
\caption{Details of printed prototypes presented in \figref{fig:prints}.}
\label{table:prints_details}
\end{table}

\begin{wrapfigure}{r}{0.35\textwidth}
    \vspace{-15pt}
    \includegraphics[width=0.35\textwidth]{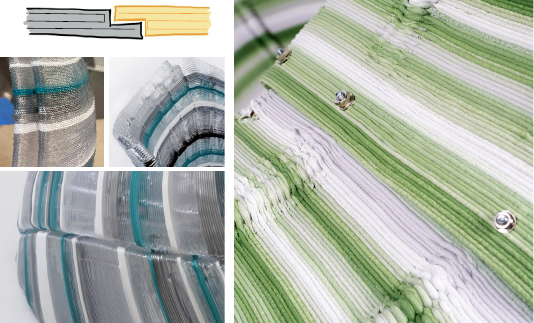}
    \vspace{-20pt}
\end{wrapfigure}
We create L-shaped connections between adjacent pieces to facilitate the assembly while allowing for tolerances. Then the pieces of smaller prototypes are connected with hot glue (inset, left), while larger prototypes are connected with screws (inset, right). 


\subsection{Relation to existing methods}

\begin{figure}[h!]
  \centering
  \includegraphics[width=1.0\linewidth]{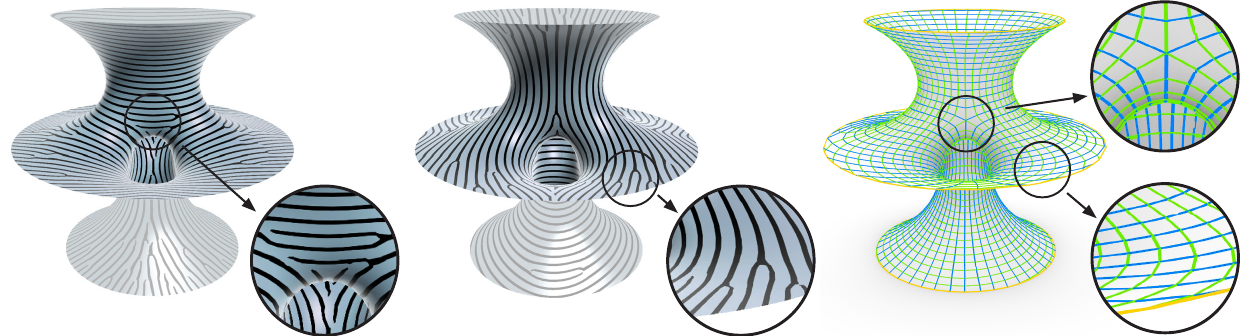}
  \caption{
      Comparison with \emph{Stripe Patterns on Surfaces} \cite{knoppel2015_stripe-patterns-on-surfaces} using alignment to principal curvature directions on the Costa minimal surface. Left and middle: their result, the magnified views show strip discontinuities. Right: our result.
    }
  \label{fig:comparison_to_stripes}
\end{figure}

In \emph{Stripe Patterns on Surfaces}, \cite{knoppel2015_stripe-patterns-on-surfaces} propose a method for synthesizing stripe patterns on triangulated surfaces, where stripe discontinuities are automatically inserted to achieve user-specified orientation and line spacing.
While the discontinuities improve the alignment and spacing between paths, they create interruptions that are problematic for various fabrication scenarios (\figref{fig:comparison_to_stripes} left, middle). In our method (\figref{fig:comparison_to_stripes} right), we sacrifice perfect alignment and uniform spacing to ensure a continuous parametrization so that no strip is interrupted in the interior of a patch. In addition, our discretization in an SDQ mesh enables the editing of the resulting strip networks. 

\begin{figure}[h!]
  \centering
  \includegraphics[width=\linewidth]{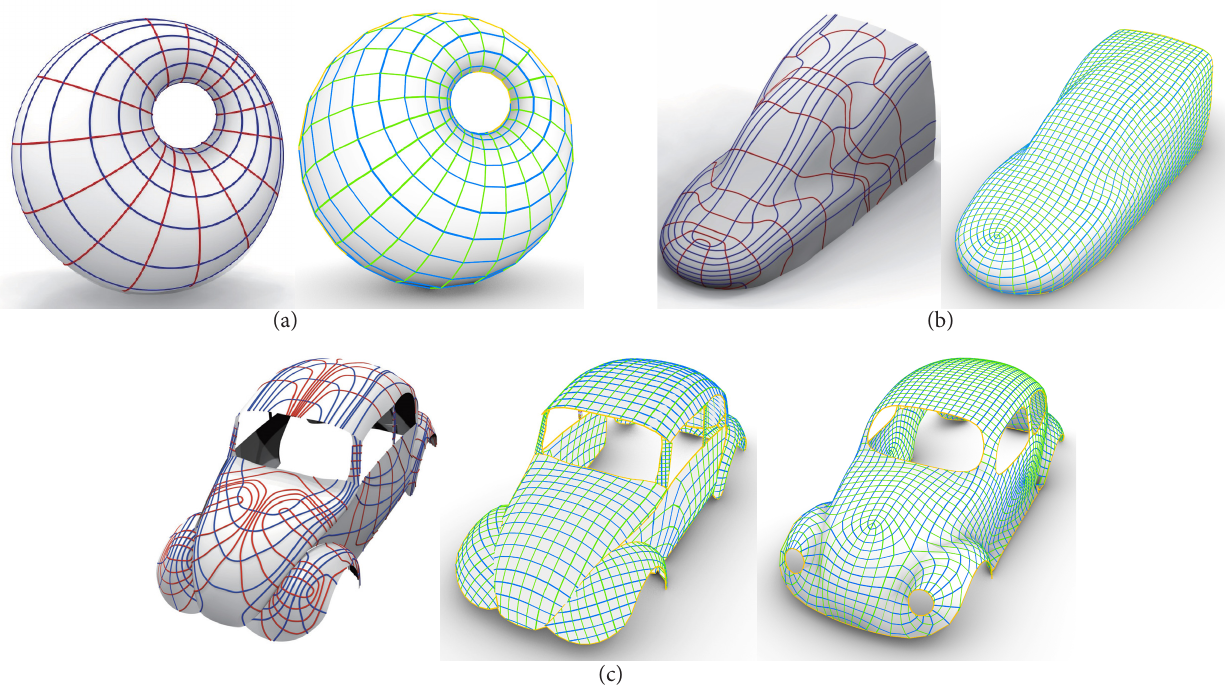}
  \caption{Comparison with \emph{Fabrication of Freeform Objects by Principal Strips} \cite{Takezawa_2016_Fabrication_of_Freeform_Objects_by_Principal_Strips}, For every examples, their result is left and our is right,  using alignment to principal curvature directions. (a) Dupin model. (b) Bullet train model. (c) Beetle model (left and middle: manually pre-segmented, right: original model). For the parametrization of (c)-right we have constrained 50\% of the model's faces.}
  \label{fig:comparison_to_principal_strips}
\end{figure}

In \emph{Fabrication of Freeform Objects by Principal Strips},
\cite{Takezawa_2016_Fabrication_of_Freeform_Objects_by_Principal_Strips} propose a method for designing orthogonal strips along the principal curvature directions of freeform surfaces, to be fabricated from flat paper sheets. In contrast to our approach, they trace the streamlines to achieve lines of curvature, which are sensitive to local bumps and need a density control method to ensure a uniform distribution over the surface. 
 In \figref{fig:comparison_to_principal_strips}, we compare visually their results to ours. In our method, we achieve a more uniform spacing of strips. In addition, complex models such as the beetle (\figref{fig:comparison_to_principal_strips}c) need to be pre-segmented manually into topological discs in \cite{Takezawa_2016_Fabrication_of_Freeform_Objects_by_Principal_Strips} before strips can be generated. In contrast, we can also parametrize such models as a whole (\figref{fig:comparison_to_principal_strips}c - right). 

\begin{figure}[h!]
  \centering
  \includegraphics[width=\linewidth]{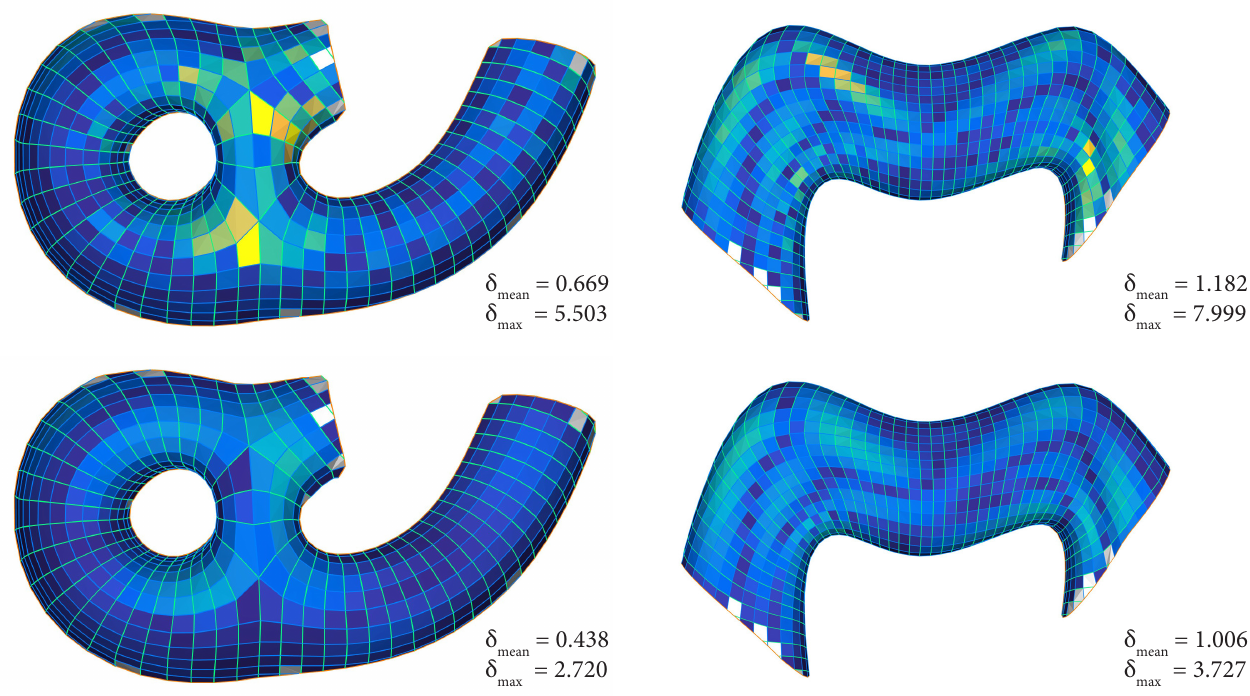}
  \caption{
  Top: Original SDQ mesh produced using alignment to principal curvature directions. Bottom: Optimized mesh for quad planarity. We measure $\delta$ by the percentage of the distance between the diagonals divided by the average diagonal length, as implemented in
\cite{vaxman2017_libhedra:-geometric-processing-and-optimization-of-polygonal}.
  }
  \label{fig:planarity}
\end{figure}

In \emph{Geometric Modeling with Conical Meshes and Developable Surfaces},
\cite{Liu_2006_Conical_meshes_and_developable_surfaces} propose a perturbation method for iteratively planarizing the faces of a quad mesh. The authors suggest that an ideal starting point for this optimization is a conjugate curve network, similar to what we produce when aligning our vector field to the principal curvature directions. Planarity is a very important property for various fabrication workflows. For example, strips that consist of planar quads are developable, meaning that they can be produced from flat sheets. When using alignment to principal curvature directions, the quads we produce can be expected to be nearly planar because we use conjugate curve networks to generate them. An additional optimization such as the one described in \cite{Liu_2006_Conical_meshes_and_developable_surfaces} can then planarize the faces further (\figref{fig:planarity}). Note that in this example, we ignore the non-quad faces that occur on the boundaries, displayed in white. 

\section{Limitations and Future Work}
\label{sec:Conclusion}

Our pipeline successfully produces quality SDQ meshes for various shapes in a semi-automatic manner that can be controlled and edited by the user. We next note some limitations and future directions.

\paragraph{Editing operations}
The topology editing operations are essential in promoting a good patch layout of the produced meshes. However, there is no guarantee that a ``good'' set of topological operations always exists for unwinding all strips and producing the desired alignments. For example, applying operations on winding strips can cause major distortion, and a mesh might even consist \emph{only} of winding strips. Another example of an undesirable solution is when all rewiring operations are blocked to maintain specific alignments, and only collapse operations on long winding strips are available. The worst-case scenario is when the only available operation is the collapse of a winding strip that covers the entire shape, which reduces the mesh to a single point, thus canceling any potential benefits of the operation. In particular, surfaces without open boundaries where strips can terminate are more likely to exhibit many winding strips.

A potential solution could be to enforce alignment already on the stage of seamless parameterization. However, this comes with the challenge of automatically deciding which singularities should be aligned, which is a difficult combinatorial problem that might result in large integration errors.

In addition, post-editing smoothing operations increase the alignment error and can create distortions close to the boundaries of shapes, especially at non-quad faces. A large number of editing operations that require a lot of smoothing afterward can lead to considerable irregularities on the boundaries. 

\paragraph{Mesh quality depends on user input}
Unanticipated inputs (e.g., shapes without open boundaries or directional constraints that produce highly noisy fields) may lead to infeasible, visually unappealing, or impractical results. Typically, this would mean many singularities, a bad patch layout, or very noisy strip networks. In addition, if the scale of the parametrization, which the user chooses, is too low, then the quads resolution might not be sufficient to capture the input surface's features. 

\paragraph{Fabrication-related properties} We are optimizing our strip networks for a selection of desirable fabrication-related properties that we have found most commonly at the state of the art. However, various other properties can be embedded in the optimization, according to the fabrication scenario that the SDQ mesh is produced for. For example, alignment of strips so that their boundaries are approximately geodesic curves is useful for weaving surfaces out of ribbons that bend more readily out-of-plane than in-plane \cite{vekhter2019_weaving-geodesic-foliations}.

\paragraph{Alignment presets} We propose three alignment presets, namely alignment to curvature, boundaries, or user-drawn directions. Another useful option would be alignment to principal stress directions or principal moments given a chosen load case, which can be instrumental for improving the structural properties of an object in various fabrication scenarios \cite{Schiftner_2010_Statics-Sensitive_Layout_of_Planar_Quadrilateral_Meshes, fang2020_reinforced-fdm:-multi-axis-filament-alignment}.


 \bibliographystyle{elsarticle-harv} 
 \bibliography{bibliography}


\end{document}